\documentclass[aps,pra,groupedaddress,showpacs,twocolumn,nofootinbib,superscriptaddress]{revtex4-1}
\usepackage{times}
\usepackage{amsmath}
\usepackage{amssymb,bm}
\usepackage{amsthm,color,xcolor,dsfont}
\usepackage{graphicx} 
\usepackage[colorlinks=true, linkcolor=blue, urlcolor=blue, citecolor=blue, hyperindex, breaklinks]{hyperref}
\usepackage{capt-of}
\newtheorem*{definition*}{Definition}

\begin{document}

\title{Benchmarking the quantum cryptanalysis of symmetric, public-key and hash-based cryptographic schemes}

\author{Vlad Gheorghiu}
\email[Electronic address: ]{vlad.gheorghiu@uwaterloo.ca}
\affiliation{Institute for Quantum Computing, University of Waterloo, 
Waterloo, ON, N2L 3G1, Canada}
\affiliation{Department of Combinatorics \& Optimization, University of Waterloo, Waterloo, ON, N2L 3G1, Canada}
\affiliation{evolutionQ Inc., Waterloo, ON, Canada}
\affiliation{softwareQ Inc., Kitchener, ON, Canada}

\author{Michele Mosca}
\email[Electronic address: ]{michele.mosca@uwaterloo.ca}
\affiliation{Institute for Quantum Computing, University of Waterloo, 
Waterloo, ON, N2L 3G1, Canada}
\affiliation{Department of Combinatorics \& Optimization, University of Waterloo, Waterloo, ON, N2L 3G1, Canada}
\affiliation{Perimeter Institute for Theoretical Physics, Waterloo, ON, N2L 6B9, Canada}
\affiliation{Canadian Institute for Advanced Research, Toronto, ON,  M5G 1Z8, Canada}
\affiliation{evolutionQ Inc., Waterloo, ON, Canada}
\affiliation{softwareQ Inc., Kitchener, ON, Canada}


\begin{abstract}
Quantum algorithms can break factoring and discrete logarithm based cryptography and weaken symmetric cryptography and hash functions.

In order to estimate the real-world impact of these attacks, apart from tracking the development of fault-tolerant quantum computers it is important to have an estimate of the resources needed to implement these quantum attacks.

For attacking symmetric cryptography and hash functions, generic quantum attacks are substantially less powerful than they are for today's public-key cryptography. So security will degrade gradually as quantum computing resources increase. At present, there is a substantial resource overhead due to the cost of fault-tolerant quantum error correction. We provide estimates of this overhead using state-of-the-art methods in quantum fault-tolerance. For example, recent lattice surgery methods reduced memory costs by roughly a factor of 5 over previous methods.  Future advances in fault-tolerance and in the quality of quantum hardware may reduce this overhead further. Another part of the cost of implementing generic quantum attacks is the cost of implementing the cryptographic functions. We use state-of-the-art optimized circuits, though further improvements in their implementation would also reduce the resources needed to implement these attacks. To bound the potential impact of further circuit optimizations we provide cost estimates assuming trivial-cost implementations of these functions.  These figures indicate the effective bit-strength of the various symmetric schemes and hash functions based on what we know today (and with various assumptions on the quantum hardware), and frame the various potential improvements that should continue to be tracked. As an example, we also look at the implications for Bitcoin's proof-of-work system.

For many of the currently used asymmetric (public-key) cryptographic schemes based on RSA and elliptic curve discrete logarithms, we again provide cost estimates based on the latest advances in cryptanalysis, circuit compilation and quantum fault-tolerance theory. These allow, for example, a direct comparison of the quantum vulnerability of RSA and elliptic curve cryptography for a fixed classical bit strength.

This analysis provides state-of-the art snap-shot estimates of the realistic costs of implementing quantum attacks on these important cryptographic algorithms, assuming quantum fault-tolerance is achieved using surface code methods, and spanning a range of potential error rates. These estimates serve as a guide for gauging the realistic impact of these algorithms and for benchmarking the impact of future advances in quantum algorithms, circuit synthesis and optimization, fault-tolerance methods and physical error rates.
\end{abstract}

\maketitle

\tableofcontents

\section{Introduction\label{sct::intro}}
Symmetric, public-key (asymmetric) and hash-based cryptography constitute a fundamental pillar of modern cryptography. 
Symmetric cryptography includes symmetric-key encryption, where a shared secret key is used for both encryption and decryption. Cryptographic hash functions map arbitrarily long strings to strings of a fixed finite length. Currently deployed public-key schemes are
used to establish a common secret key between two remote parties. They are based on factoring large numbers or solving the discrete logarithm problem over a finite group. For more details about modern cryptography the interested reader can consult one of the many excellent references on the topic, e.g.~\cite{Katz:2007:IMC:1206501}.

In contrast to asymmetric schemes based on factoring or solving the discrete logarithm problem and which are completely broken by a quantum adversary via Shor's algorithm~\cite{SJC.26.1484}, symmetric schemes and hash functions are less vulnerable to quantum attacks. The best known quantum attacks against them are based on Grover's quantum search algorithm~\cite{PhysRevLett.79.325}, which offers a quadratic speedup compared to classical brute force searching. Given a search space of size $N$, Grover's algorithm finds, with high probability, an element $x$ for which a certain property such as $f(x)=1$ holds, for some function $f$ we know how to evaluate (assuming such a solution exists). The algorithm evaluates $f$ a total of $\mathcal{O}(\sqrt{N})$ times. It applies a simple operation in between the evaluations of $f$, so the $\mathcal{O}(\sqrt{N})$ evaluations of $f$ account for most of the complexity. In contrast, any classical algorithm that evaluates $f$ in a similar ``black-box'' way requires on the order of $N$ evaluations of $f$ to find such an element.

Any quantum algorithm can be mapped to a quantum circuit, which can be implemented on a quantum computer. The quantum circuit represents what we call the ``logical layer". Such a circuit can always be decomposed in a sequence of ``elementary 
gates", such as Clifford gates (CNOT, Hadamard etc.~\cite{NC00}) augmented by a non-Clifford gate such as the T gate.

Running a logical circuit on a full fault-tolerant quantum computer is highly non-trivial. The sequence of logical gates have to be mapped to 
sequences of surface code measurement cycles (see e.g.~\cite{PhysRevA.86.032324} for extensive details). By far, the most resource-consuming (in 
terms of number of qubits required and time) is the T gate\footnote{Clifford gates are ``cheap", i.e. they require relatively small overhead for implementation in the surface code, but are not universals, hence a non-Clifford gate is required. One such gate is the T gate. There are other possible choices, however all of the non-Clifford gates require special techniques such as magic state distillation~\cite{1367-2630-14-12-123011,PhysRevA.86.052329} and significant overhead (order of magnitudes higher than Clifford gates) to be implemented in the surface code. In fact, to a first order approximation, for the purpose of resource estimation, one can simply ignore the overhead introduced by the Clifford gates and simply focus only on the T gates.}. 
In comparison with surface code defects and braiding techniques~\cite{PhysRevA.86.032324}, novel lattice surgery 
techniques~\cite{2018arXiv180806709F,1808.02892,1367-2630-14-12-123011} reduce the spatial overhead required for implementing T gates via magic state distillation by approximately a factor of 5, while also modestly improving the running time. 

In this paper we first analyze the security of symmetric schemes and hash functions against large-scale fault-tolerant quantum adversaries, using surface code defects and braiding techniques. We take into account the time-space trade-offs with parallelizing quantum search, down to the fault-tolerant layer. Naively, one might hope that $K$ quantum computers (or quantum ``processors'', as we will call them later in the paper) running in parallel reduce the number the circuit depth down to $\mathcal{O}(\sqrt{N})/K$ steps, similar to the classical case of distributing a search space across $K$ classical processors. However quantum searching does not parallelize so well, and the required number of steps
for parallel quantum searching is of the order $\mathcal{O}(\sqrt{N/K})$~\cite{quantph.9711070}. This is a factor of $\sqrt{K}$ larger than $\mathcal{O}(\sqrt{N})/K$ . As shown in~\cite{quantph.9711070}, the optimal way of doing parallel quantum search is to partition the search space into $N/K$ parts, and to perform independent quantum searches on each part.

Secondly, we investigate the security of public-key cryptographic schemes such as RSA and ECC against 
quantum attacks, using the latest developments in theory of fault-tolerant quantum error correction, i.e. novel lattice surgery 
techniques~\cite{2018arXiv180806709F,1808.02892,1367-2630-14-12-123011}.

The remainder of this paper is organized as follows. In Sec.~\ref{sct::method}, we provide an overview of the methodology used in our analysis. In Sec.~\ref{sct::ciphers} we investigate the security of the AES family of modern symmetric ciphers. In Sec.~\ref{sct::hash} we analyze the security of the SHA family of hash functions. In Sec.~\ref{sct::bitcoin} we investigate the security of Bitcoin's~\cite{satoshi:bitcoin} proof-of-work consensus mechanism. We conclude our investigation of symmetric and hash-based cryptographic schemes in Sec.~\ref{sct::intrinsic_parallel_grover}, where we evaluate the intrinsic cost of running the Grover algorithm with a trivial oracle (i.e., an oracle with a unit cost of 1 for each invocation).

In the subsequent sections  we analyze public-key cryptographic schemes. In Sec.~\ref{sct::rsa} and Sec.~\ref{sct::ecc} we examine the most common public-key establishment schemes, such as RSA and ECC, respectively. In the subsequent sections  we analyze public-key cryptographic schemes. In Sec.~\ref{sct::rsa} and Sec.~\ref{sct::ecc} we examine the most common public-key establishment schemes, such as RSA and ECC, respectively. Finally we summarize our findings and conclude in Sec.~\ref{sct::conclusion}.
\section{Methodology\label{sct::method}}

\subsection{Symmetric cryptography and hash functions\label{sct::symmetric}}
\begin{figure}[htb]
	\centering
         \includegraphics[width=0.35\textwidth]{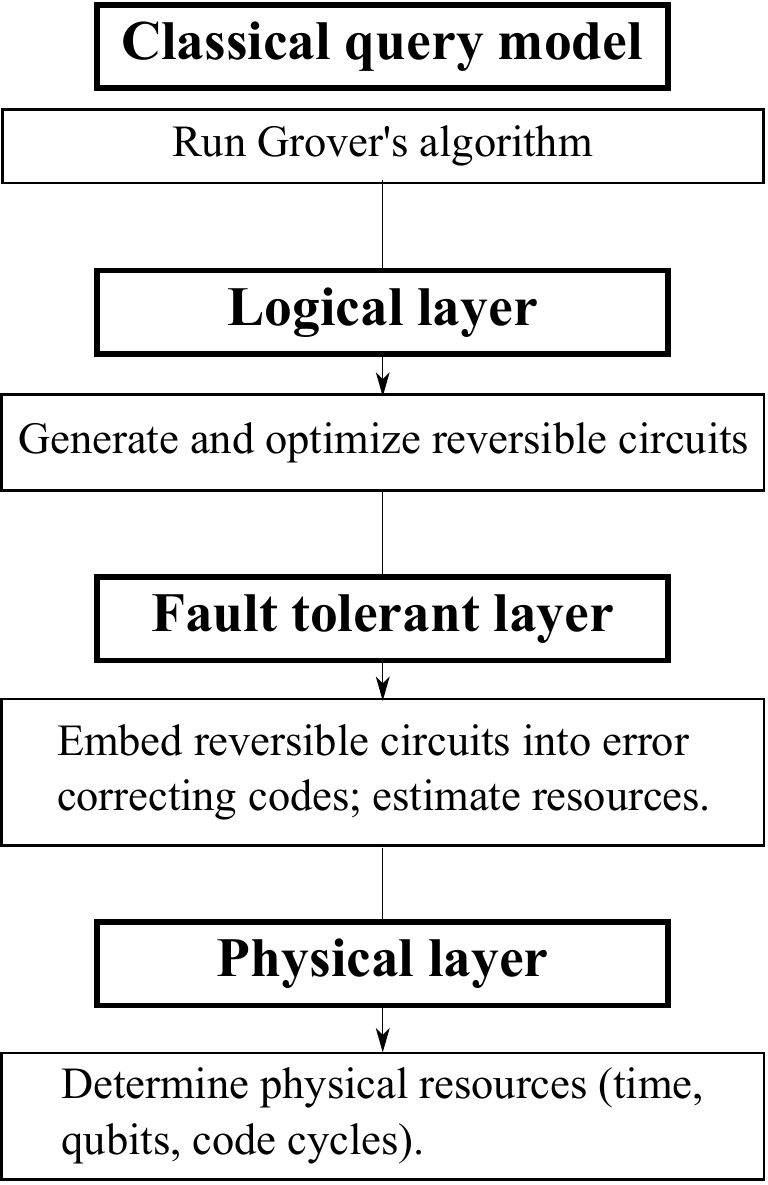}
         \caption{Analyzing an attack against a symmetric cryptographic function with a fault-tolerant quantum adversary. Our resource estimation methodology takes into account several of the layers between the high level description of an algorithm and the physical hardware required for its execution. Our approach is modular should assumptions about any of these layers change, and hence it allows one to calculate the impact of improvements in any particular layer.}
         \label{fgr:flowchart_lite}
\end{figure}

\begin{figure}
	\centering
	 \includegraphics[width=0.46\textwidth]{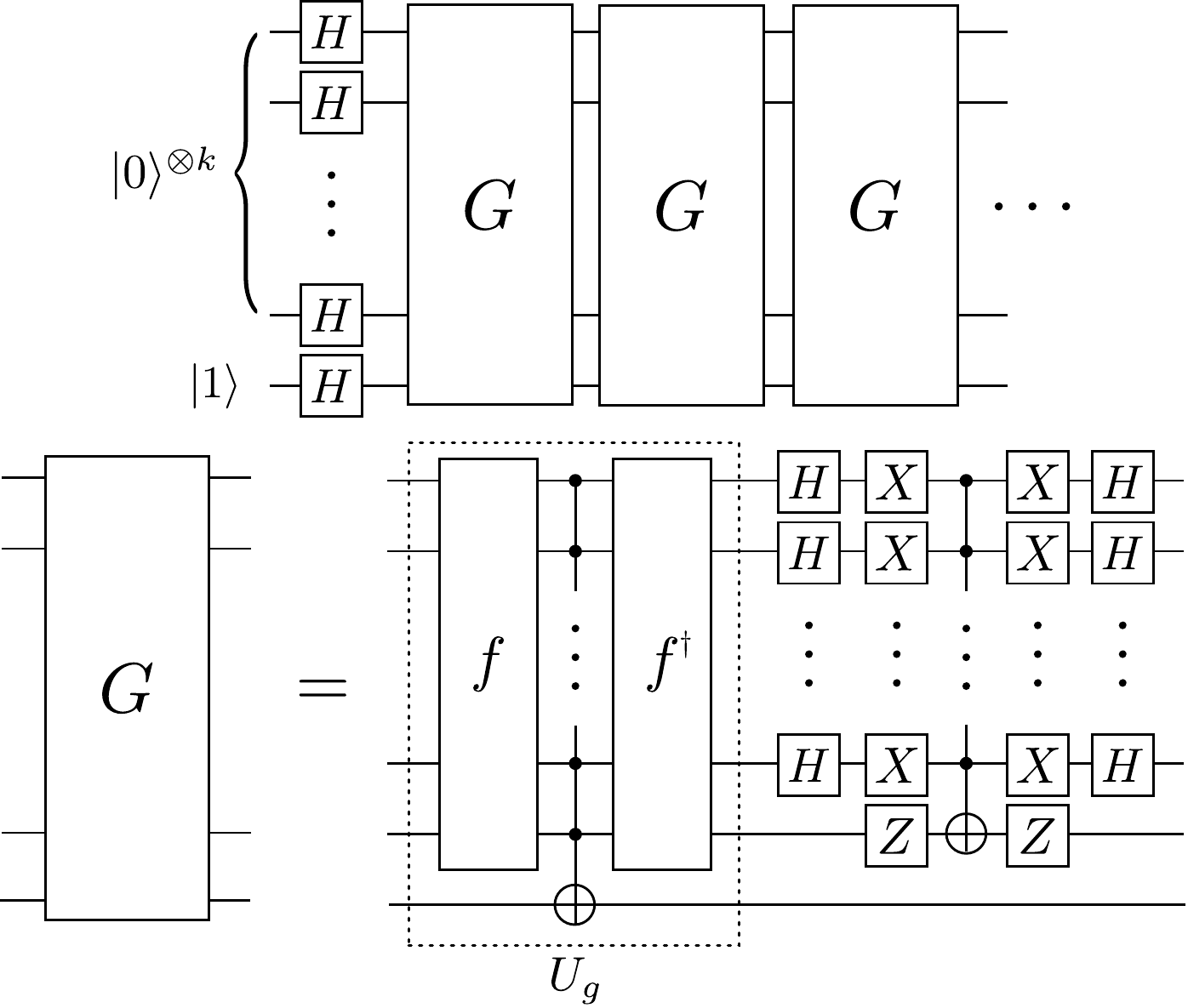}
          \caption{Grover searching with an oracle for $f : \{0,1\}^k \rightarrow \{0,1\}^k$. The algorithm makes $\lfloor \frac{\pi}{4} 2^{N/2}\rfloor$ calls to
$G$, the \emph{Grover iteration}, or, if parallelized on $K$ processors, $\lfloor \frac{\pi}{4} 2^{N/(2K)}\rfloor$ calls to $G$. The Grover iteration has two
subroutines. The first, $U_g$, implements the predicate $g : \{0,1\}^k
\rightarrow \{0,1\}$ that maps $x$ to $1$ if and only if $f(x) = y$. Each call to $U_g$ involves two calls to a reversible implementation of $f$ and one call to a comparison circuit that checks whether $f(x) = y$.}
          \label{fgr:full_algorithm}
\end{figure}

The methodology, sketched in Fig.~\ref{fgr:flowchart_lite} and Fig.~\ref{fgr:full_algorithm}, follows the same lines as the one described in great detail in our earlier paper~\cite{10.1007/978-3-319-69453-5_18}, which we refer the interested reader to for more details.

We assume a surface-code based fault-tolerant architecture~\cite{PhysRevA.86.032324}, using Reed-Muller distillation schemes~\cite{Fowler:2013aa}. For each scheme we vary the possible physical error rates per gate from $10^{-4}$ to $10^{-7}$. We believe that this range of physical error rates is wide enough to cover both first generation quantum computers as well as more advanced future machines. 
In comparison to surface code defects and braiding methods~\cite{PhysRevA.86.032324}, lattice surgery 
techniques~\cite{2018arXiv180806709F,1808.02892,1367-2630-14-12-123011} mostly impact the physical footprint of the fault-tolerant layer required to run a specific quantum algorithm, reducing the distillation overhead by approximately a factor of 5. The temporal overhead (i.e. the number of surface code cycles) is reduced less drastically. For this reason, lattice surgery has less significant effects in estimating the security of symmetric schemes or hash functions, reducing the security parameter\footnote{The security parameter is defined as the logarithm base two of the number of fundamental operations (in our case surface code cycles) required to break the scheme.} by at most 1 and decreasing the spatial overhead by at most a factor of 5. Therefore when estimating the security of symmetric and hash-based cryptographic schemes we use surface code defects and braiding techniques.

For each cryptographic primitive, we display four plots, in the following order:
\begin{enumerate}
\item We plot the total number of surface code cycles per CPU (where a CPU is a quantum computer capable of executing a single instance of Grover's quantum search algorithm) as a function of the number of CPUs. We directly tie the quantum security parameter to the total number of surface code cycles (see~\cite{10.1007/978-3-319-69453-5_18} for more details). We also add to the plot the theoretical lower bound achievable by quantum search in the cases of: a) considering the oracle a black box of unit cost (lower line), and b) considering the oracle as composed of ideal quantum gates, each of unit cost (upper line). Note that the difference between b) and a) represents the intrinsic cost of logical overhead (i.e. the overhead introduced by treating the oracle as a logical circuit and not a blackbox), whereas the difference between the upper lines and b) represents the intrinsic cost introduced by the fault-tolerant layer.

\item We plot the total wall-time per CPU (i.e. how long will the whole computation take on a parallel quantum architecture) as a function of the number of CPUs. The horizontal dashed line represents the one-year time line, i.e. the $x$ coordinate of the intersection point between the ``Total time per CPU'' line and the one-year time line provides the number of processors required to break the system within one year (in $\log_2$ units).

\item We plot the total physical footprint (number of qubits) per CPU, as a function of the number of CPUs.
\item Finally we plot the total physical footprint (number of qubits) of all quantum search machines (CPUs) running in parallel.
\end{enumerate}

In the following sections we proceed to analyze symmetric ciphers (AES, Sec.~\ref{sct::ciphers}), hash functions (SHA-256, SHA3-256, Sec.~\ref{sct::hash}, Bitcoin's hash function, Sec.~\ref{sct::bitcoin}), and finally the minimal resources required for running Grover's algorithm with a trivial oracle~\ref{sct::intrinsic_parallel_grover} (e.g. the identity gate) on search spaces of various sizes.

Note that in some ranges of the plots from sections~\ref{sct::ciphers},~\ref{sct::hash},~\ref{sct::intrinsic_parallel_grover} and~\ref{sct::bitcoin} the total physical footprint increases slightly with the number of processors, which may seem counter-intuitive. This happens due to the fact that with more processors the required code distances decrease, and in some instances one can pipeline more magic states factories in parallel into the surface code, which in effect causes an increase in the overall physical footprint. Note that the total time per CPU is monotonically decreasing, as parallelizing distilleries does not increase the wall time. For more details see~\cite{10.1007/978-3-319-69453-5_18}. 

\subsection{Public-key cryptography\label{sct::pk}}

Most of the recent progress in quantum cryptanalysis is related to the fault-tolerant layer in Fig.~\ref{fgr:flowchart_lite}. New methods and techniques
based on surface code lattice surgery~\cite{2018arXiv180806709F,1808.02892,1367-2630-14-12-123011} allow a significant decrease of the overall 
footprint (number of qubits, or space) taken by the quantum computation, and also a relatively modest decrease in time, in comparison with  methods based on surface code defects and braiding~\cite{PhysRevA.86.032324,Fowler:2013aa}.

We consider the best up-to-date optimized logical quantum circuits for attacking RSA and ECC public-key 
schemes~\cite{1706.06752,PhysRevA.52.3457,cuccaro04,Beauregard:2003:CSA:2011517.2011525} then perform a physical footprint resource estimation
analysis using lattice surgery techniques. We remark that the overall time required to run the algorithm depends on the level of parallelization 
for the magic state factories\footnote{Every T gate in the circuit must be implemented by a specialized magic state factory, each of which occupies a 
significant physical footprint. One can implement more magic states in parallel if one is willing to increase the physical footprint of the computation.}. 

For each public-key cryptogrpric scheme, we analyze the space/time tradeoffs and plot the results on a double logarithmic scale. We fit the data using a third degree 
polynomial\footnote{A third degree polynomial fits the data very precisely, providing a coefficient of determination $R^2$ greater than 0.997.} and obtain an analytical closed-form formula for the relation between the time and the number of qubits required to attack the scheme, in 
the form

\begin{equation}\label{eqn1}
y(x) = \alpha x^3 + \beta x^2 + \gamma x + \delta,
\end{equation}
where $y$ represents logarithm base 2 of the number of qubits and $x$ represents the logarithm base 2 of the time (in seconds). For example,
the quantity 
\begin{equation}\label{eqn2}
y\left(\log_2(24\times 3600)\right) \approx y(16.3987)
\end{equation}
represents how many qubits are required to break the scheme in one day (24 hours) for a fixed physical error rate per gate $p_g$, assuming a 
surface code cycle time of 200ns. Note that the computation time scales linearly with the surface code cycle time, e.g. a 1000ns surface code cycle 
time will result in a computation that is 5 times longer than a $200ns$ surface code cycle time. Therefore, for a specific cryptographic scheme for 
which we plotted the space/time tradeoffs using a surface code cycle time of $200ns$ and a fixed physical error rate per gate $p_g$, the number of 
qubits required to break a specific scheme in a time $t$ using an alternative surface code cycle time $t_c$ is given by

\begin{equation}\label{eqn3}
y\left(\log_2\left(\frac{200ns}{t_c}t\right)\right),
\end{equation}
where $t$ is expressed in seconds and $t_c$ is expressed in nanoseconds.

We assume a surface code cycle time of 200ns, in conformance with~\cite{PhysRevA.86.032324}. For each scheme we analyze, we compare its security using the more conservative (and realistic in the short term) $p_g=10^{-3}$ and also the more optimistic  $p_g=10^{-5}$. Note that assuming the more optimistic assumption from a quantum computing perspective is the more conservative assumption from a cybersecurity perspective.

Furthermore, in this analysis, we are reporting the full physical footprint, including the memory required for magic state distillation.
Using present-day techniques, the memory required for generating these generic input states accounts for a substantial fraction of the total memory cost and thus we are including these in the total cost estimate and will track the impact of improved methods.

\section{Symmetric ciphers\label{sct::ciphers}}
Below we analyze the security of AES family of symmetric ciphers against large-scale fault-tolerant quantum adversaries. We used the highly optimized logical circuits produced in
\cite{10.1007/978-3-319-29360-8_3}. 

\subsection{AES-128}

        \includegraphics[width=0.429\textwidth]{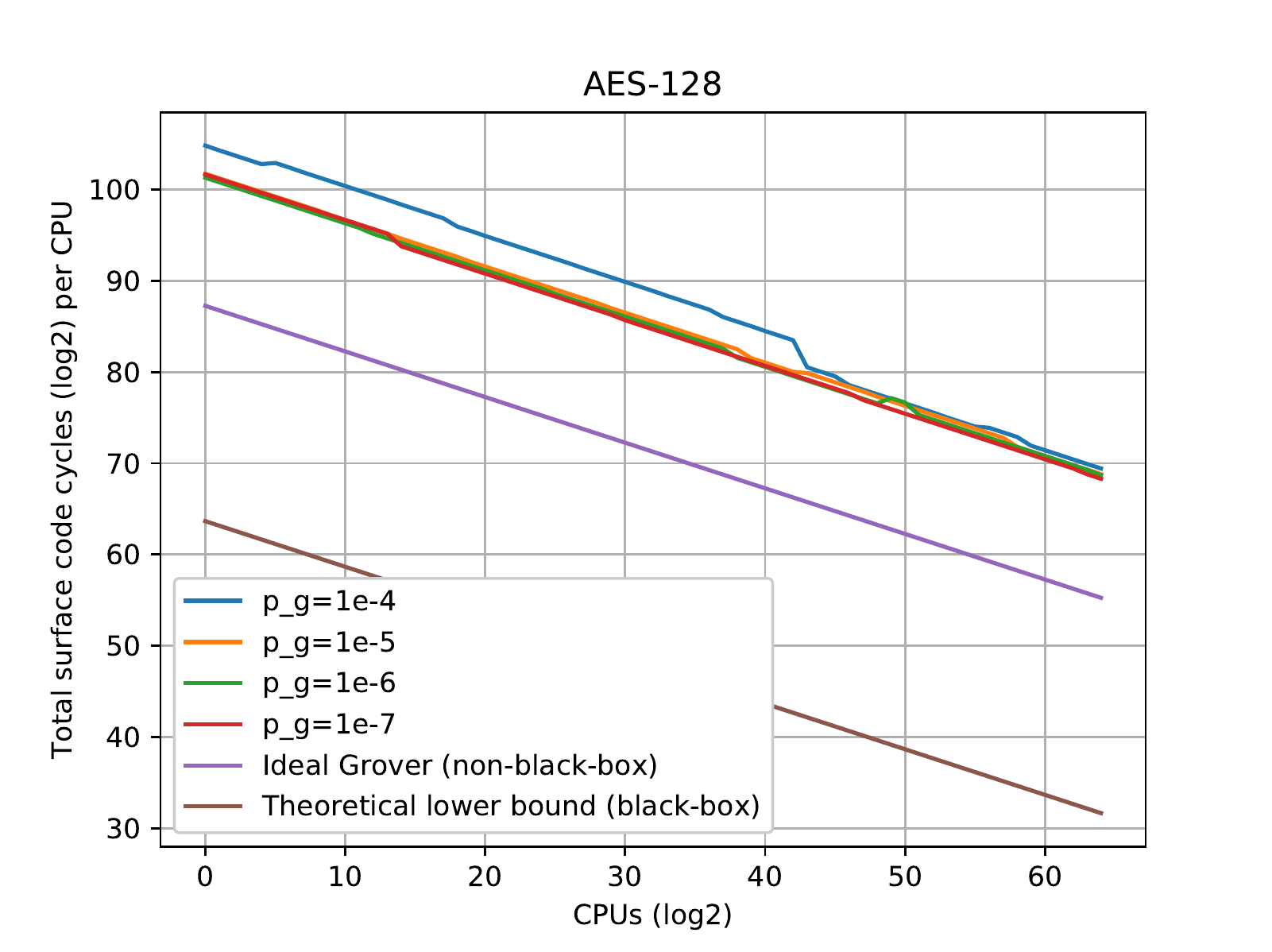}
      	\captionof{figure}{AES-128 block cipher. Required surface clock cycles per processor, as a function of the  number of processors ($\log_2$ scale). The bottom brown line (theoretical lower bound, black box) represents the minimal number of queries required
	by Grover's algorithm, the cost function being the total number of queries to a black-box oracle, each query assumed to have unit cost, and a completely error-free circuit. The purple line (ideal grover, non-black-box) takes into consideration the structure of the oracle, the cost function being the total number of gates in the circuit, each gate having unit cost; the quantum circuit is assumed error-free as well. Both brown and magenta lines are displayed only for comparisons; for both of them, the $y$ axis should be interpreted as number of logical queries (operations, respectively).	
The curves above the purple line show the overhead introduced by fault tolerance (in terms of required surface code cycles, each surface code cycle assumed to have unit cost). More optimization at the logical layer will shift the purple line down, whereas more optimization at the fault-tolerant layer will move the upper curves closer to the purple line. Similar remarks to the above hold for the remaining plots in this manuscript.}
      	\label{fgr:aes_128_cycles}
	
	For example, the plots in Fig.~\ref{fgr:aes_128_cycles} tells us that if we have $2^{50}$ quantum computers running Grover's algorithm in parallel, with no physical errors, then it would take about $2^{63}$ gate calls (where the purple line intersects the vertical line at $50$), where we assume each gate to have unit cost. Still with no errors, a trivial cost for implementing the cryptographic function (oracle) would bring the cost down to about $2^{38}$ oracle calls per quantum computer. Keeping the actual function implementation, but adding the fault-tolerant layer with a physical error rate of $10^{-7}$ (with appropriate assumptions and using state-of-the-art quantum error correction) pushes the cost up to around $2^{76}$ surface code cycles per quantum computer (where now each code cycle is assumed to have unit cost). Similar remarks hold for the remaining plots in this manuscript.
        \includegraphics[width=0.429\textwidth]{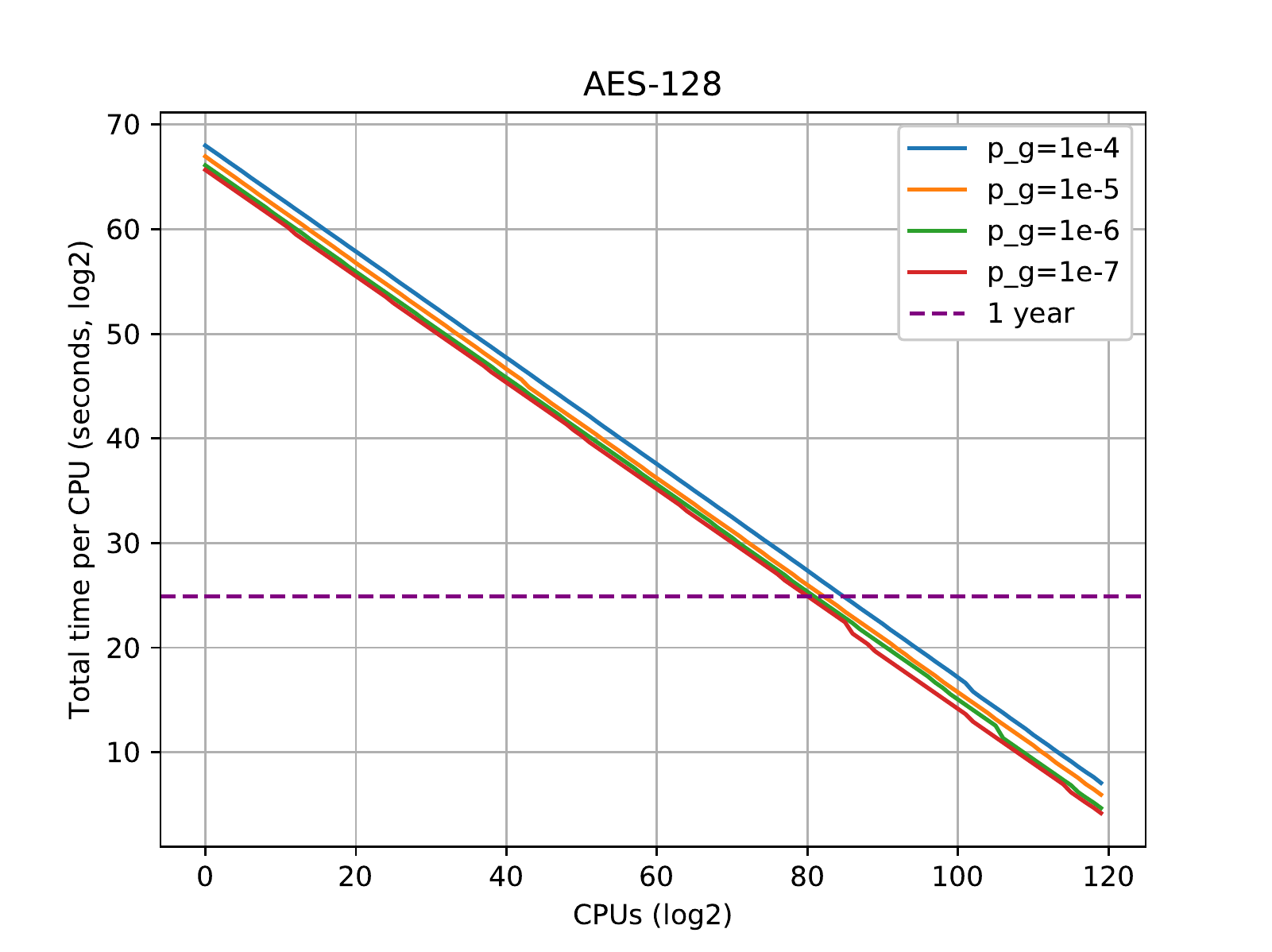}
      	\captionof{figure}{AES-128 block cipher. Required time per processor, as a function of the  number of processors ($\log_2$ scale). The horizontal dotted line indicates one year. The $x$-axis is deliberately extended to show the necessary number of CPUs for a total time of one year. Thus the figure shows that it would take, with the stated assumptions, over $2^{80}$ parallel quantum searches to break AES-128 in a year. Similar remarks to the above hold for the remaining plots in this manuscript.}
      	\label{fgr:aes_128_time}
        \includegraphics[width=0.429\textwidth]{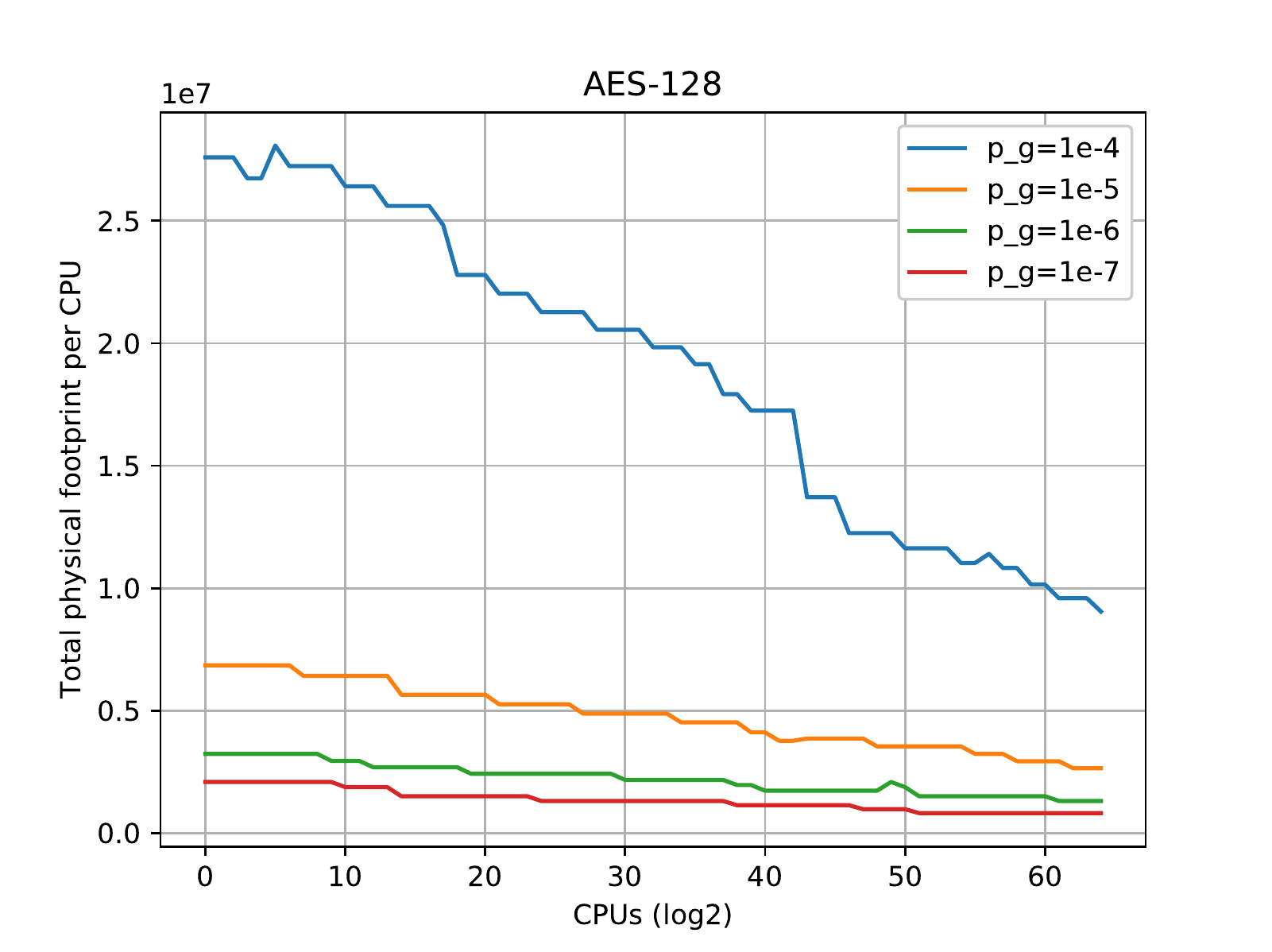}
	\captionof{figure}{AES-128 block cipher. Physical footprint (physical qubits) per processor, as a function of the number of processors ($\log_2$ scale).}
      	\label{fgr:aes_128_phys}
        \includegraphics[width=0.429\textwidth]{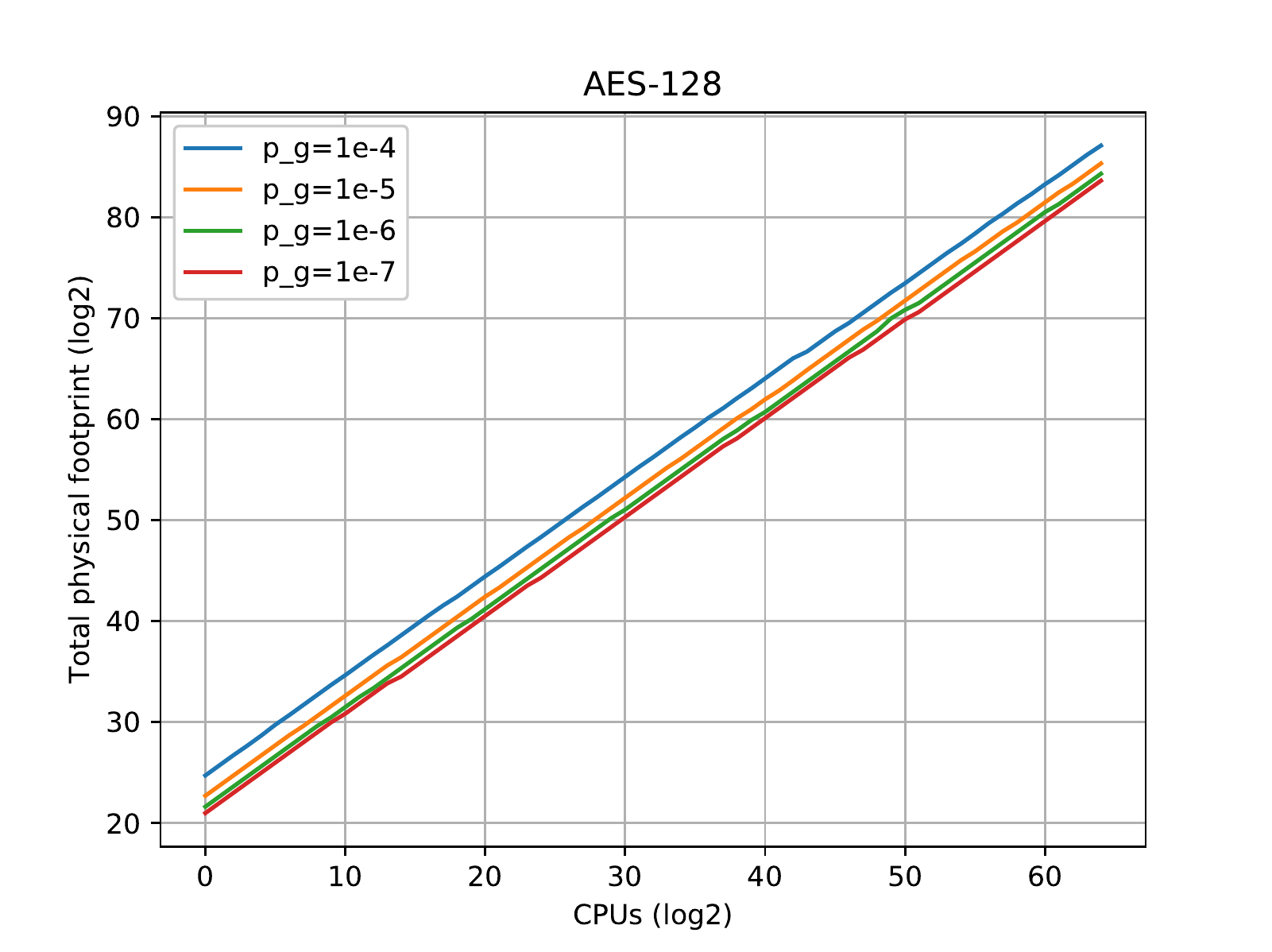}
	\captionof{figure}{AES-128 block cipher. Total physical footprint (physical qubits), as a function of the number of processors ($\log_2$ scale). Note that the qubits are not correlated across processors.}
      	\label{fgr:aes_128_phys_total}

\subsection{AES-192}

	
        \includegraphics[width=0.429\textwidth]{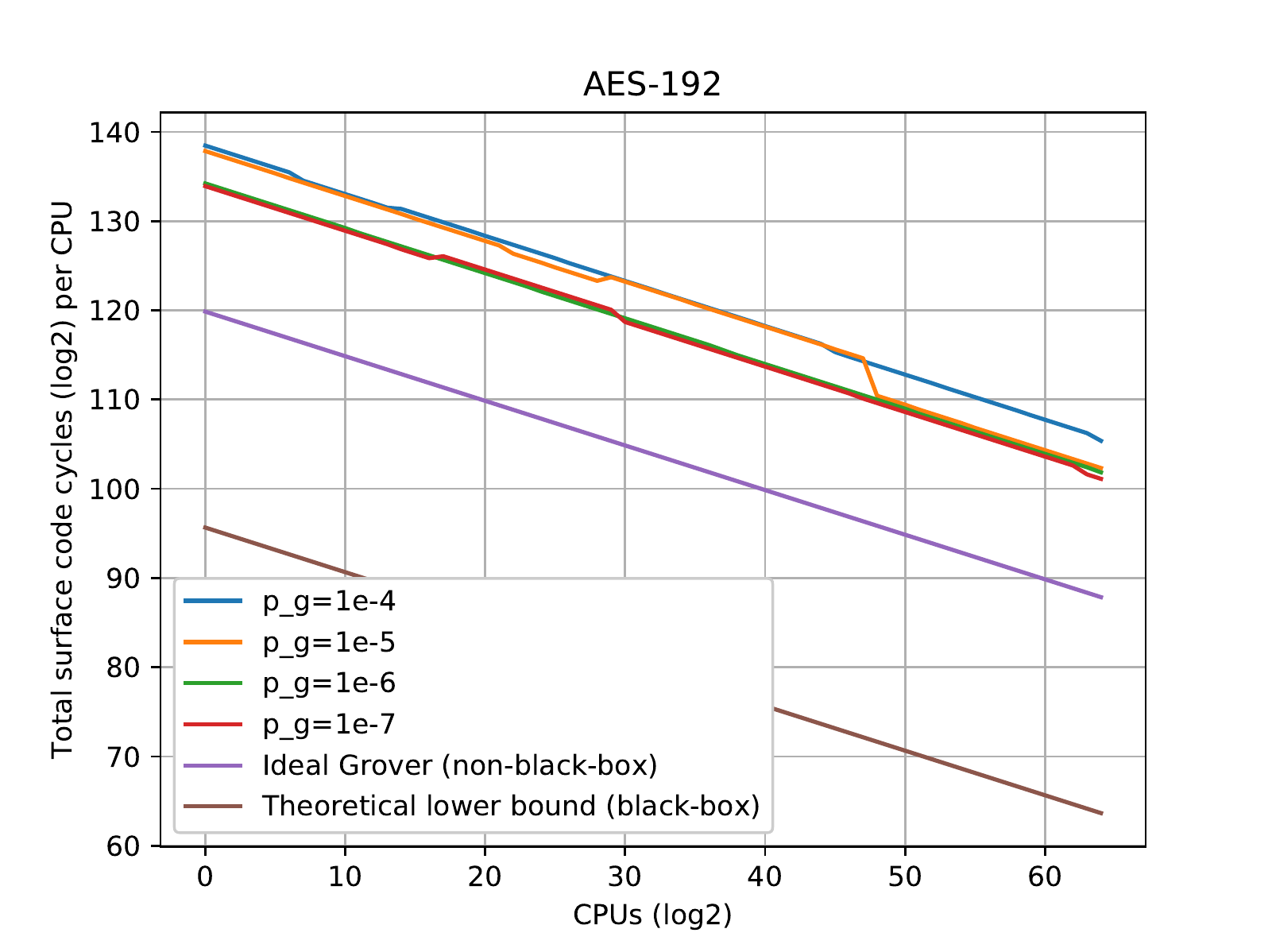}
      	\captionof{figure}{AES-192 block cipher. Required surface clock cycles per processor, as a function of the  number of processors ($\log_2$ scale).}
      	\label{fgr:aes_192_cycles}
	
        \includegraphics[width=0.429\textwidth]{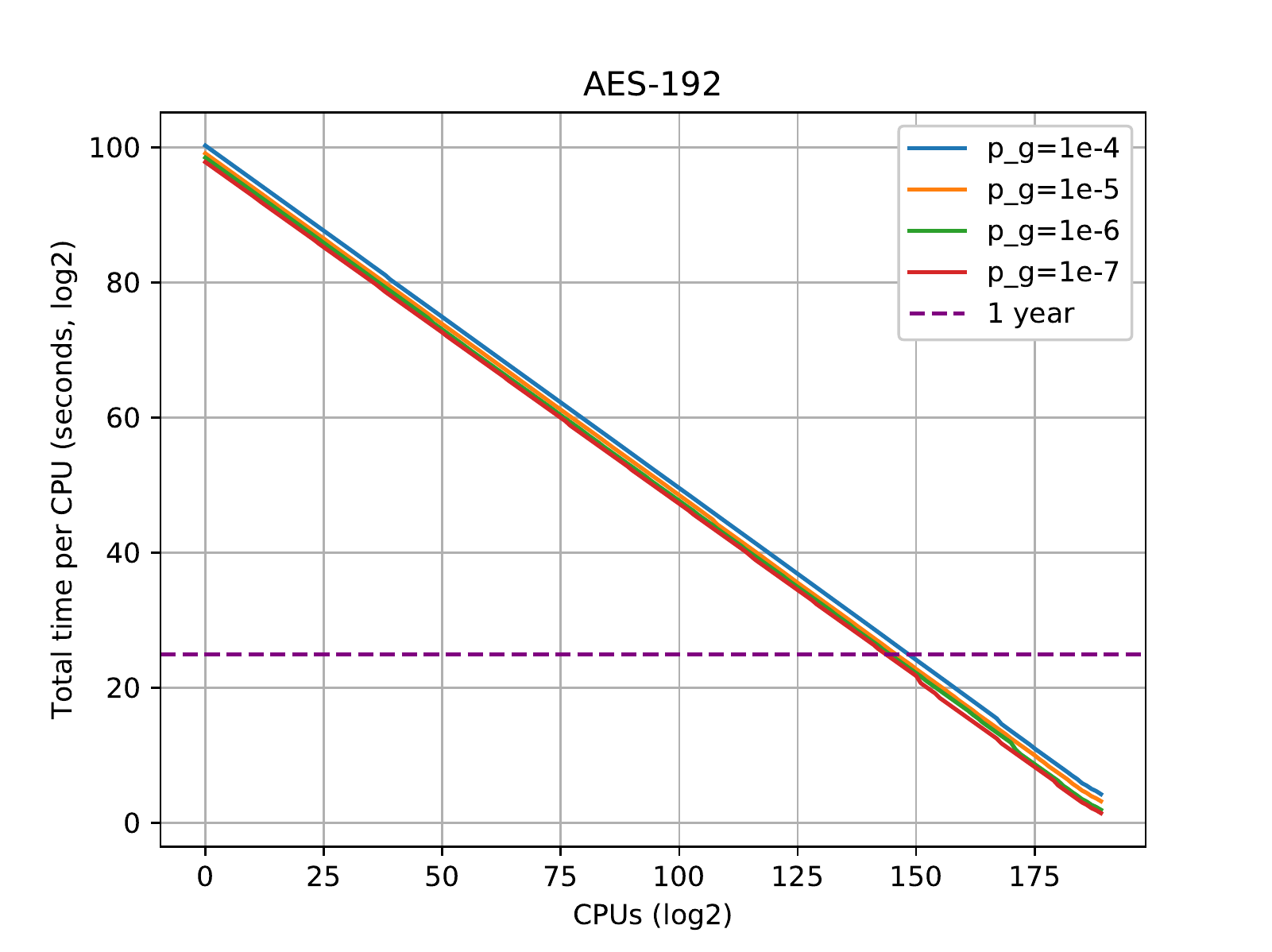}
      	\captionof{figure}{AES-192 block cipher. Required time per processor, as a function of the  number of processors ($\log_2$ scale).}
      	\label{fgr:aes_192_time}
	
        \includegraphics[width=0.429\textwidth]{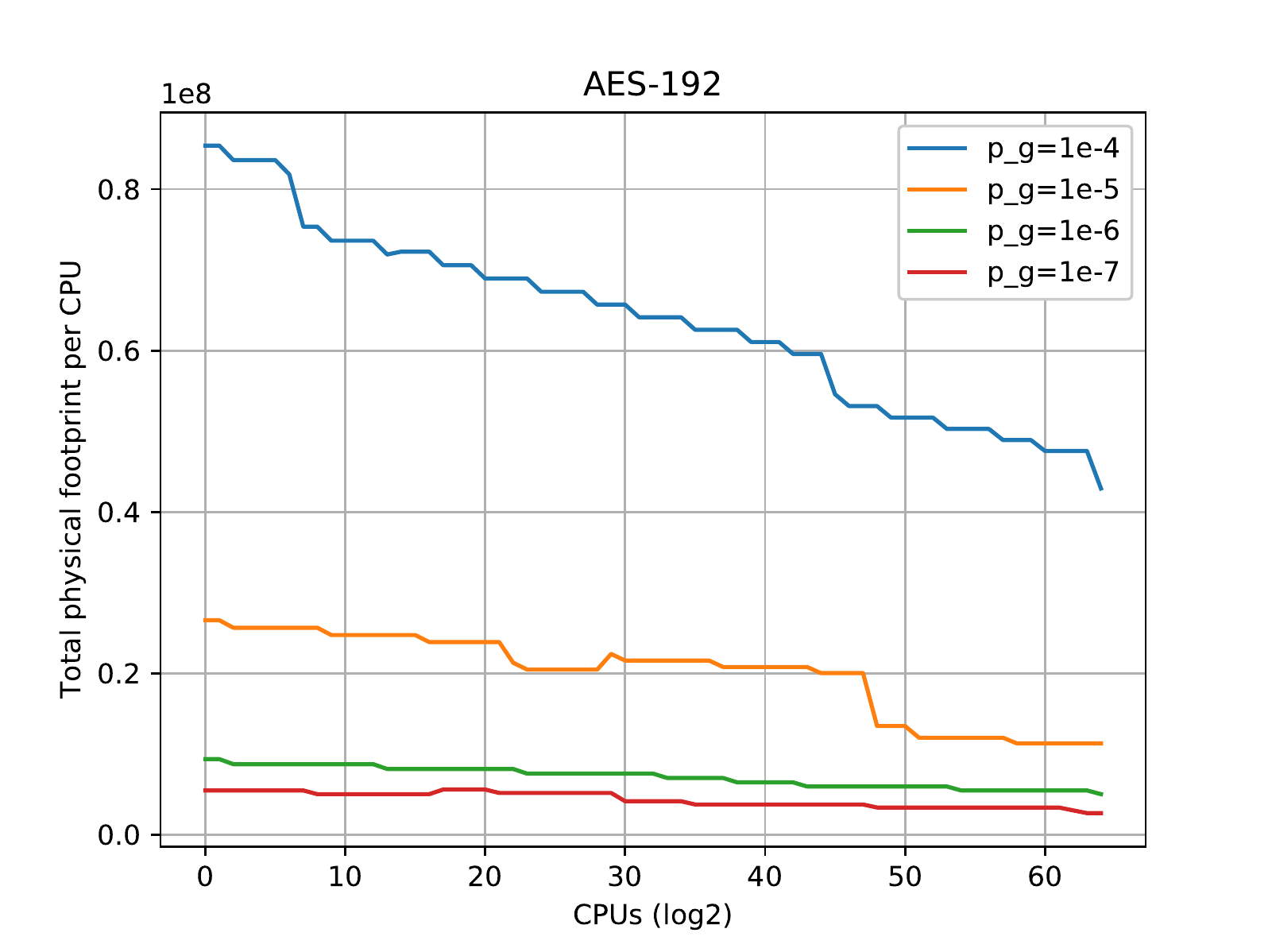}
	\captionof{figure}{AES-192 block cipher. Physical footprint (physical qubits) per processor, as a function of the number of processors ($\log_2$ scale).}
      	\label{fgr:aes_192_phys}
	
        \includegraphics[width=0.429\textwidth]{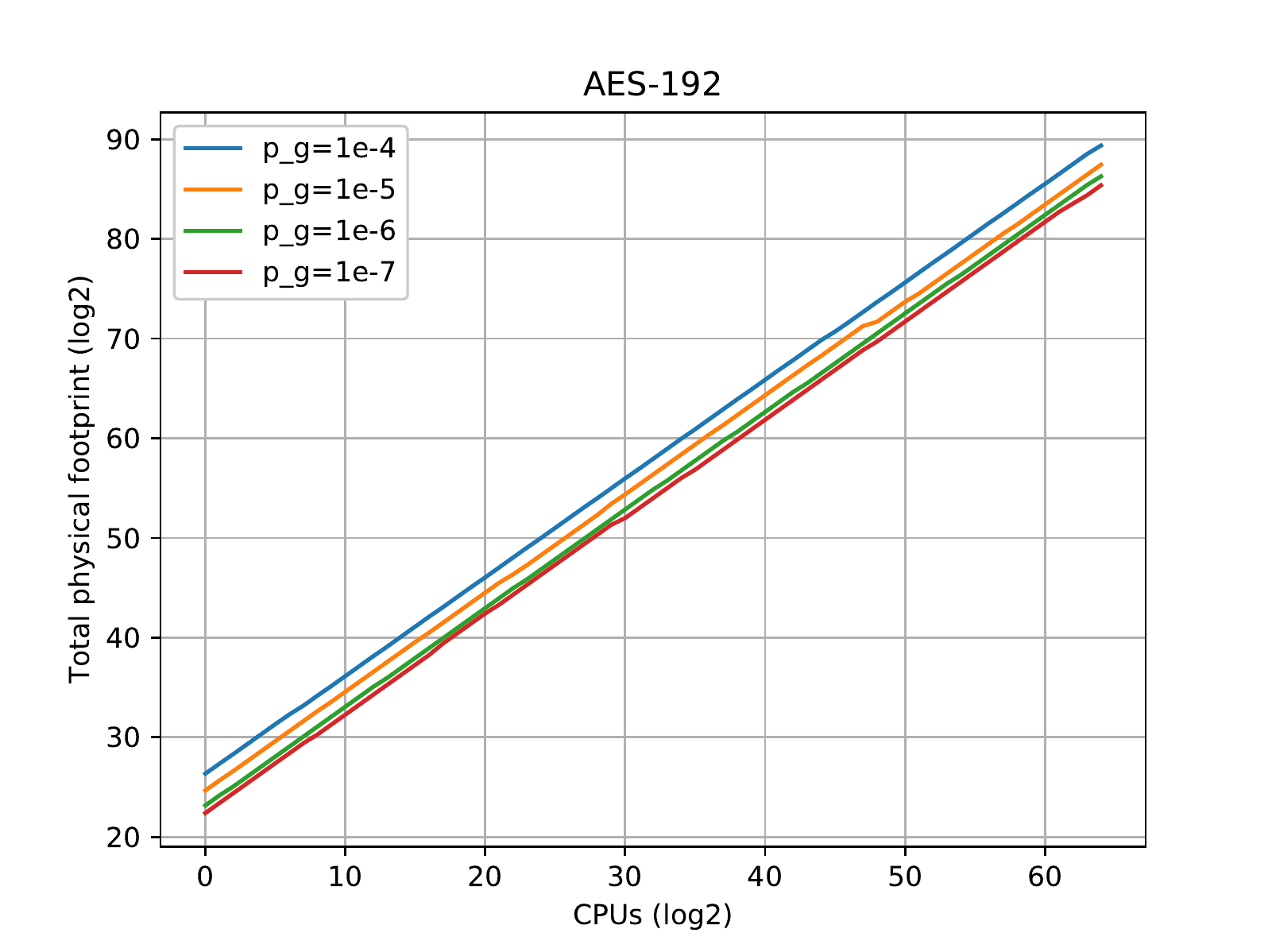}
	\captionof{figure}{AES-192 block cipher. Total physical footprint (physical qubits), as a function of the number of processors ($\log_2$ scale). Note that the qubits are not correlated across processors.}
      	\label{fgr:aes_192_phys_total}

\subsection{AES-256}

	
        \includegraphics[width=0.429\textwidth]{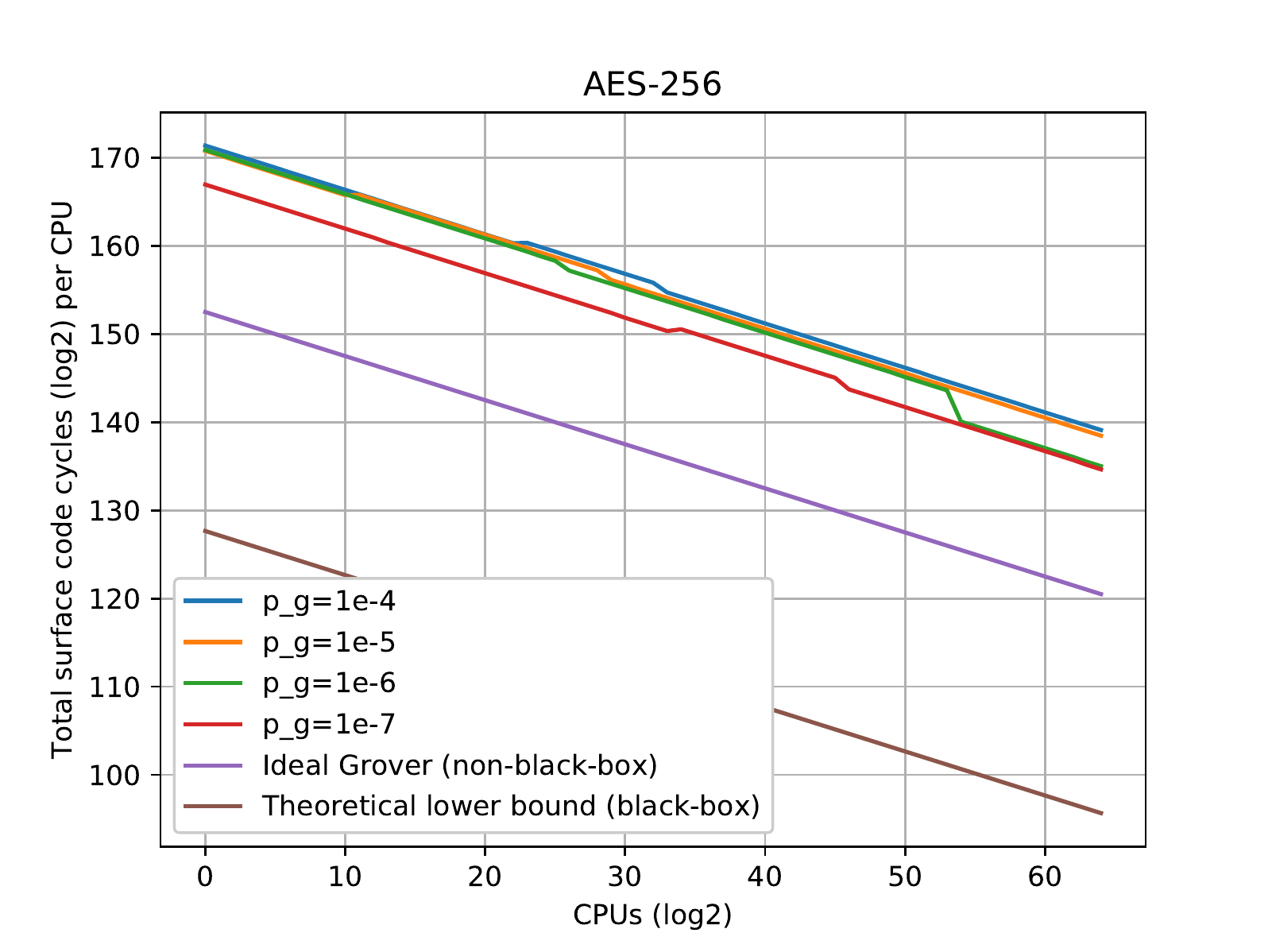}
      	\captionof{figure}{AES-256 block cipher. Required surface clock cycles per processor, as a function of the  number of processors ($\log_2$ scale).}
      	\label{fgr:aes_256_cycles}
	
        \includegraphics[width=0.429\textwidth]{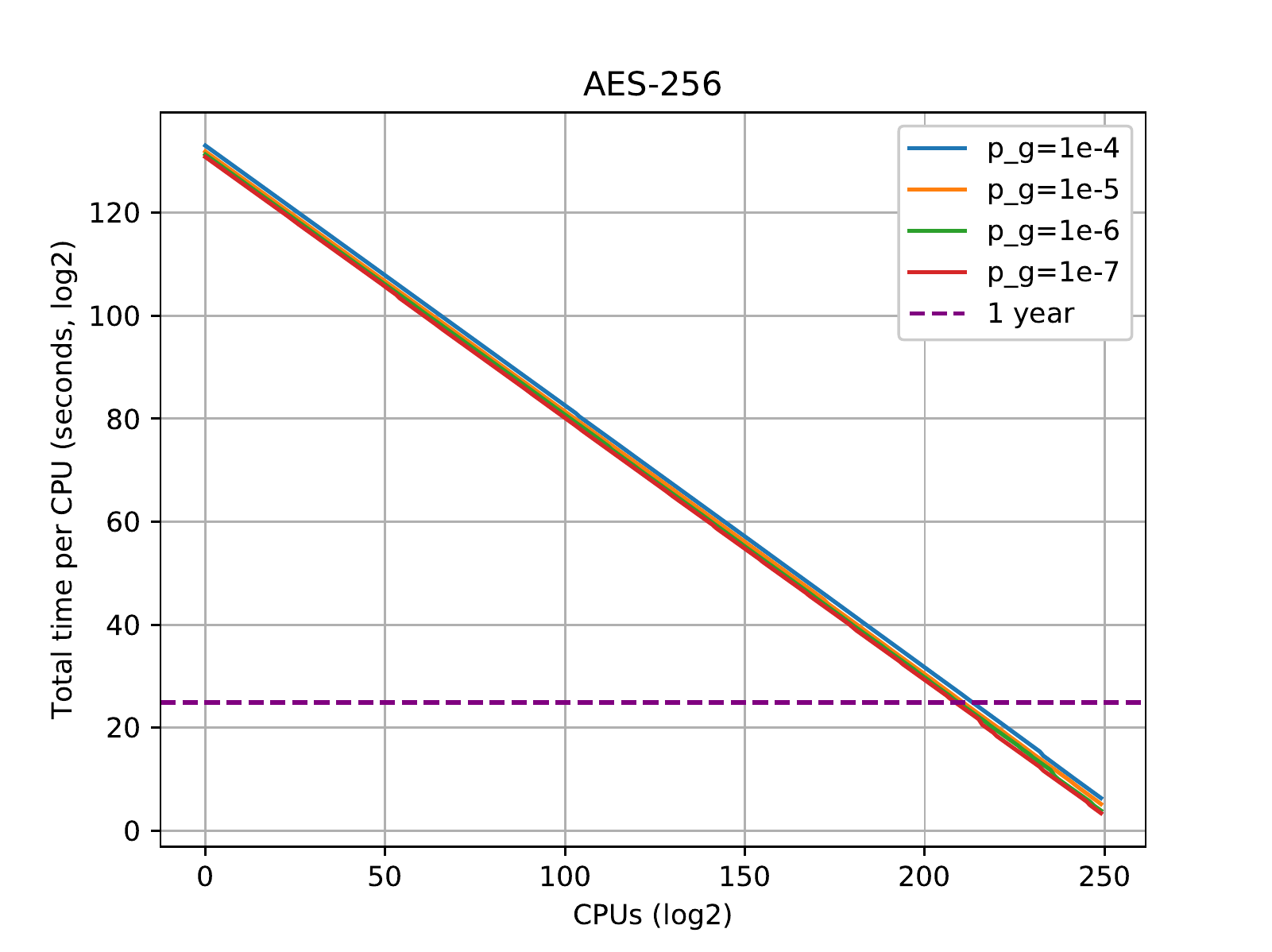}
      	\captionof{figure}{AES-256 block cipher. Required time per processor, as a function of the  number of processors ($\log_2$ scale).}
      	\label{fgr:aes_256_time}
	
        \includegraphics[width=0.429\textwidth]{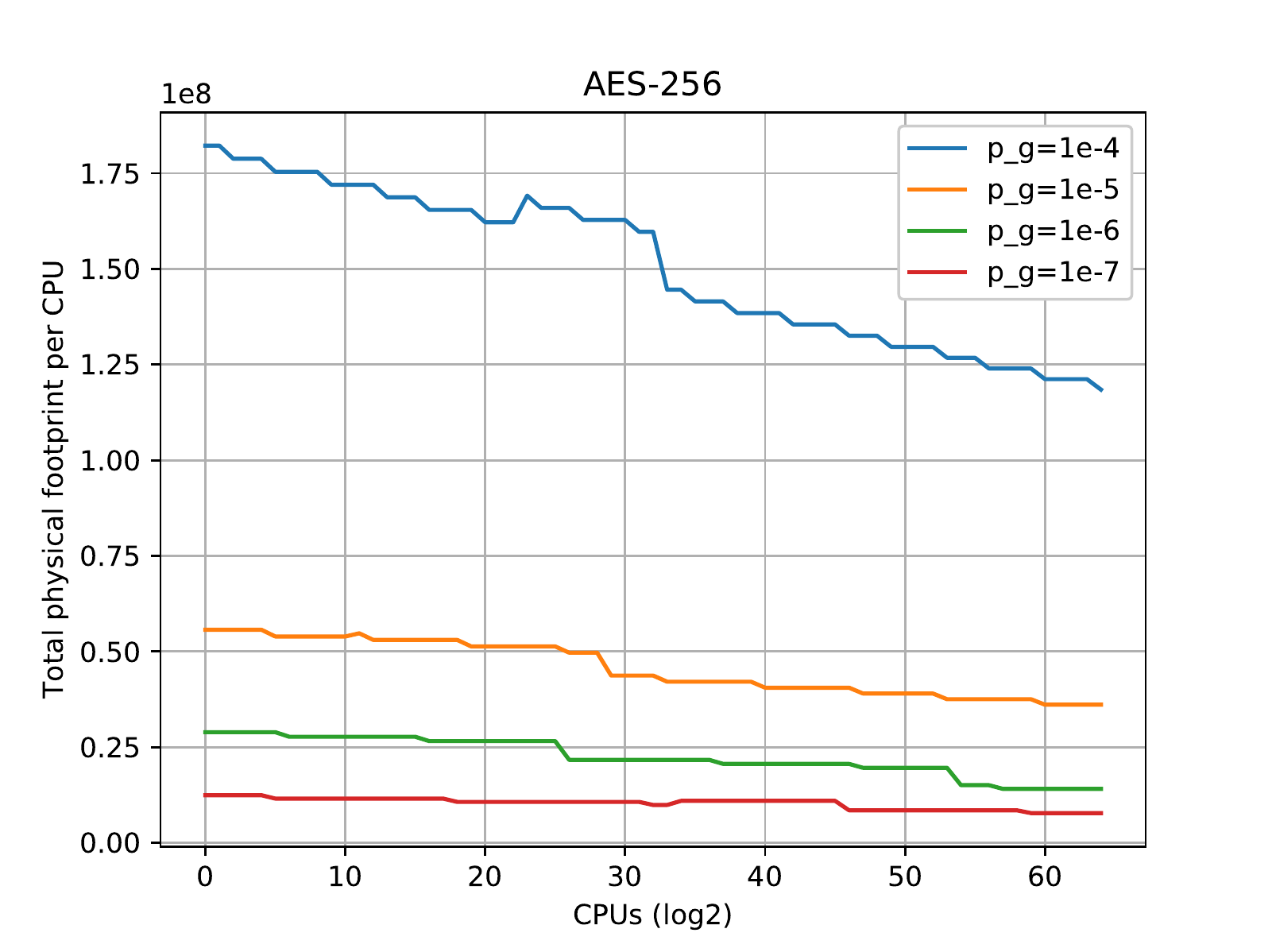}
	\captionof{figure}{AES-256 block cipher. Physical footprint (physical qubits) per processor, as a function of the number of processors ($\log_2$ scale).}
      	\label{fgr:aes_256_phys}
	
        \includegraphics[width=0.429\textwidth]{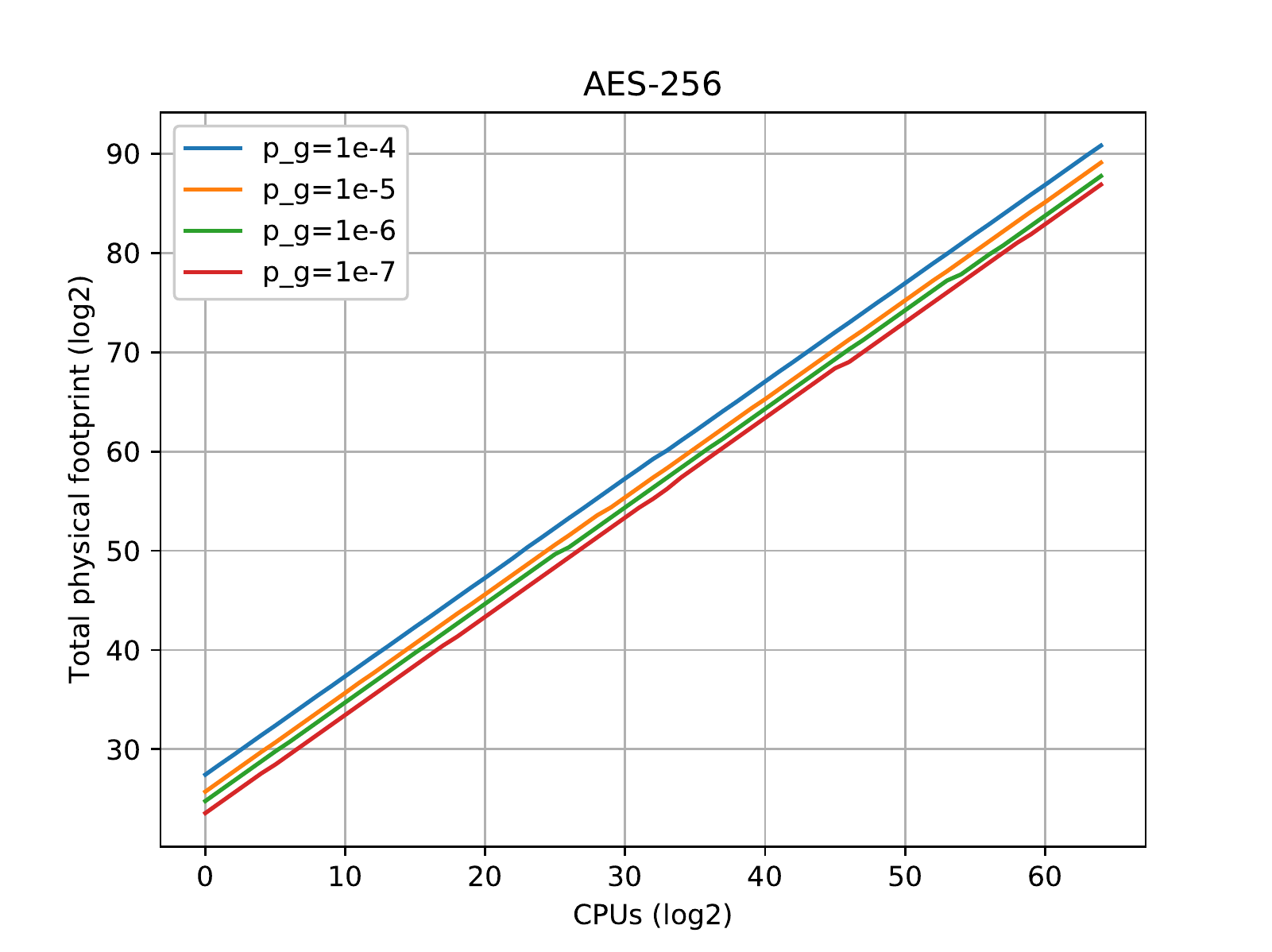}
	\captionof{figure}{AES-256 block cipher. Total physical footprint (physical qubits), as a function of the number of processors ($\log_2$ scale). Note that the qubits are not correlated across processors.}
      	\label{fgr:aes_256_phys_total}

\section{Hash functions\label{sct::hash}}
In this section we study the effect of parallelized Grover attacks on the SHA-256~\cite{SHA2} snd SHA3-256~\cite{SHA3} family of hash functions. We used the highly optimized logical circuits produced in~\cite{10.1007/978-3-319-69453-5_18}.

\subsection{SHA-256}

        \includegraphics[width=0.429\textwidth]{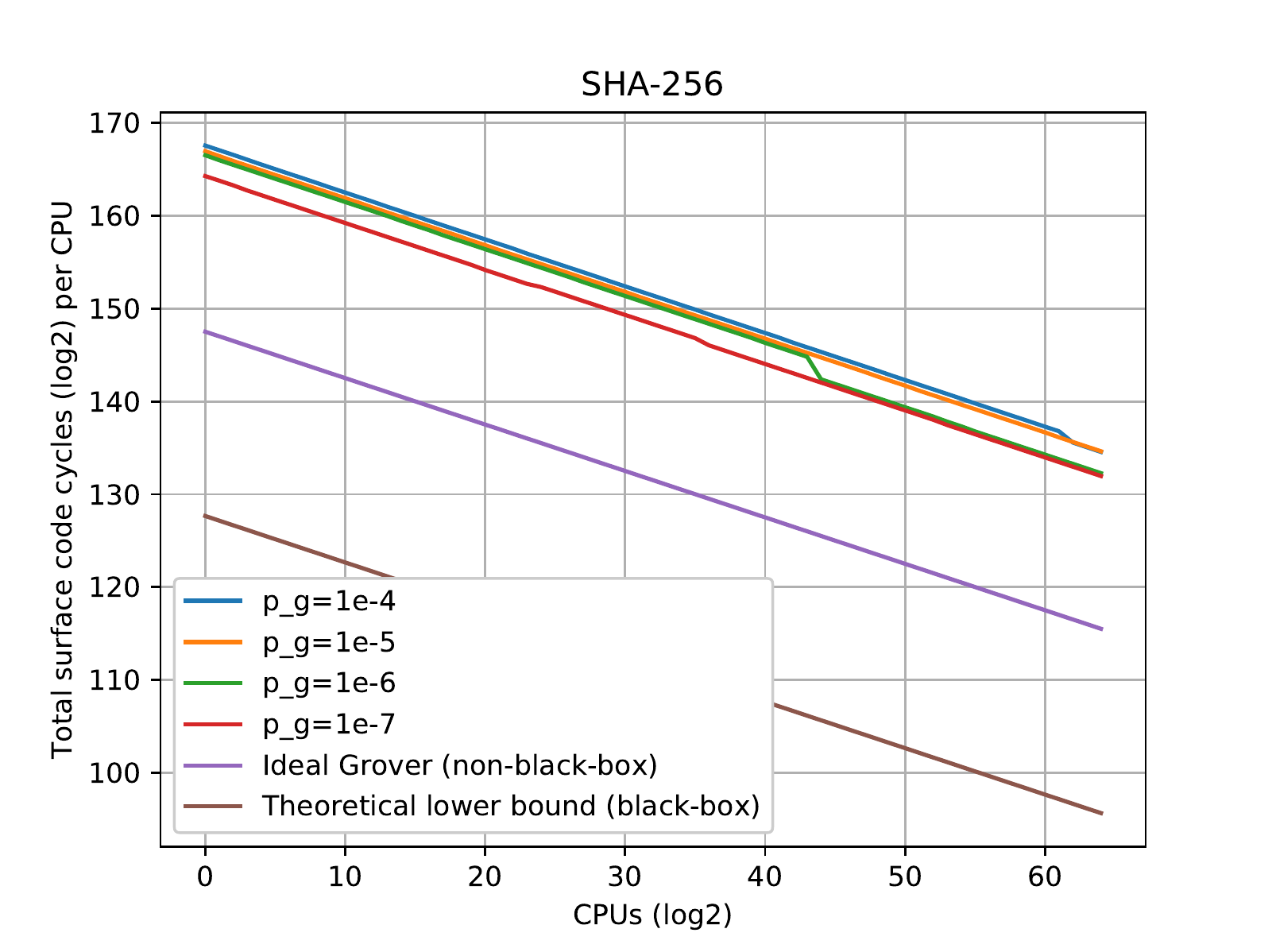}
      	\captionof{figure}{SHA-256 cryptographic hash function. Required surface clock cycles per processor, as a function of the  number of processors ($\log_2$ scale).}
      	\label{fgr:sha_256_cycles}
        \includegraphics[width=0.429\textwidth]{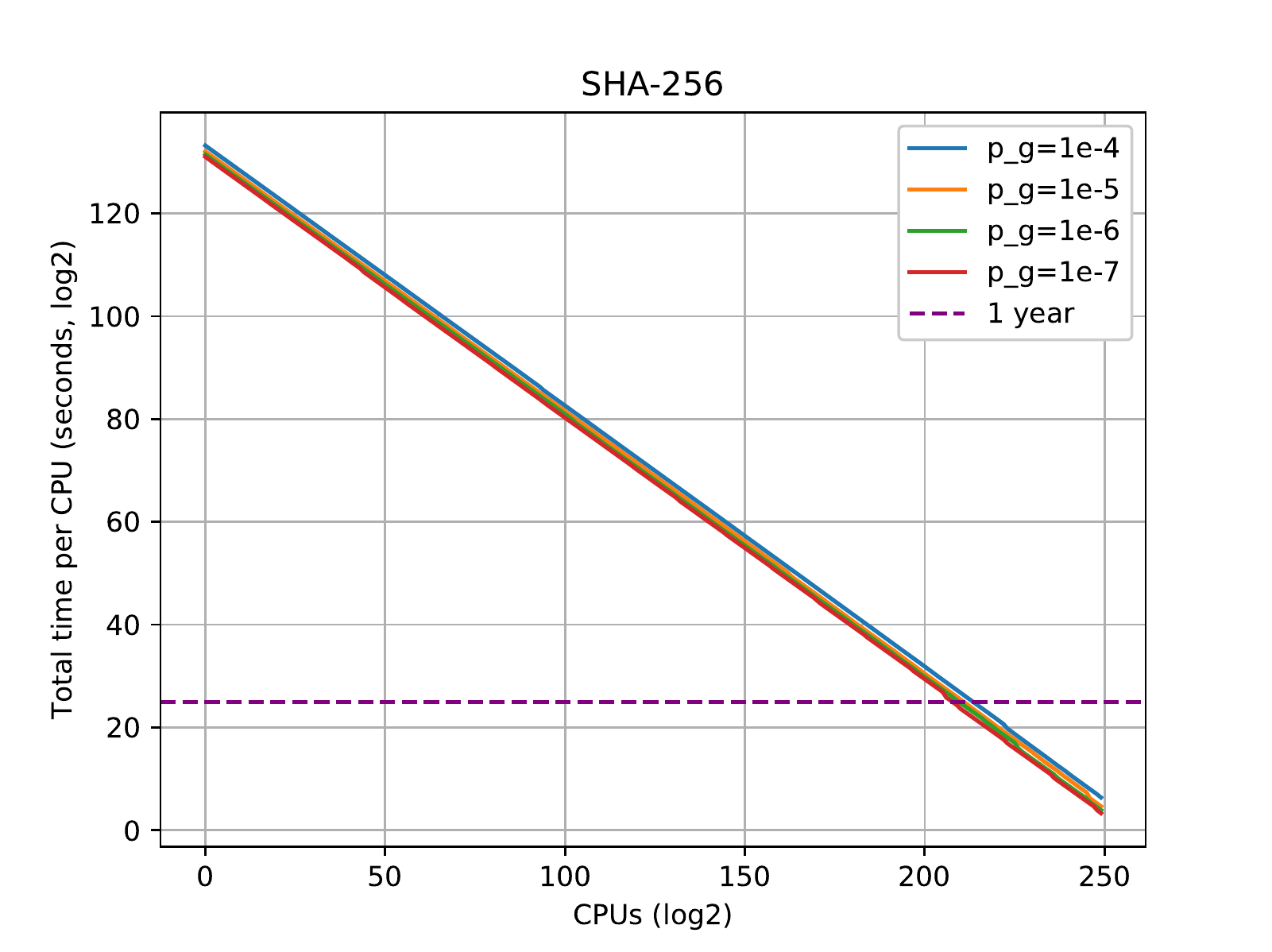}
      	\captionof{figure}{SHA-256 cryptographic hash function. Required time per processor, as a function of the  number of processors ($\log_2$ scale).}
      	\label{fgr:sha_256_time}
        \includegraphics[width=0.429\textwidth]{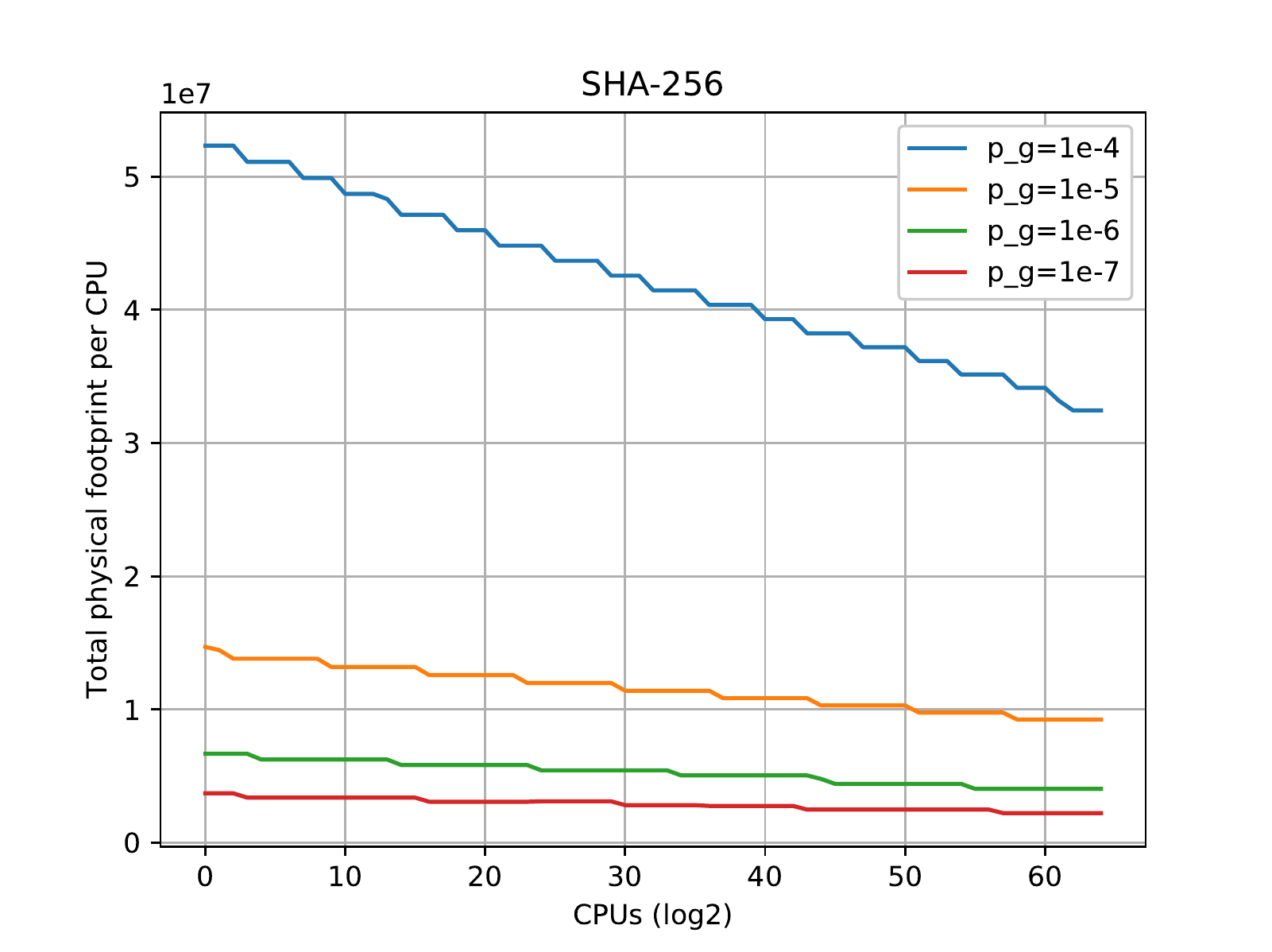}
	\captionof{figure}{SHA-256 cryptographic hash function. Physical footprint (physical qubits) per processor, as a function of the number of processors ($\log_2$ scale).}
      	\label{fgr:sha_256_phys}
        \includegraphics[width=0.429\textwidth]{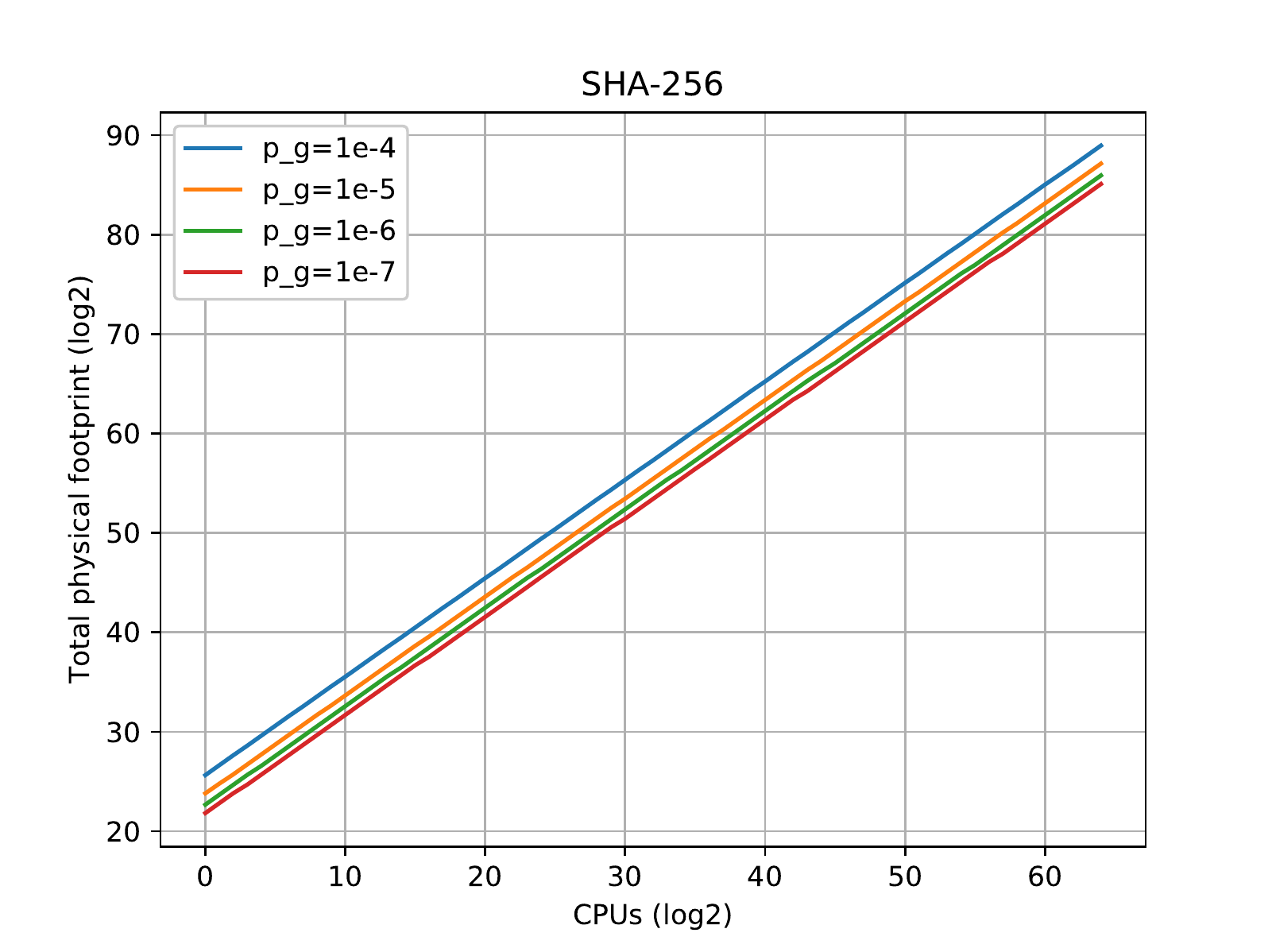}
	\captionof{figure}{SHA-256 cryptographic hash function. Total physical footprint (physical qubits), as a function of the number of processors ($\log_2$ scale). Note that the qubits are not correlated across processors.}
      	\label{fgr:sha_256_phys_total}

\subsection{SHA3-256}

        \includegraphics[width=0.429\textwidth]{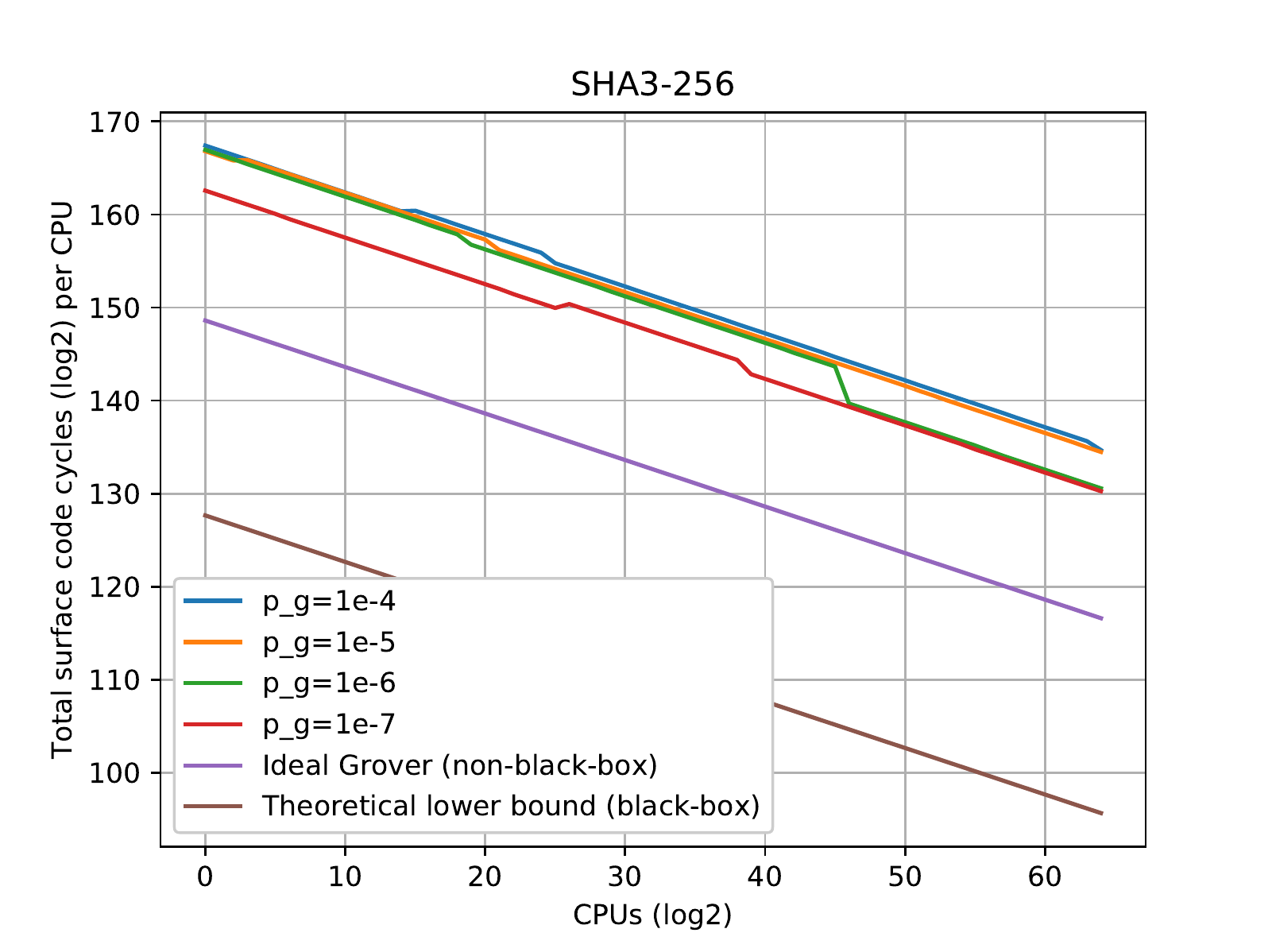}
      	\captionof{figure}{SHA3-256 cryptographic hash function. Required surface clock cycles per processor, as a function of the  number of processors ($\log_2$ scale).}
      	\label{fgr:sha3_256_cycles}
        \includegraphics[width=0.429\textwidth]{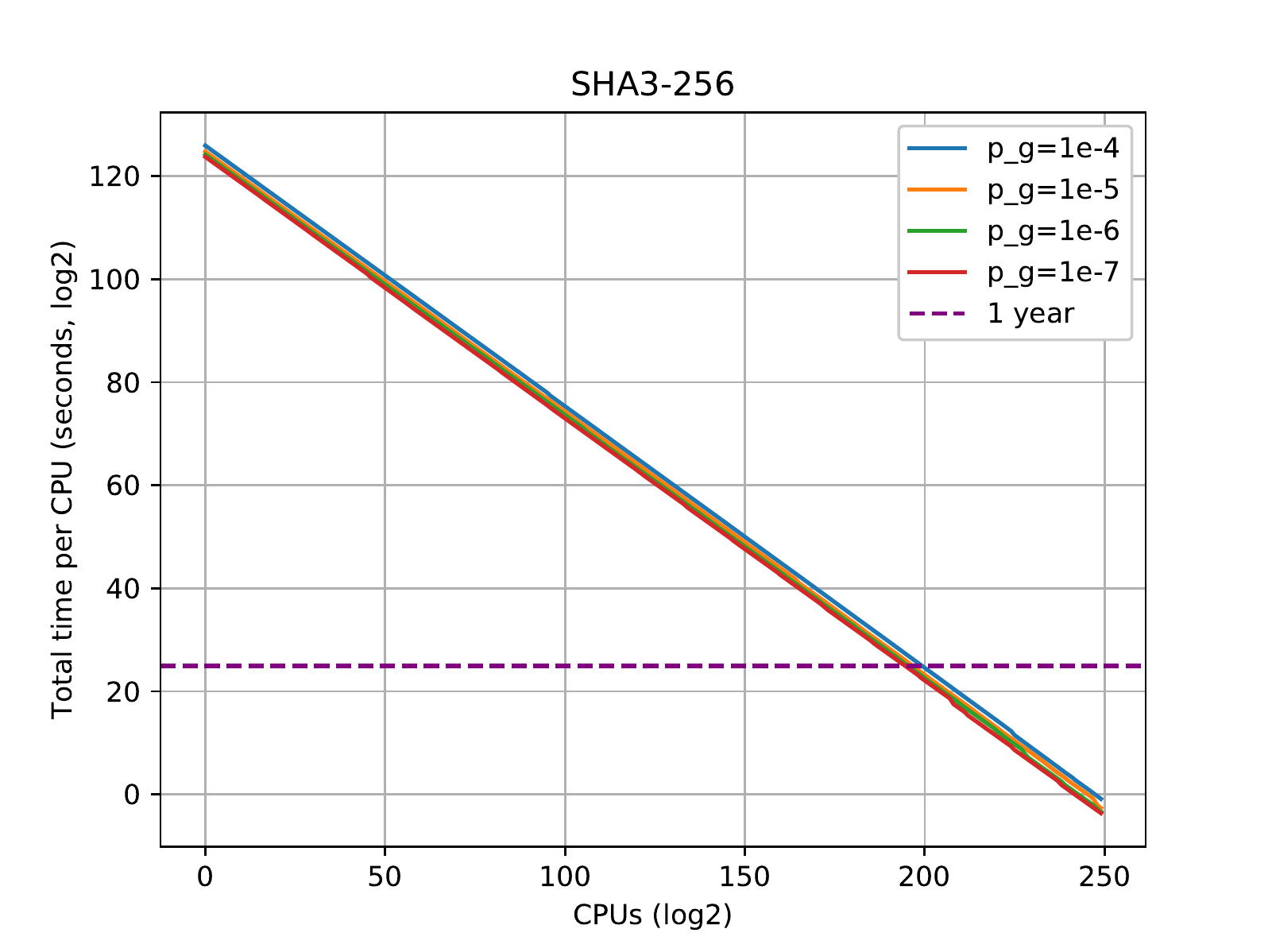}
      	\captionof{figure}{SHA3-256 cryptographic hash function. Required time per processor, as a function of the  number of processors ($\log_2$ scale).}
      	\label{fgr:sha3_256_time}
        \includegraphics[width=0.429\textwidth]{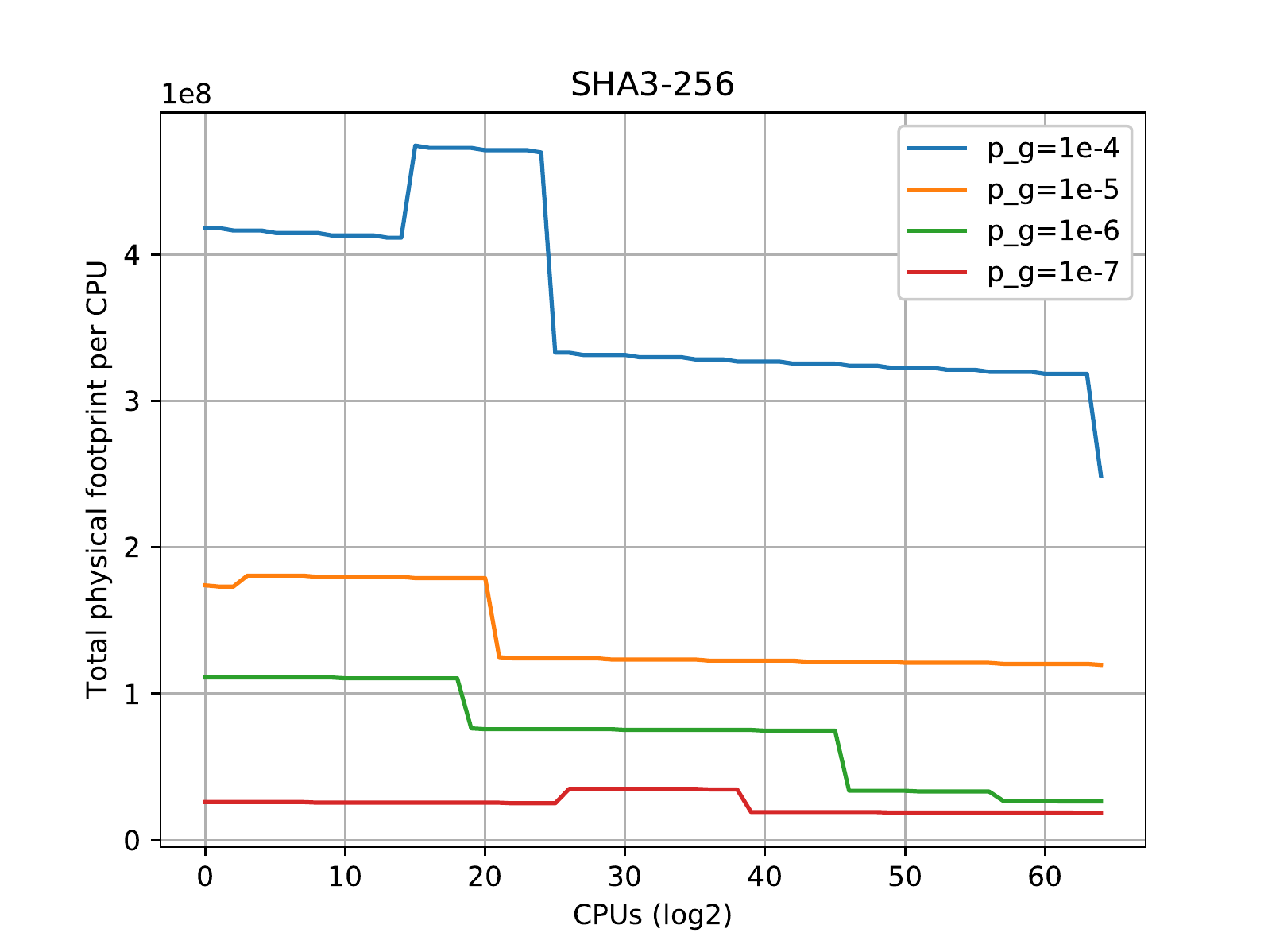}
	\captionof{figure}{SHA3-256 cryptographic hash function. Physical footprint (physical qubits) per processor, as a function of the number of processors ($\log_2$ scale).}
      	\label{fgr:sha3_256_phys}
        \includegraphics[width=0.429\textwidth]{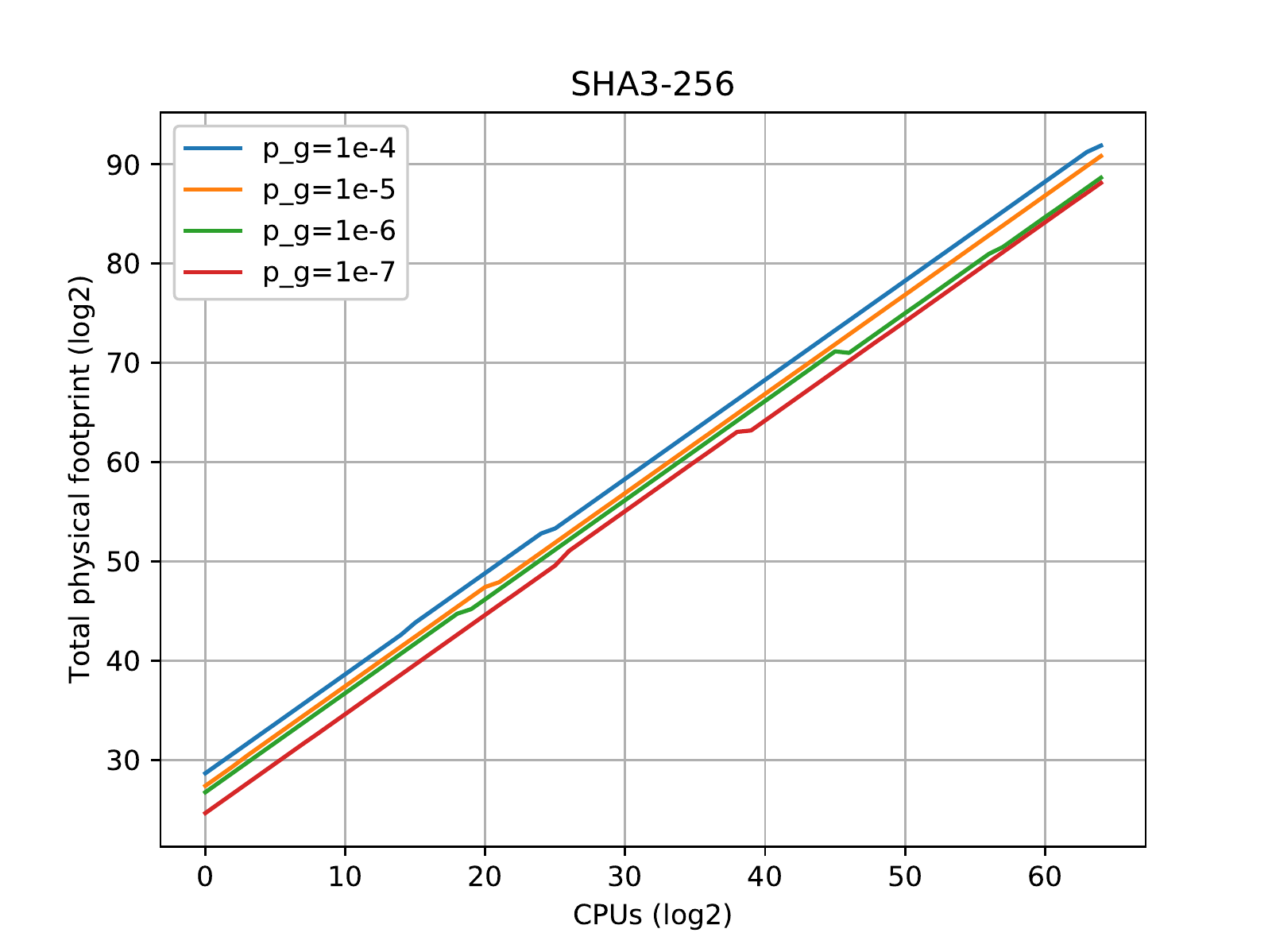}
	\captionof{figure}{SHA3-256 cryptographic hash function. Total physical footprint (physical qubits), as a function of the number of processors ($\log_2$ scale). Note that the qubits are not correlated across processors.}
      	\label{fgr:sha3_256_phys_total}
\section{Bitcoin~\label{sct::bitcoin}}
In this section we analyze the security of Bitcoin's~\cite{satoshi:bitcoin} proof-of-work protocol, which is based on finding a hash\footnote{The hash function being used by the protocol is H($x$) := SHA-256(SHA-256($x$).} pre-image which that starts
with a certain number of zeros. The latter is dynamically adjusted by the protocol so that the problem is on average solved by
the whole network in 10 minutes. Currently, it takes around $2^{75}$ classical hashing operations~\cite{btc_difficulty} for finding a desired hash pre-image successfully via brute-force search with specialized hardware.
        \includegraphics[width=0.429\textwidth]{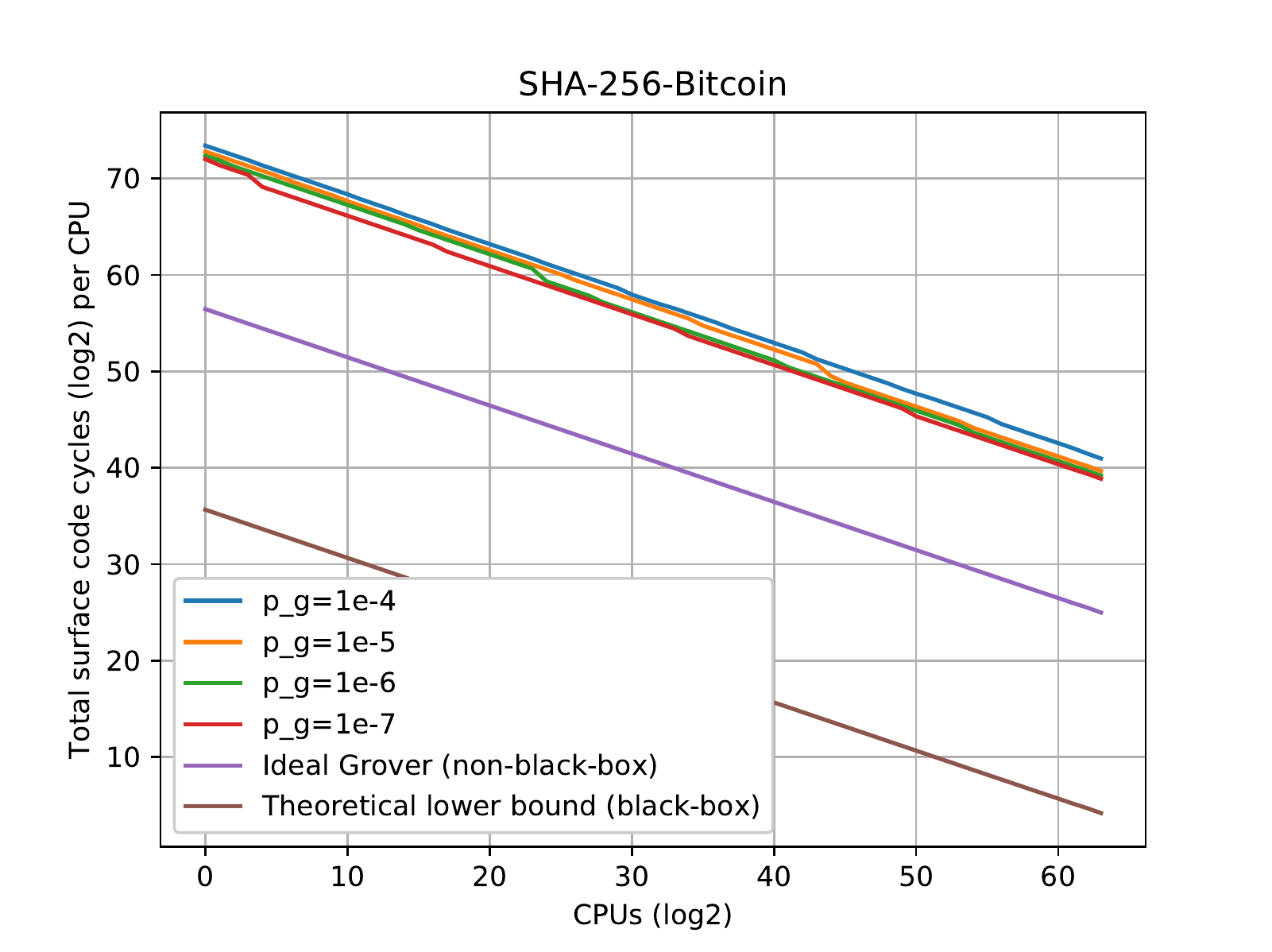}
      	\captionof{figure}{Bitcoin's cryptographic hash function H($x$) := SHA-256(SHA-256($x$)). Required surface clock cycles per processor, as a function of the  number of processors ($\log_2$ scale).}
      	\label{fgr:sha_256_bitcoin_cycles}
        \includegraphics[width=0.429\textwidth]{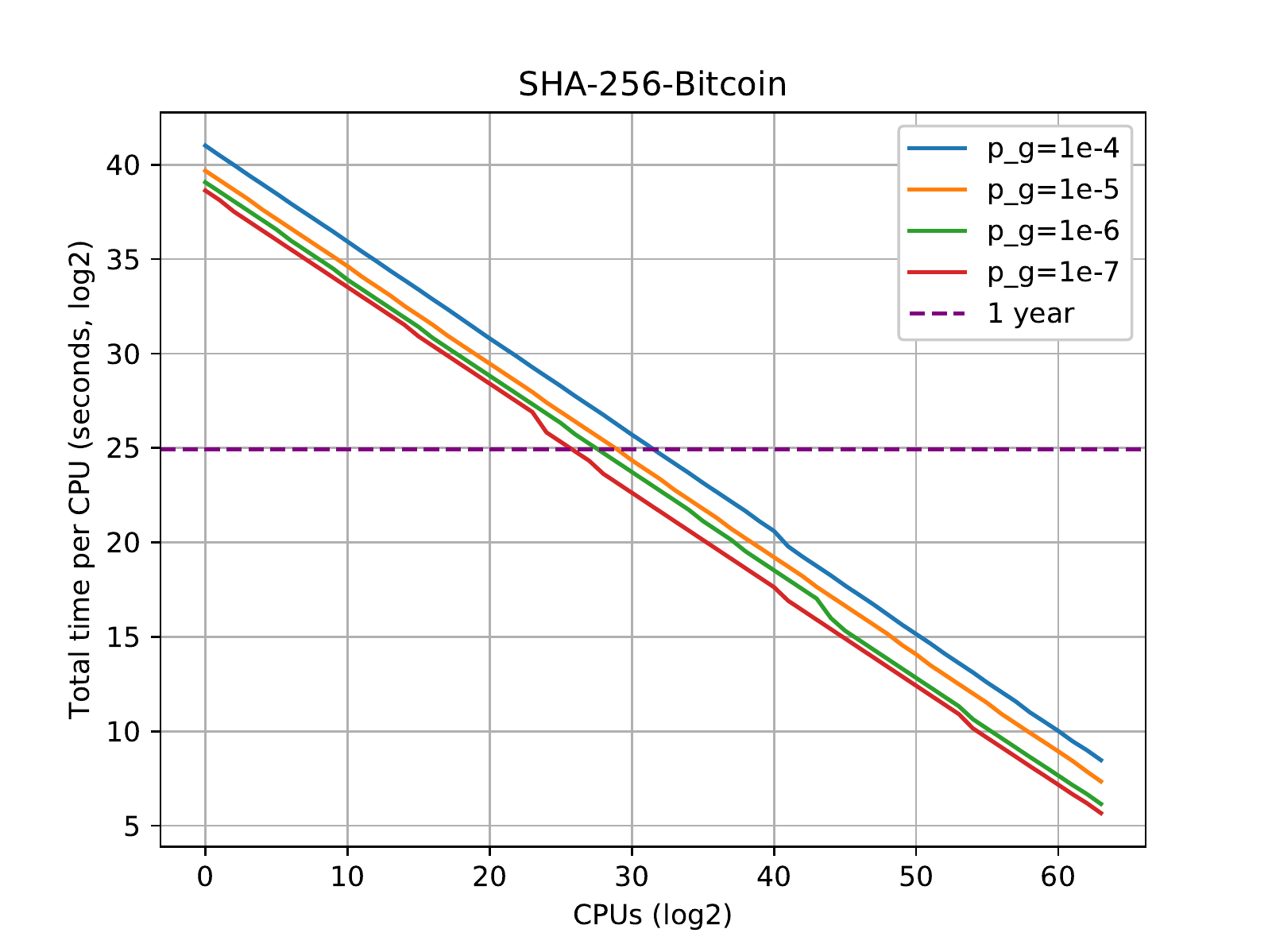}
      	\captionof{figure}{Bitcoin's cryptographic hash function H($x$) := SHA-256(SHA-256($x$)). Required time per processor, as a function of the  number of processors ($\log_2$ scale).}
      	\label{fgr:sha_256_bitcoin_time}
        \includegraphics[width=0.429\textwidth]{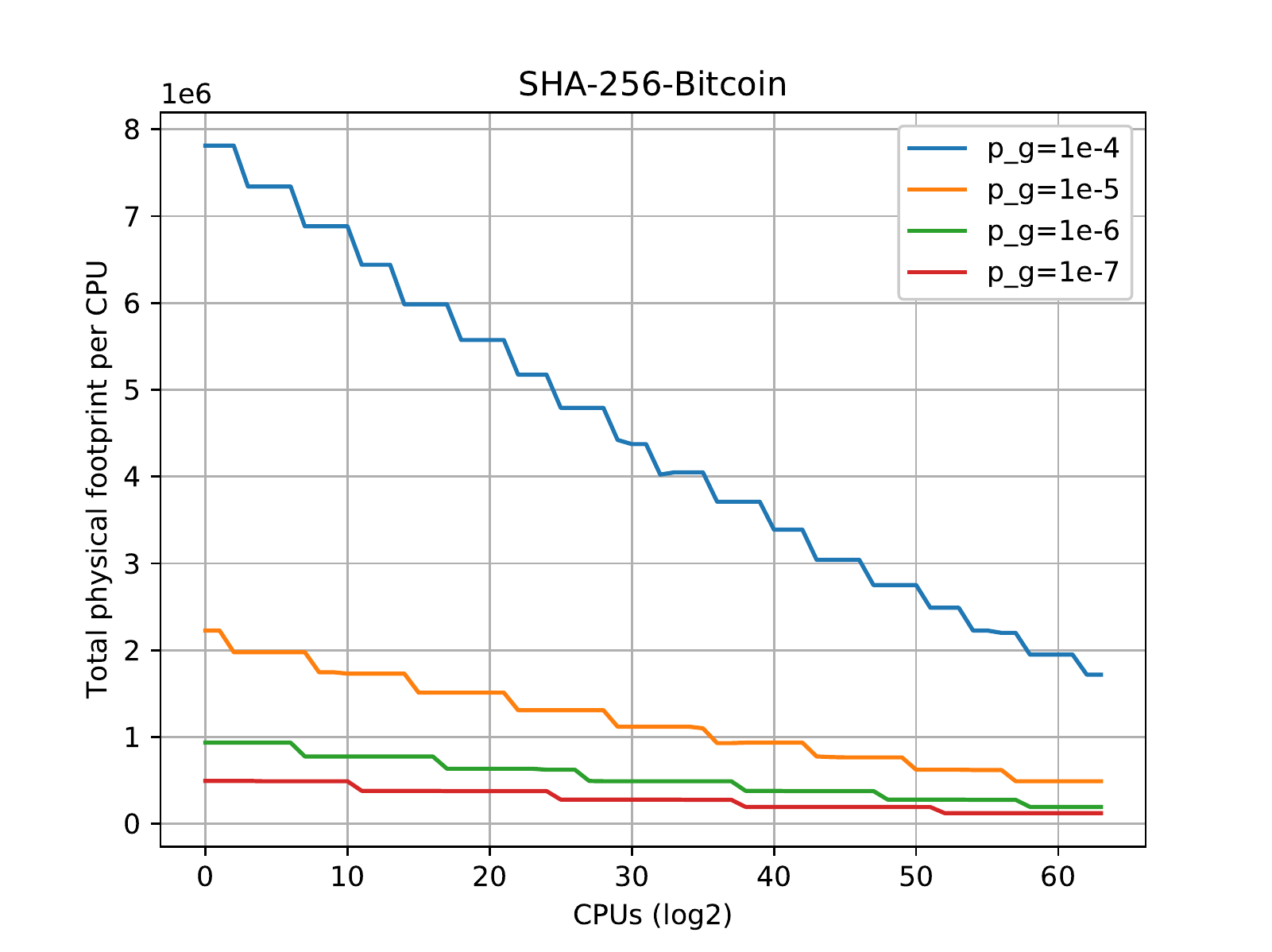}
	\captionof{figure}{Bitcoin's cryptographic hash function H($x$) := SHA-256(SHA-256($x$)). Physical footprint (physical qubits) per processor, as a function of the number of processors ($\log_2$ scale).}
      	\label{fgr:sha_256_bitcoin_phys}
        \includegraphics[width=0.429\textwidth]{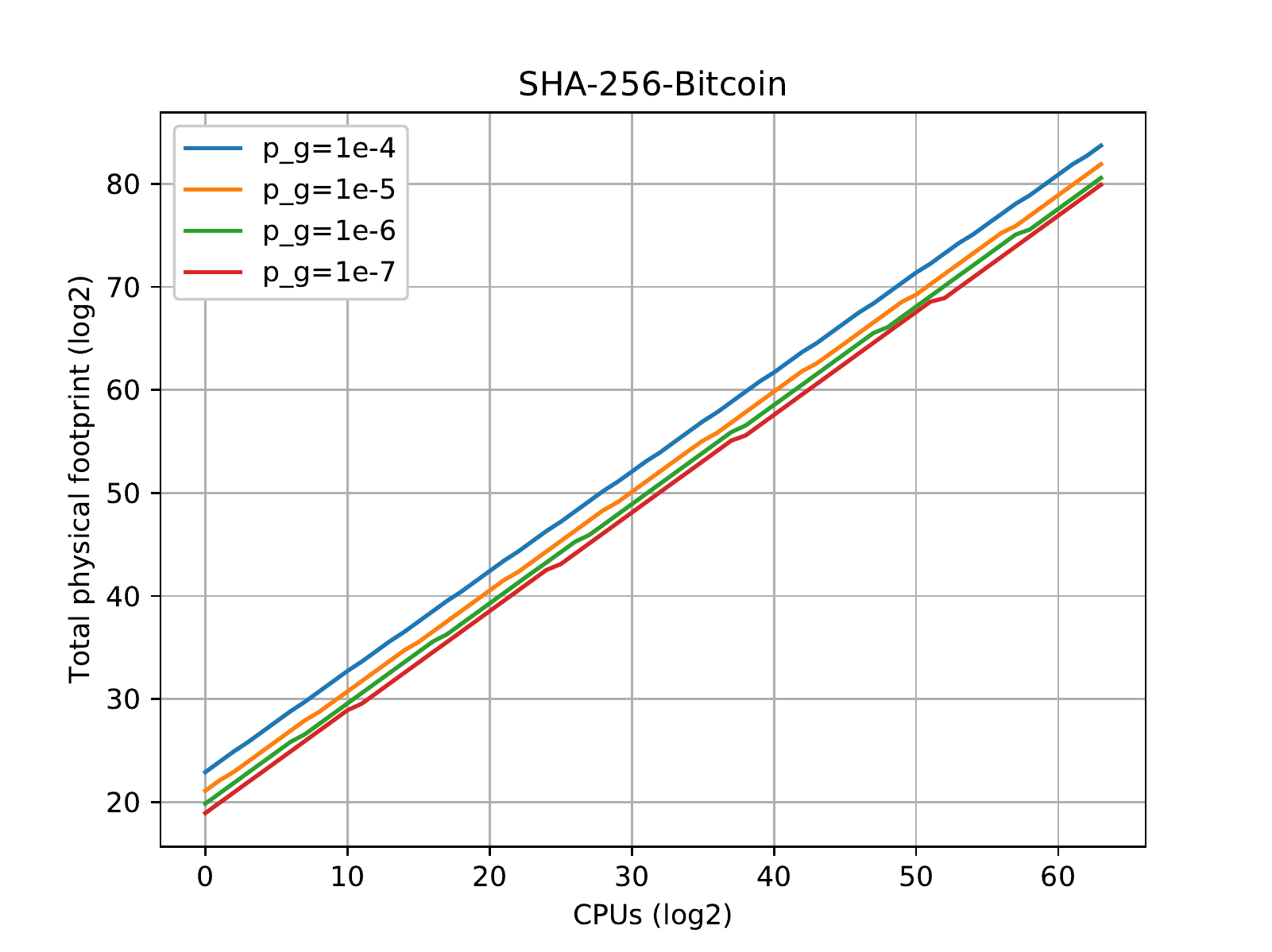}
	\captionof{figure}{Bitcoin's cryptographic hash function H($x$) := SHA-256(SHA-256($x$)). Total physical footprint (physical qubits), as a function of the number of processors ($\log_2$ scale). Note that the qubits are not correlated across processors.}
      	\label{fgr:sha_256_bitcoin_phys_total}

\section{Intrinsic cost of parallelized Grover's algorithm\label{sct::intrinsic_parallel_grover}}

More efficient quantum implementations of AES and SHA imply more efficient cryptanalysis. In this section, we aim to bound how much further optimized implementations of these cryptographic functions could help. We do so by assuming a trivial cost of $1$ for each function evaluation.

\subsection{Searching space of size $2^{56}$}

        \includegraphics[width=0.429\textwidth]{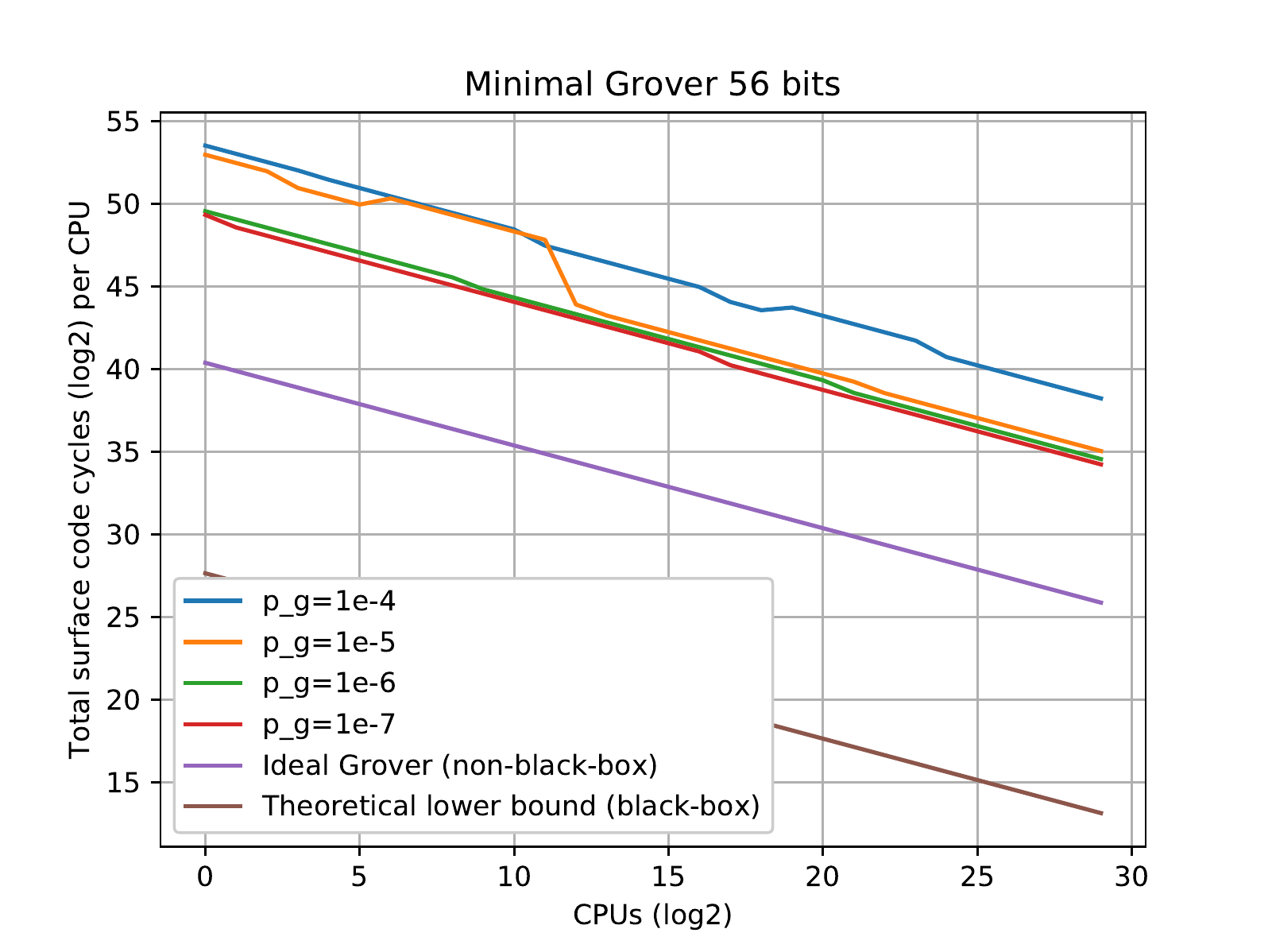}
      	\captionof{figure}{Running Grover's algorithm with a trivial oracle, for a searching space of size $2^{56}$. Required surface clock cycles per processor, as a function of the  number of processors ($\log_2$ scale).}
      	\label{fgr:minimal_grover_56_cycles}
        \includegraphics[width=0.429\textwidth]{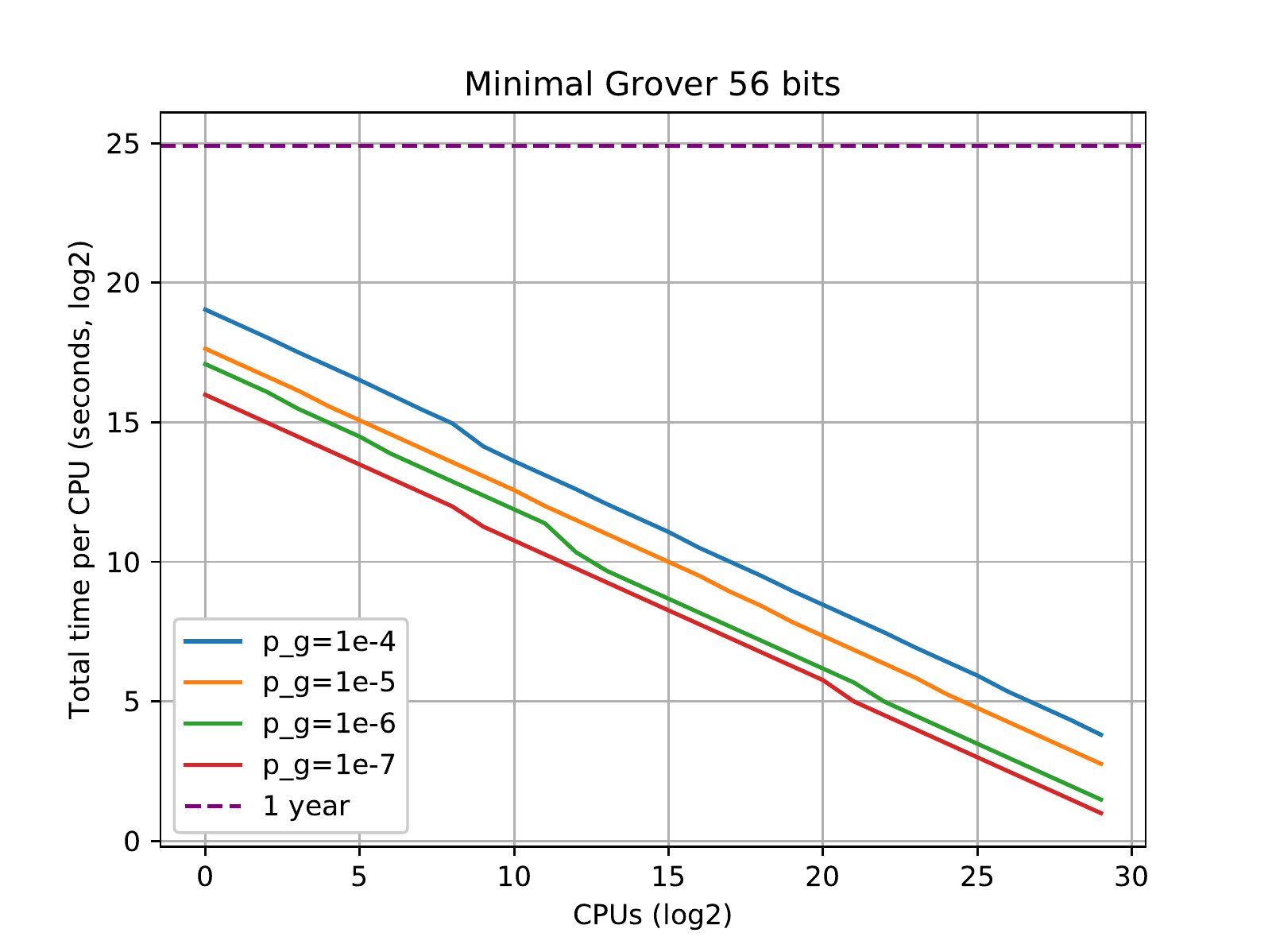}
      	\captionof{figure}{Running Grover's algorithm with a trivial oracle, for a searching space of size $2^{56}$. Required time per processor, as a function of the  number of processors ($\log_2$ scale). The dotted horizontal line indicates one year. }
      	\label{fgr:minimal_grover_56_time}
        \includegraphics[width=0.429\textwidth]{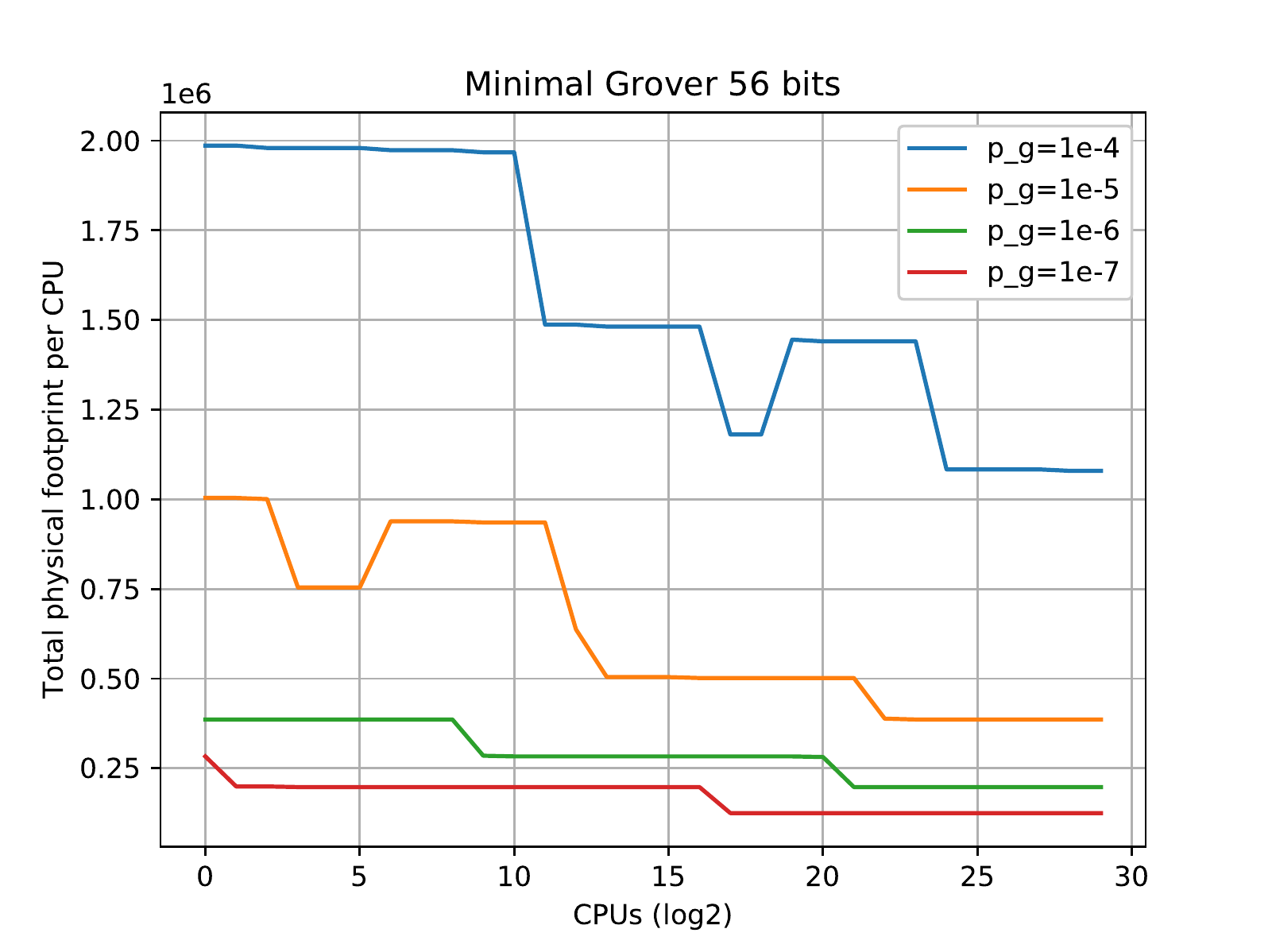}
	\captionof{figure}{Running Grover's algorithm with a trivial oracle, for a searching space of size $2^{56}$. Physical footprint (physical qubits) per processor, as a function of the number of processors ($\log_2$ scale).}
      	\label{fgr:minimal_grover_56_phys}
        \includegraphics[width=0.429\textwidth]{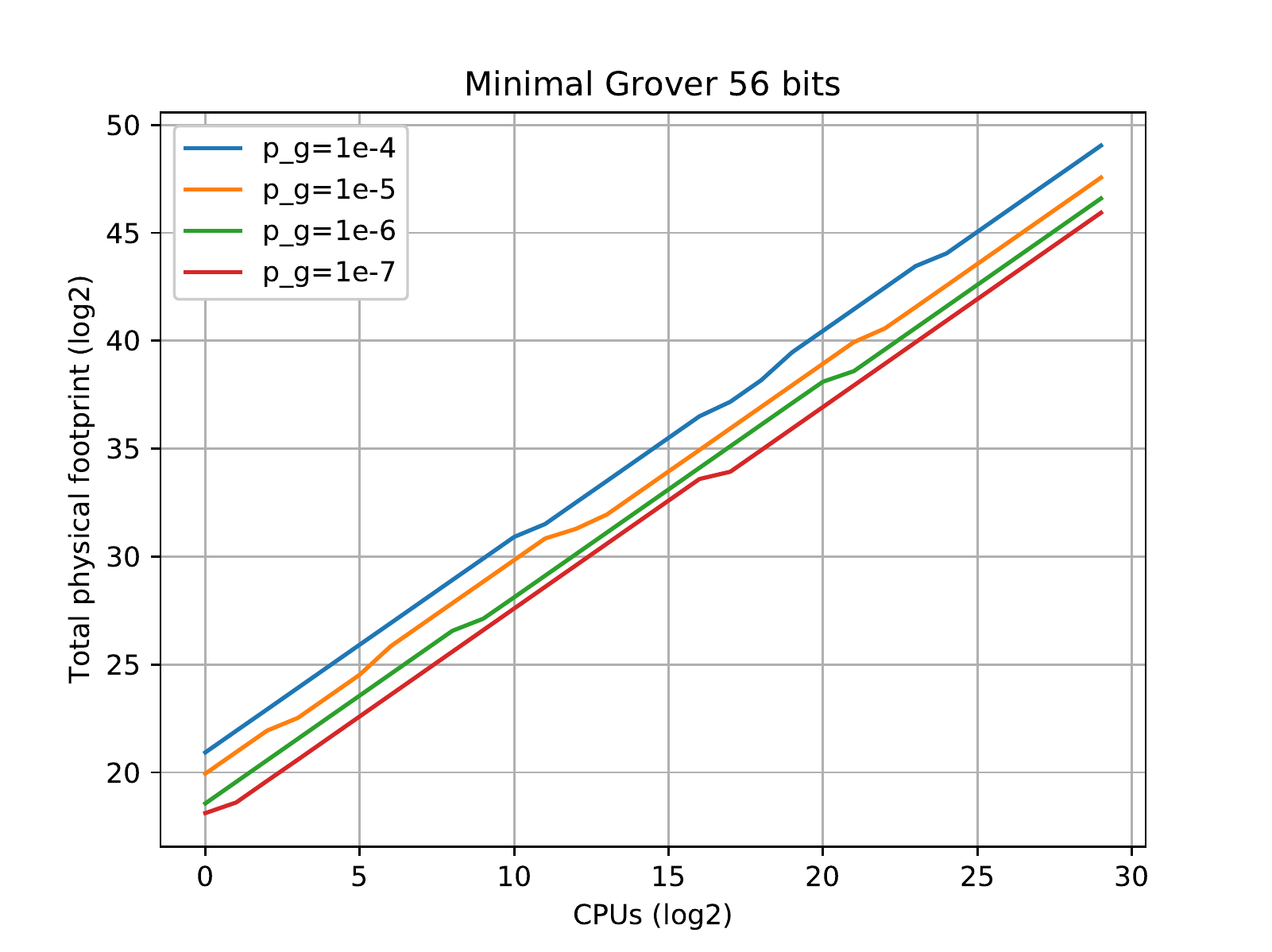}
	\captionof{figure}{Running Grover's algorithm with a trivial oracle, for a searching space of size $2^{56}$. Total physical footprint (physical qubits), as a function of the number of processors ($\log_2$ scale). Note that the qubits are not correlated across processors.}
      	\label{fgr:minimal_grover_56_phys_total}

\subsection{Searching space of size $2^{64}$}

        \includegraphics[width=0.429\textwidth]{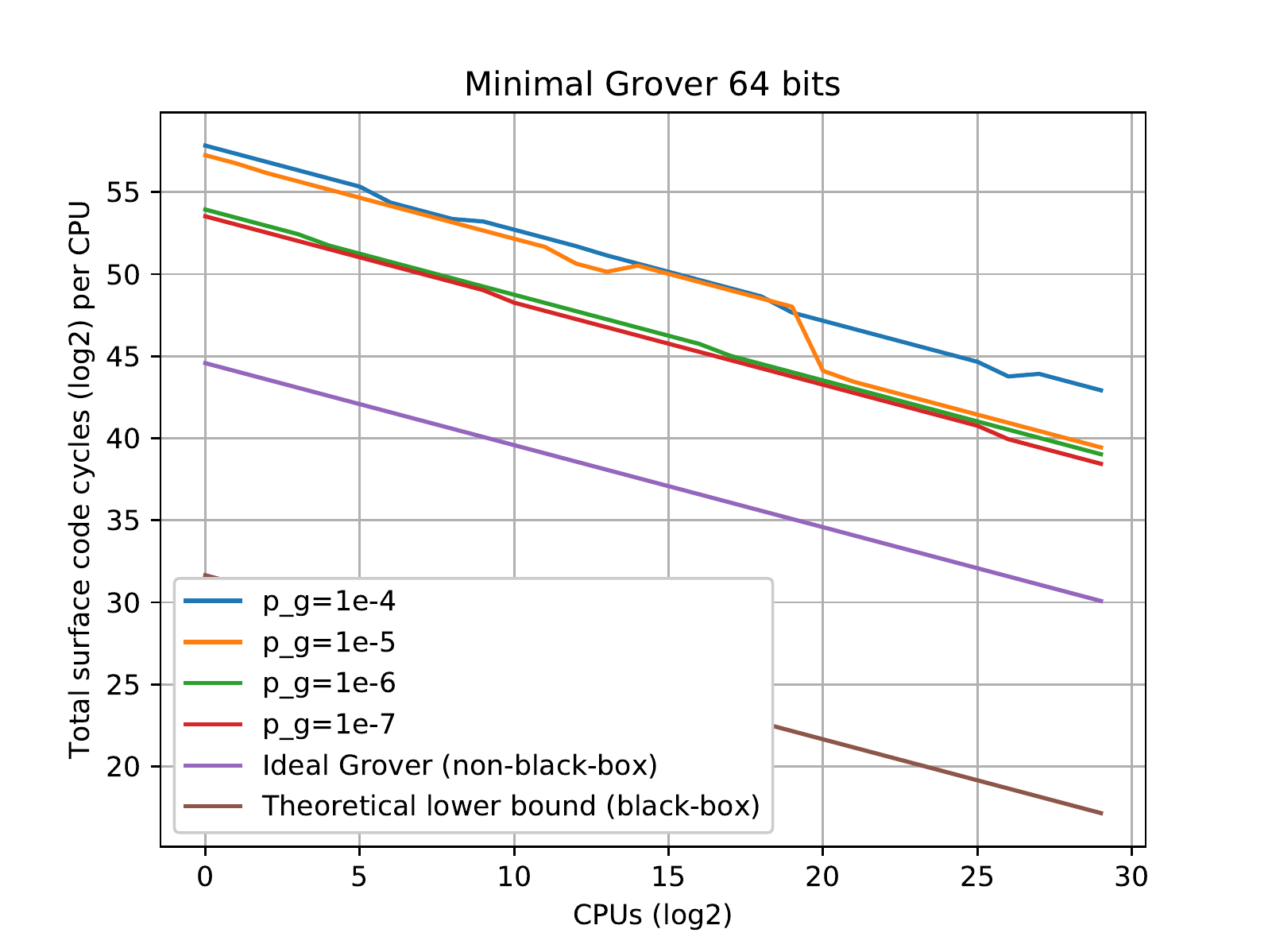}
      	\captionof{figure}{Running Grover's algorithm with a trivial oracle, for a searching space of size $2^{64}$. Required surface clock cycles per processor, as a function of the  number of processors ($\log_2$ scale).}
      	\label{fgr:minimal_grover_64_cycles}
        \includegraphics[width=0.429\textwidth]{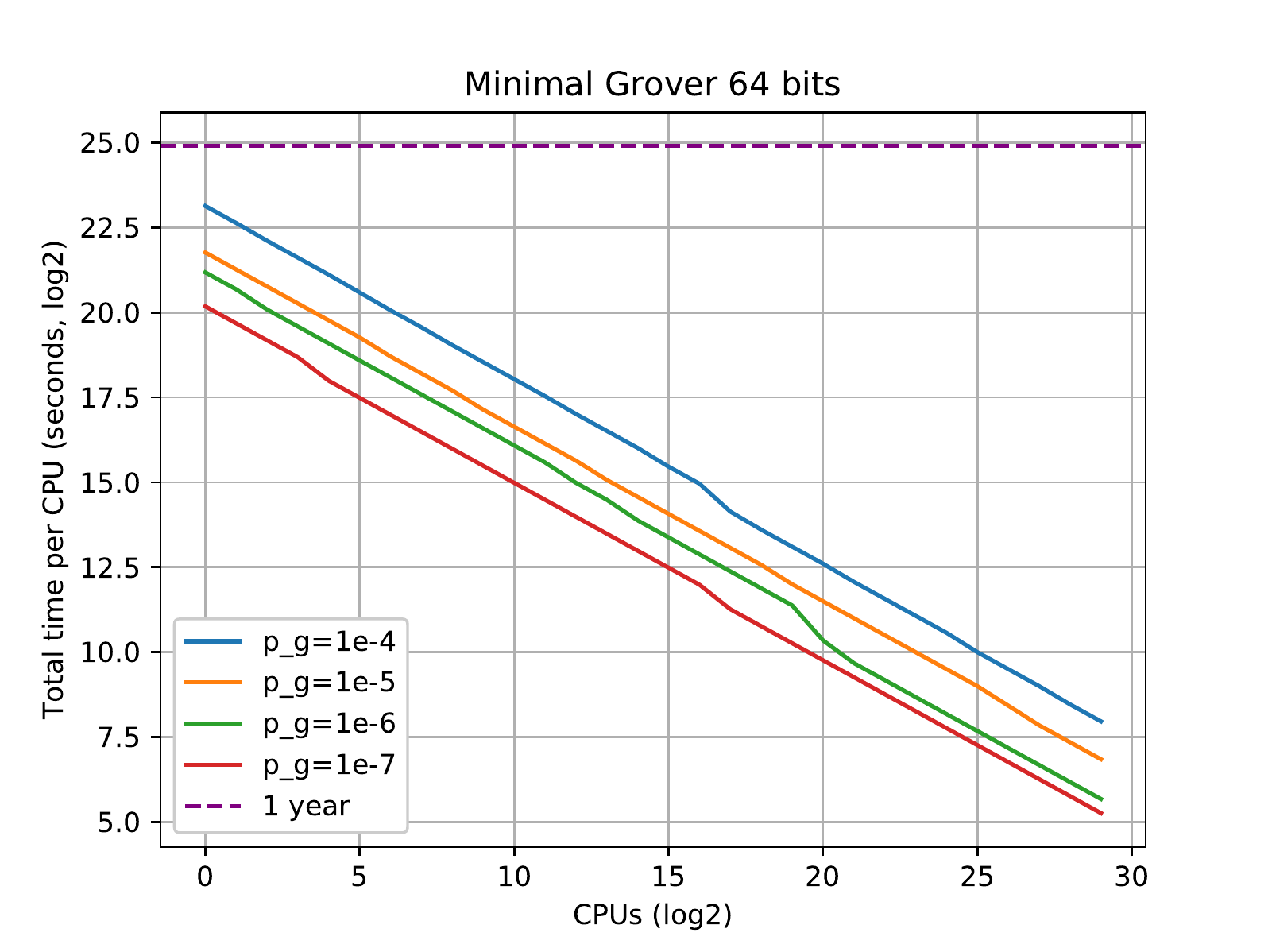}
      	\captionof{figure}{Running Grover's algorithm with a trivial oracle, for a searching space of size $2^{64}$. Required time per processor, as a function of the  number of processors ($\log_2$ scale).}
      	\label{fgr:minimal_grover_64_time}
        \includegraphics[width=0.429\textwidth]{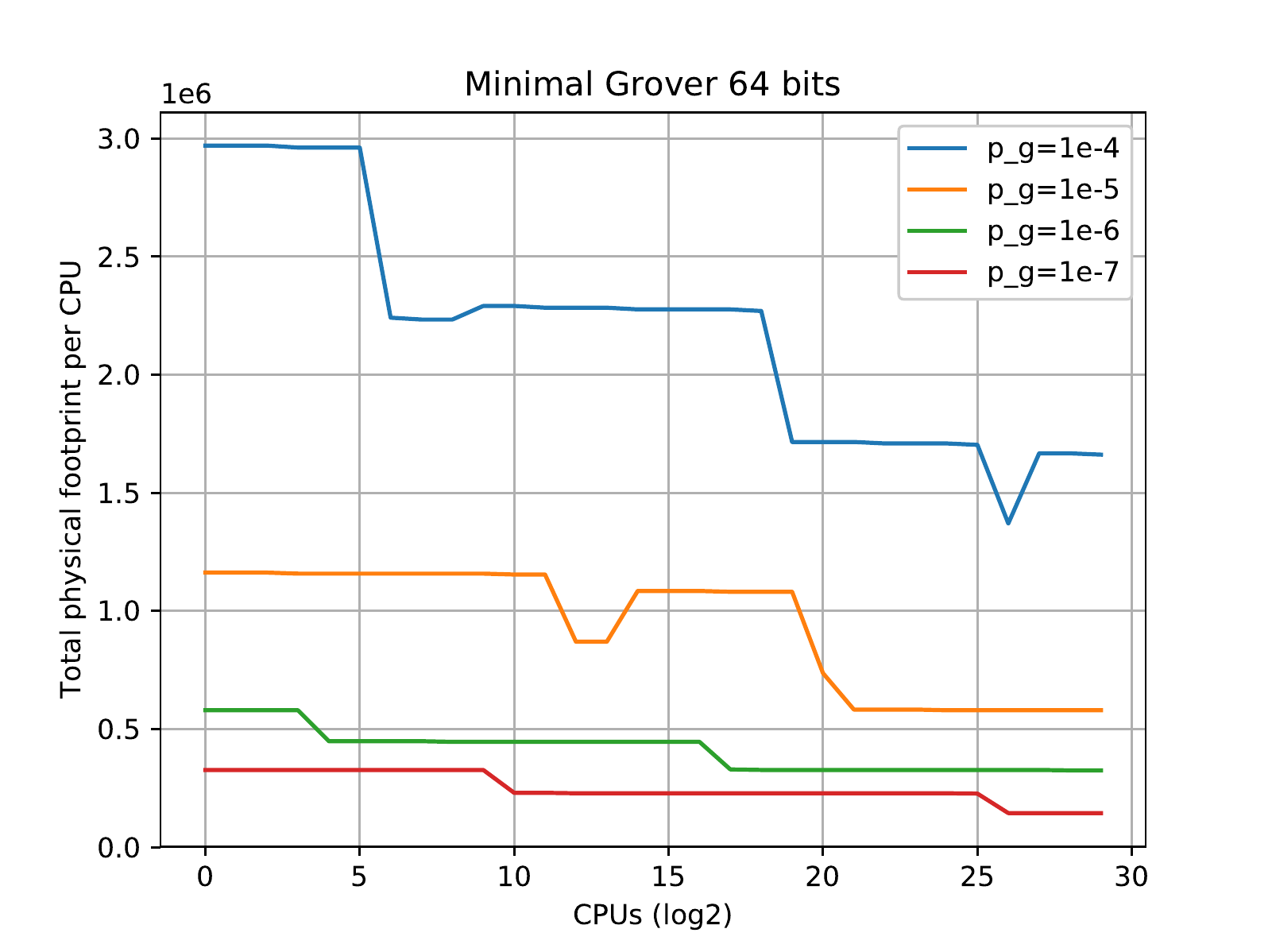}
	\captionof{figure}{Running Grover's algorithm with a trivial oracle, for a searching space of size $2^{64}$. Physical footprint (physical qubits) per processor, as a function of the number of processors ($\log_2$ scale).}
      	\label{fgr:minimal_grover_64_phys}
        \includegraphics[width=0.429\textwidth]{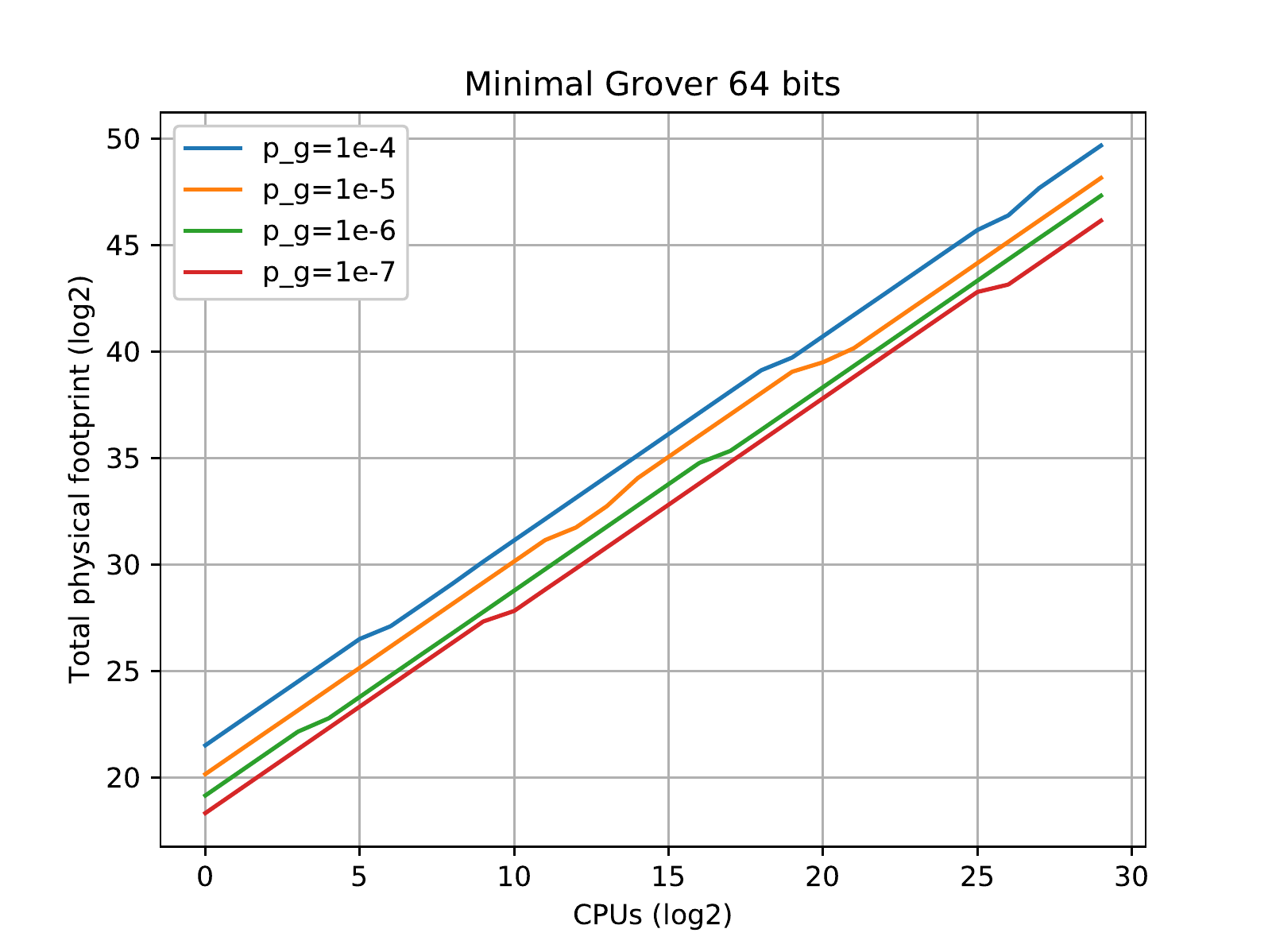}
	\captionof{figure}{Running Grover's algorithm with a trivial oracle, for a searching space of size $2^{64}$. Total physical footprint (physical qubits), as a function of the number of processors ($\log_2$ scale). Note that the qubits are not correlated across processors.}
      	\label{fgr:minimal_grover_64_phys_total}

\subsection{Searching space of size $2^{128}$}

        \includegraphics[width=0.429\textwidth]{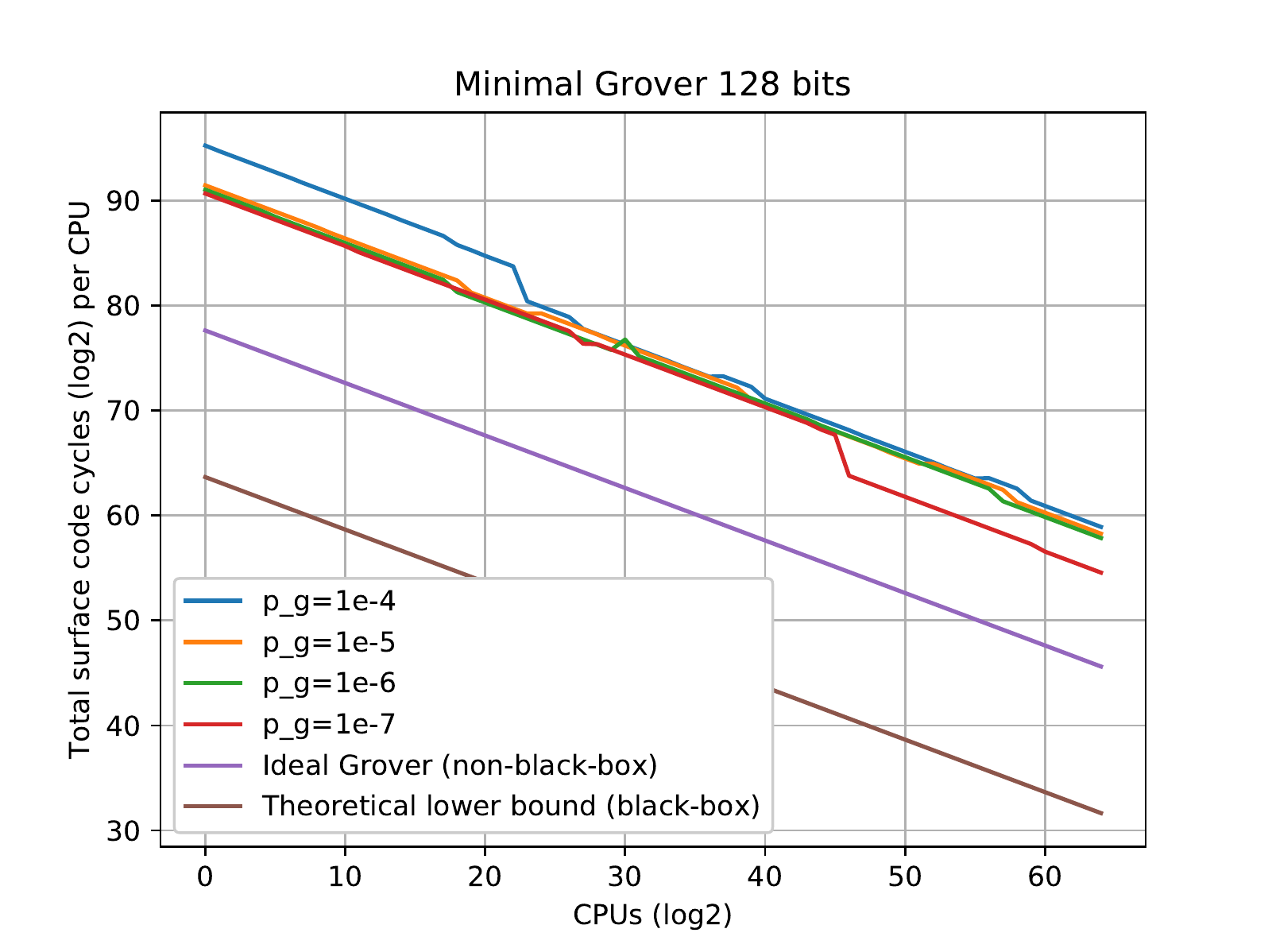}
      	\captionof{figure}{Running Grover's algorithm with a trivial oracle, for a searching space of size $2^{128}$. Required surface clock cycles per processor, as a function of the  number of processors ($\log_2$ scale).}
      	\label{fgr:minimal_grover_128_cycles}
        \includegraphics[width=0.429\textwidth]{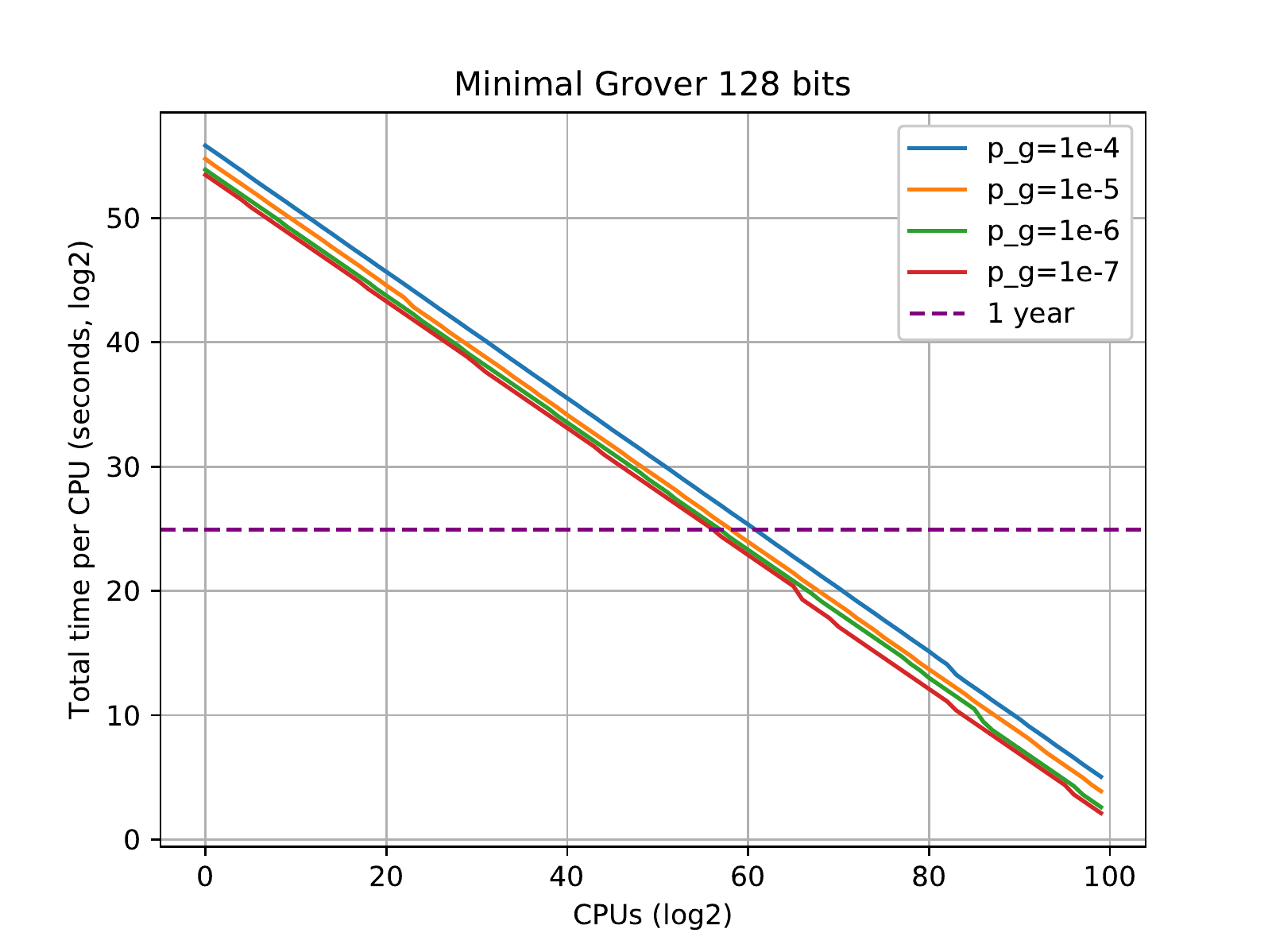}
      	\captionof{figure}{Running Grover's algorithm with a trivial oracle, for a searching space of size $2^{128}$. Required time per processor, as a function of the  number of processors ($\log_2$ scale).}
      	\label{fgr:minimal_grover_128_time}
        \includegraphics[width=0.429\textwidth]{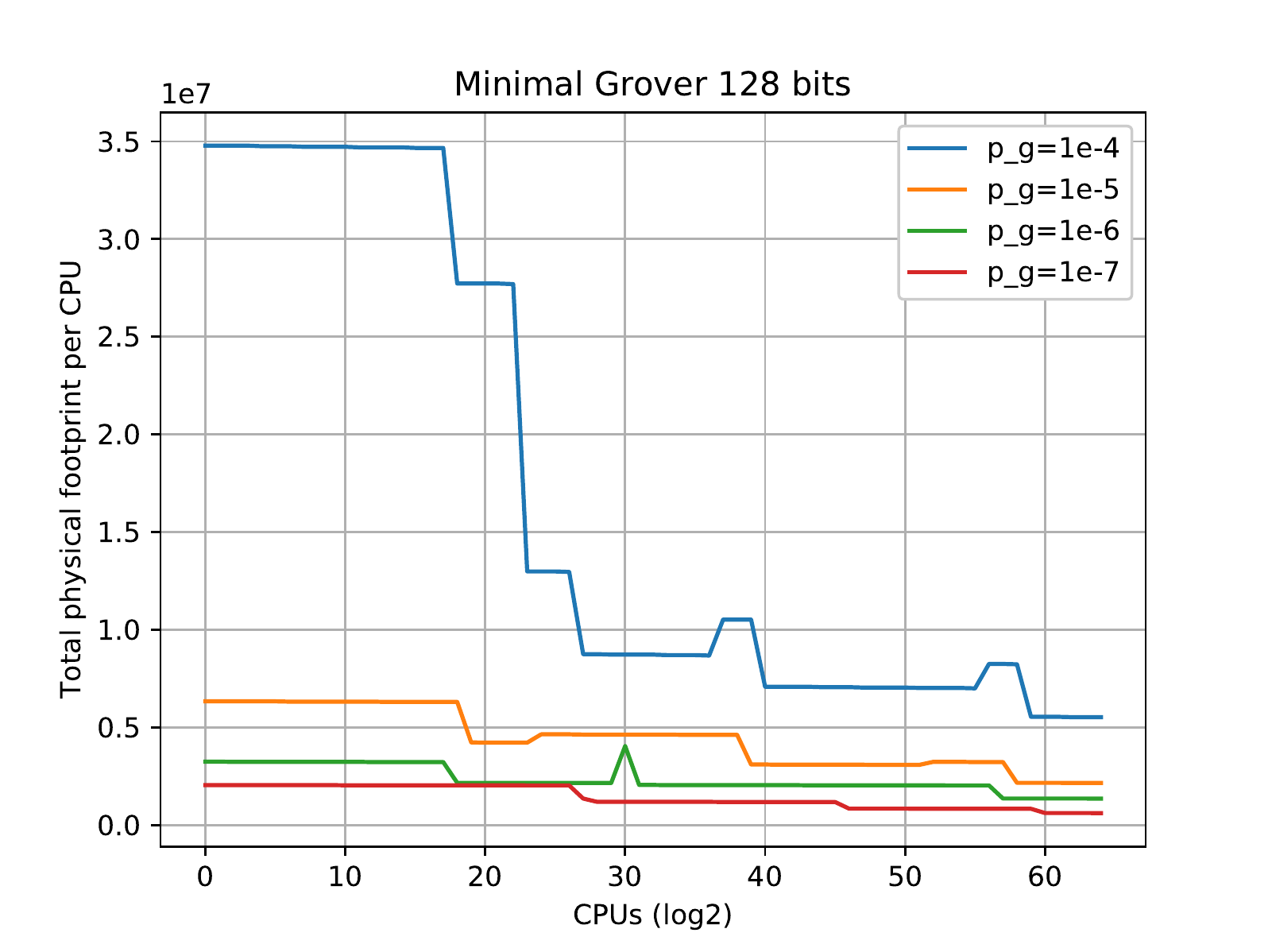}
	\captionof{figure}{Running Grover's algorithm with a trivial oracle, for a searching space of size $2^{128}$. Physical footprint (physical qubits) per processor, as a function of the number of processors ($\log_2$ scale).}
      	\label{fgr:minimal_grover_128_phys}
        \includegraphics[width=0.429\textwidth]{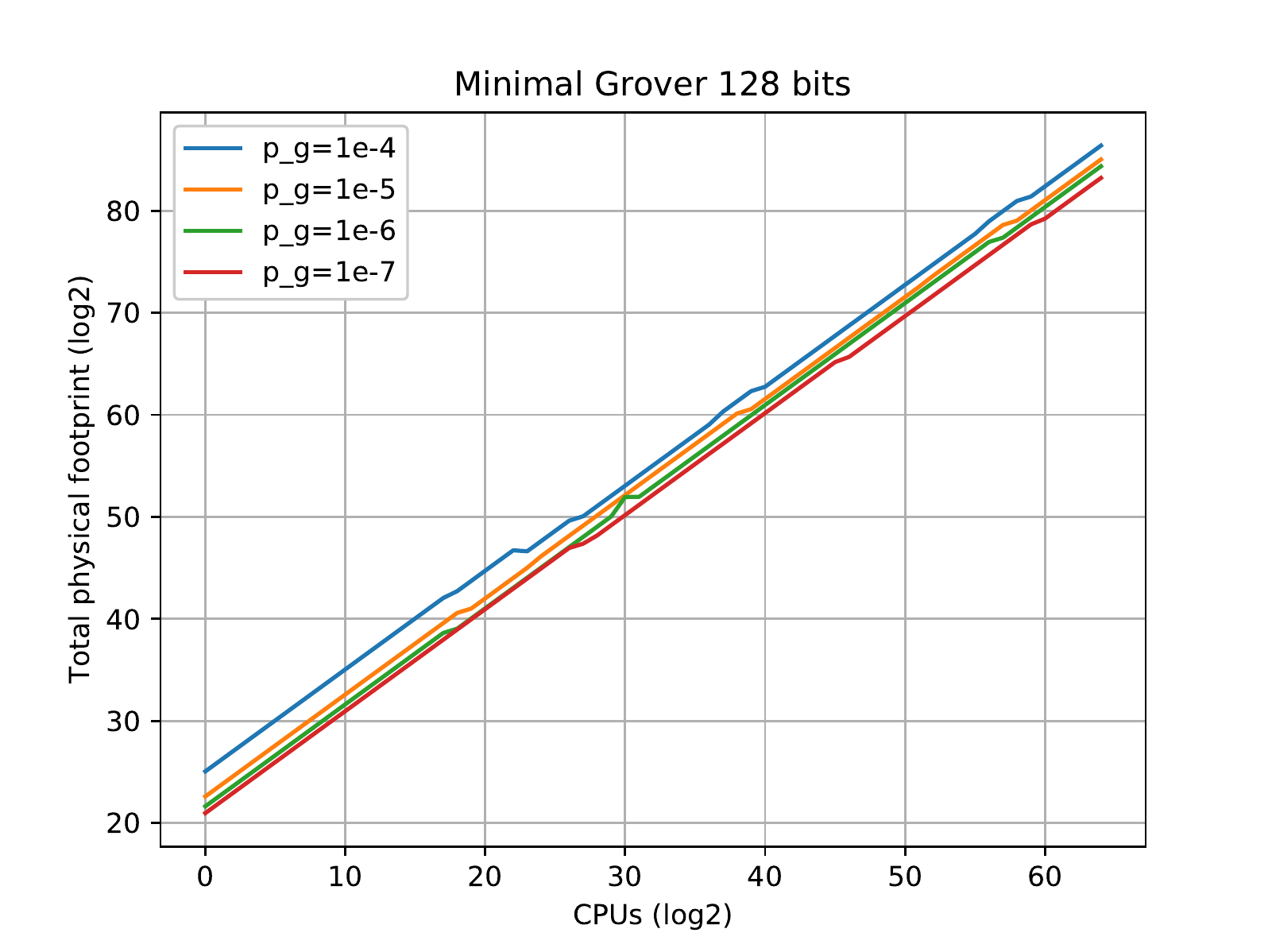}
	\captionof{figure}{Running Grover's algorithm with a trivial oracle, for a searching space of size $2^{128}$. Total physical footprint (physical qubits), as a function of the number of processors ($\log_2$ scale). Note that the qubits are not correlated across processors.}
      	\label{fgr:minimal_grover_128_phys_total}

\subsection{Searching space of size $2^{256}$}

        \includegraphics[width=0.429\textwidth]{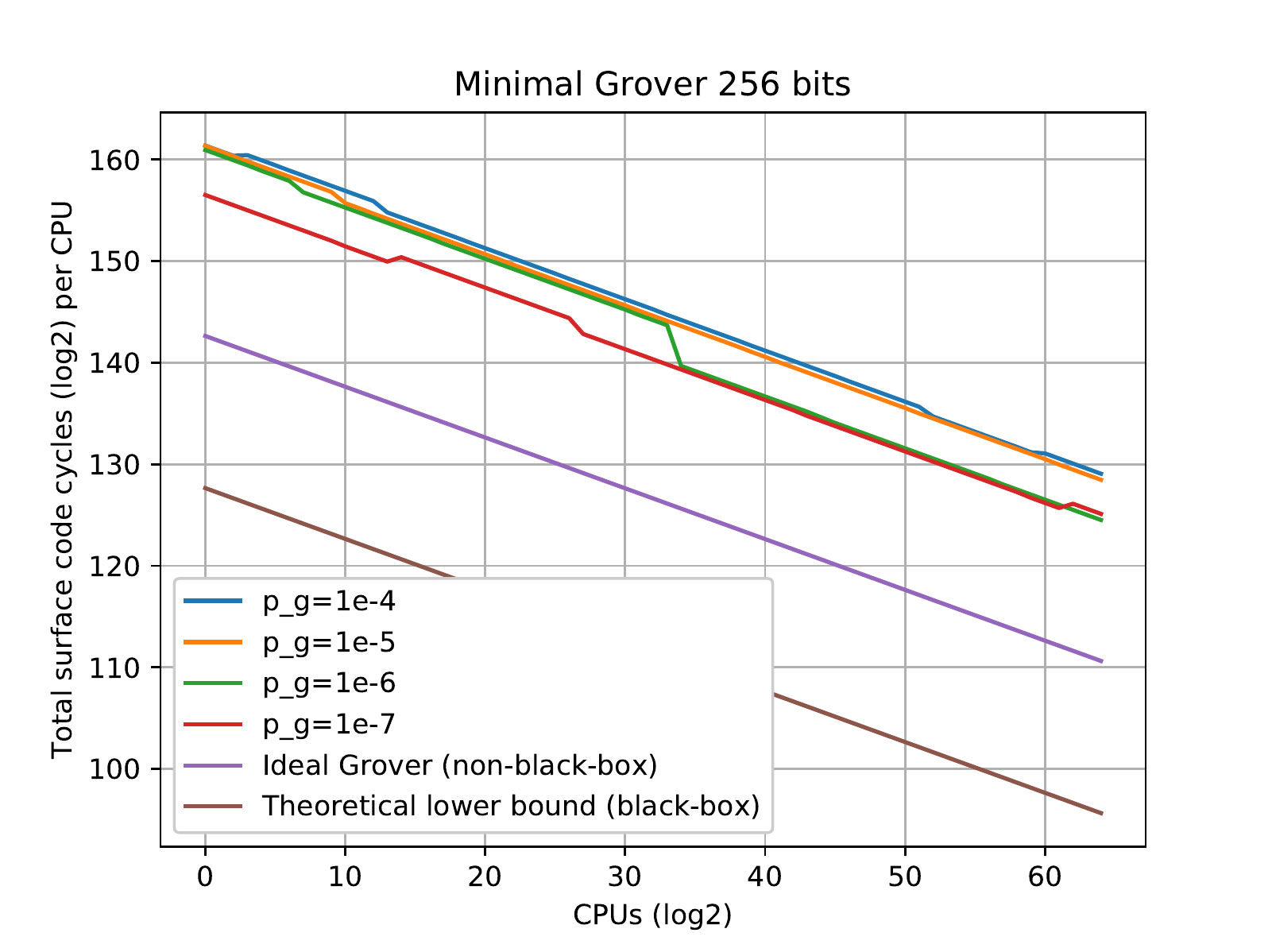}
      	\captionof{figure}{Running Grover's algorithm with a trivial oracle, for a searching space of size $2^{256}$. Required surface clock cycles per processor, as a function of the  number of processors ($\log_2$ scale).}
      	\label{fgr:minimal_grover_256_cycles}
        \includegraphics[width=0.429\textwidth]{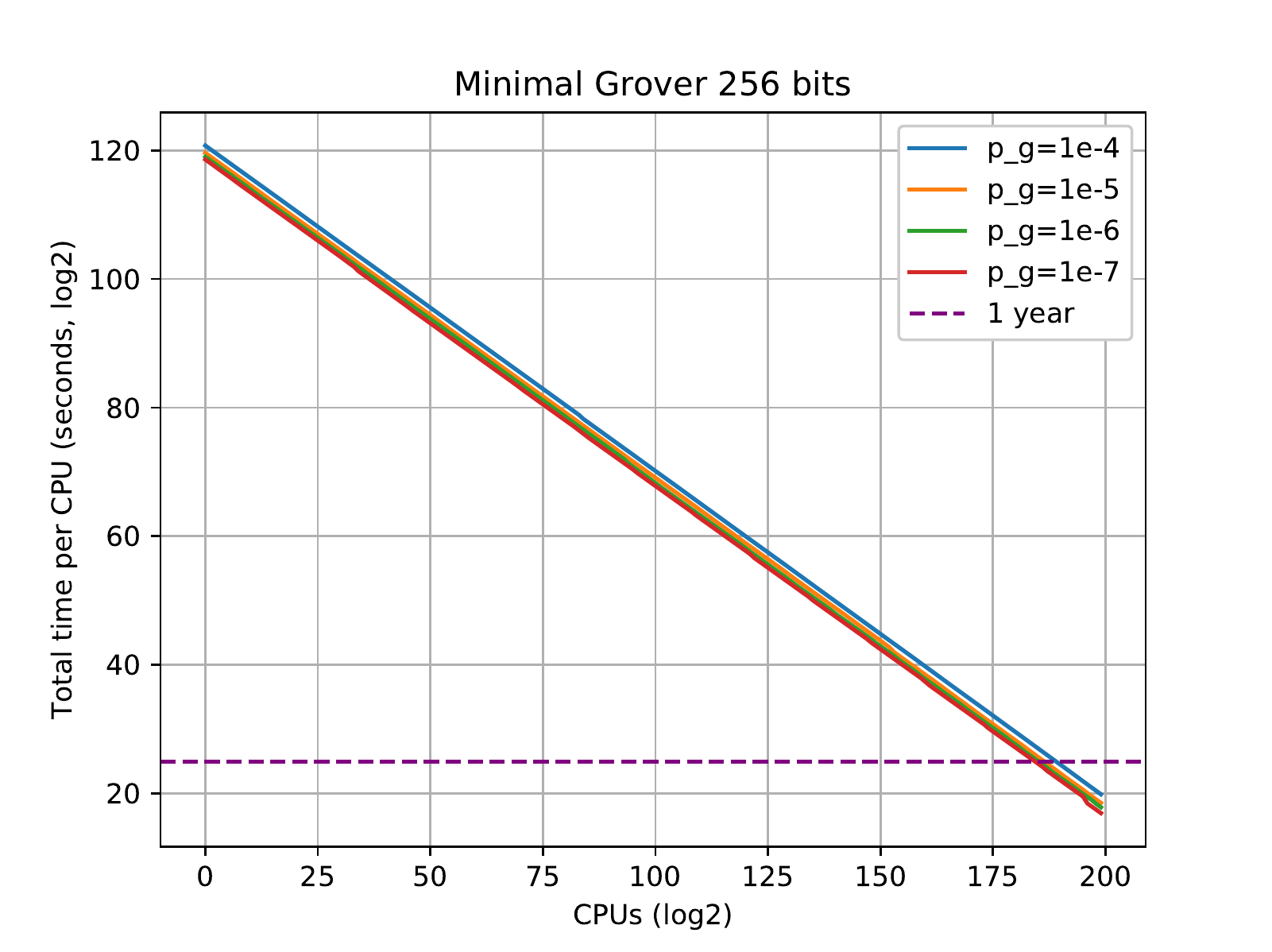}
      	\caption{Running Grover's algorithm with a trivial oracle, for a searching space of size $2^{256}$. Required time per processor, as a function of the  number of processors ($\log_2$ scale).}
      	\label{fgr:minimal_grover_256_time}
        \includegraphics[width=0.429\textwidth]{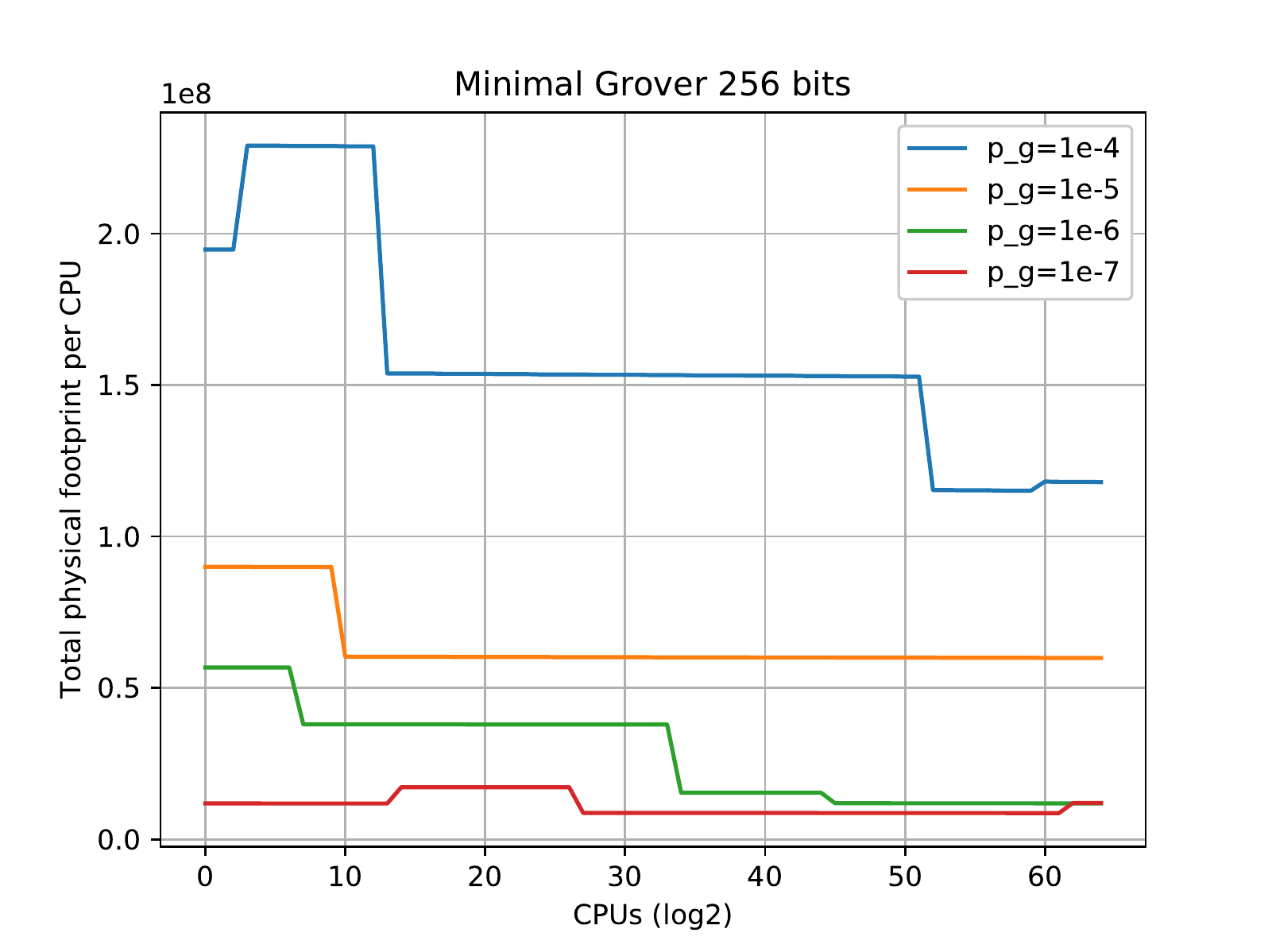}
	\caption{Running Grover's algorithm with a trivial oracle, for a searching space of size $2^{256}$. Physical footprint (physical qubits) per processor, as a function of the number of processors ($\log_2$ scale).}
      	\label{fgr:minimal_grover_256_phys}
        \includegraphics[width=0.429\textwidth]{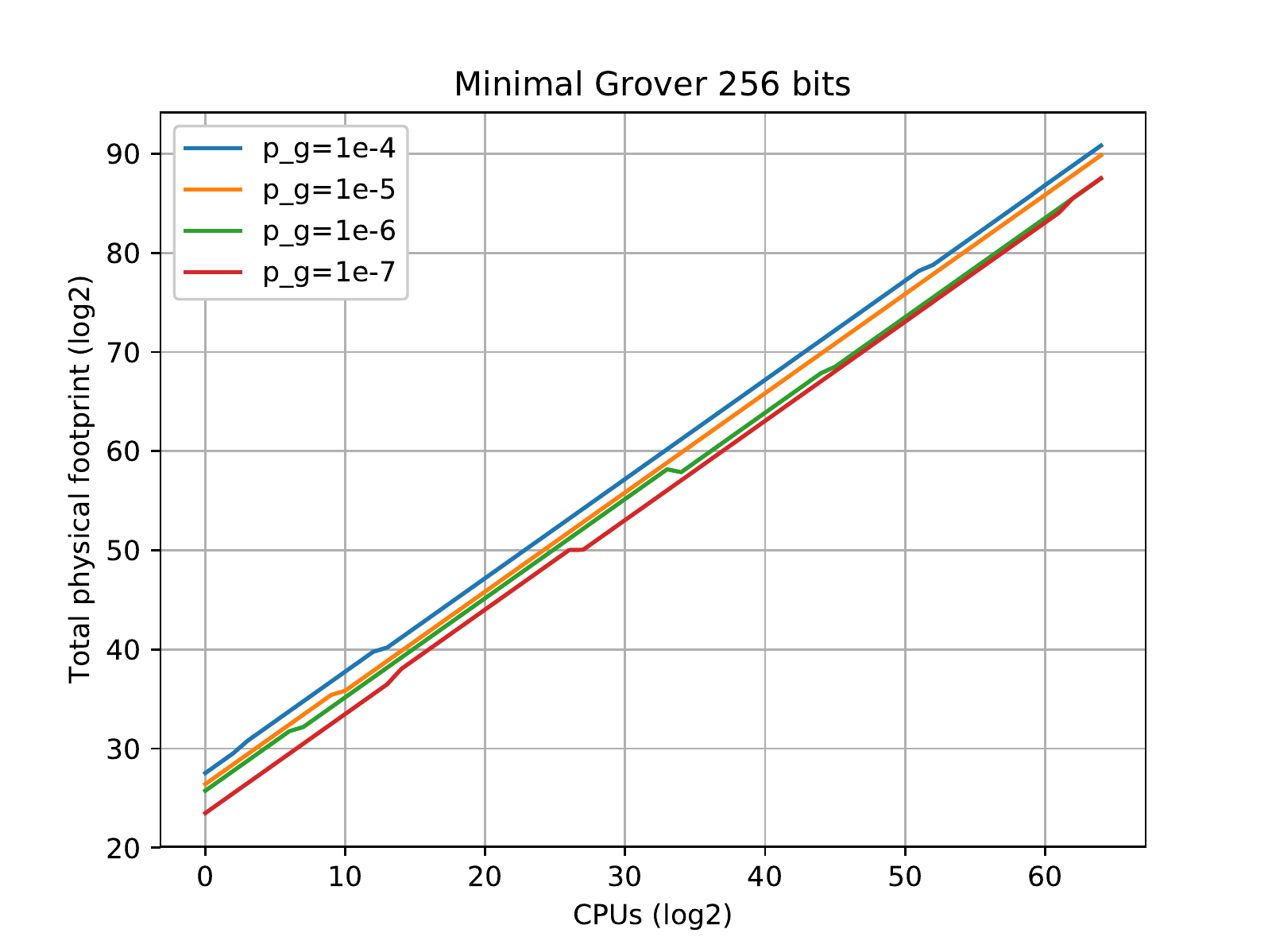}
	\caption{Running Grover's algorithm with a trivial oracle, for a searching space of size $2^{256}$. Total physical footprint (physical qubits), as a function of the number of processors ($\log_2$ scale). Note that the qubits are not correlated across processors.}
      	\label{fgr:minimal_grover_256_phys_total}

\section{RSA schemes\label{sct::rsa}}
In the following section we compute the space/time tradeoffs for attacking public-key cryptographic schemes based on factoring large numbers, 
namely RSA-1024, RSA-2048, RSA-3072, RSA-4096, RSA-7680 and RSA-15360.
For each scheme, we plot the space/time tradeoff points then fit it with a third degree polynomial, for $p_g=10^{-3}$ and $p_g=10^{-5}$, respectively.

\subsection{RSA-1024}

\includegraphics[width=0.475\textwidth]{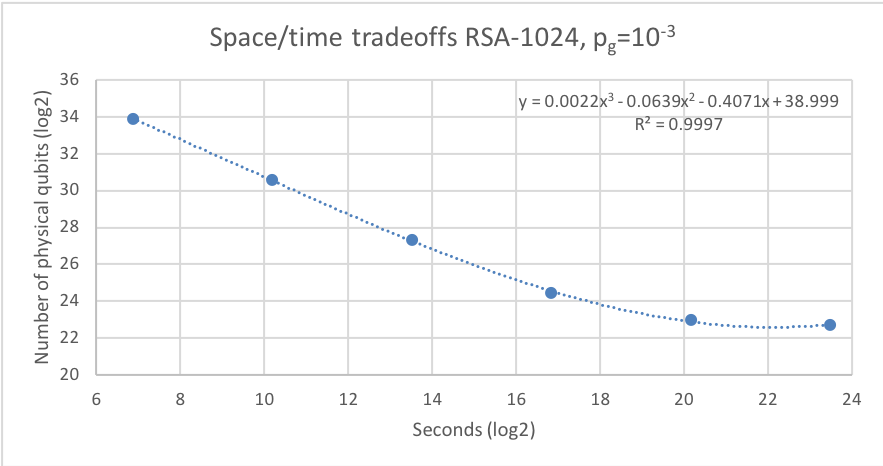}
\captionof{figure}{RSA-1024 space/time tradeoffs with physical error rate per gate $p_g=10^{-3}$. The scale is logarithmic (base 2). Approximately $y(16.3987) \approx 3.01\times 10^7$ physical qubits are required to break the scheme in one day (24 hours). The number of T gates in the circuit is $3.01\times 10^{11}$, the corresponding number of logical qubits is 2050, and the total number of surface code cycles is $5.86\times 10^{13}$. The quantity $R^2$ represents the coefficient of determination (closer to 1, better the fitting). The classical security parameter is approximately 80 bits.}
\label{fgr:rsa1024a} 

\includegraphics[width=0.475\textwidth]{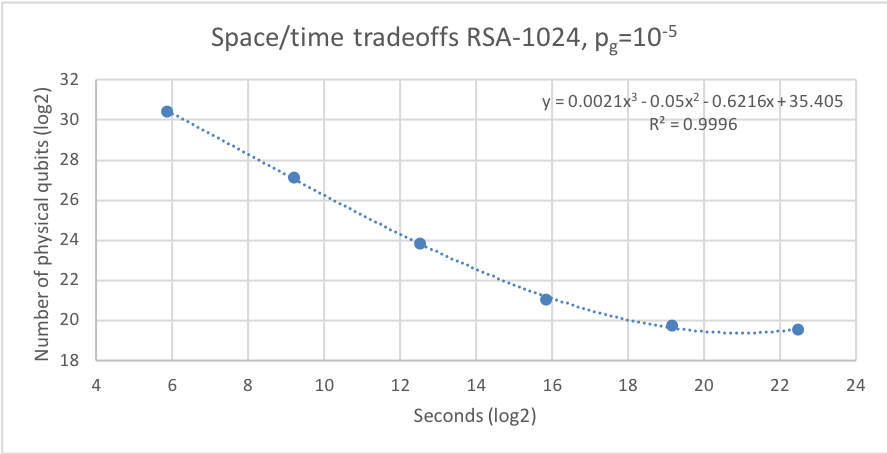}
\captionof{figure}{RSA-1024 space/time tradeoffs with physical error rate per gate $p_g=10^{-5}$. The scale is logarithmic (base 2). Approximately $y(16.3987) \approx 2.14\times 10^6$ physical qubits are required to break the scheme in one day (24 hours). The number of T gates in the circuit is $3.01\times 10^{11}$, the corresponding number of logical qubits is 2050, and the total number of surface code cycles is $2.93\times 10^{13}$. The classical security parameter is approximately 80 bits.}
\label{fgr:rsa1024b}

\subsection{RSA-2048}

\includegraphics[width=0.475\textwidth]{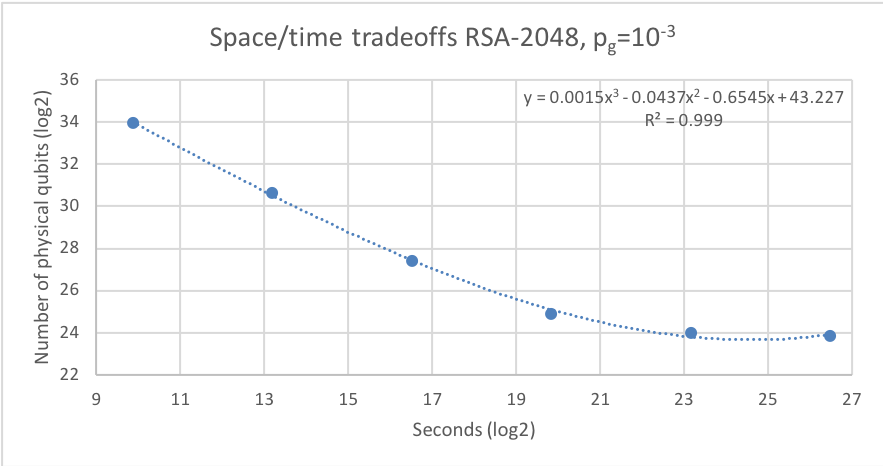}
\captionof{figure}{RSA-2048 space/time tradeoffs with physical error rate per gate $p_g=10^{-3}$. The scale is logarithmic (base 2). Approximately $y(16.3987) \approx 1.72\times 10^8$ physical qubits are required to break the scheme in one day (24 hours). The number of T gates in the circuit is $2.41\times 10^{12}$, the corresponding number of logical qubits is 4098, and the total number of surface code cycles is $4.69\times 10^{14}$. The classical security parameter is approximately 112 bits.}
\label{fgr:rsa2048a}



\includegraphics[width=0.475\textwidth]{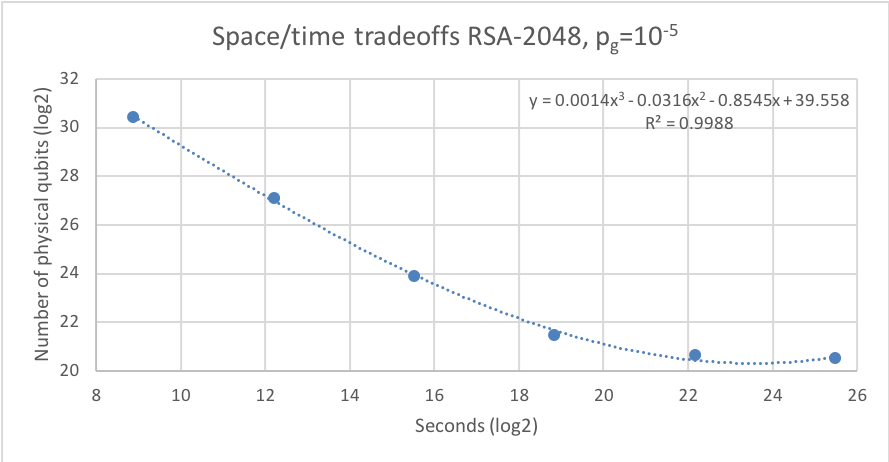}
\captionof{figure}{RSA-2048 space/time tradeoffs with physical error rate per gate $p_g=10^{-5}$. The scale is logarithmic (base 2). Approximately $y(16.3987) \approx 9.78\times 10^6$ physical qubits are required to break the scheme in one day (24 hours). The number of T gates in the circuit is $2.41\times 10^{12}$, the corresponding number of logical qubits is 4098, and the total number of surface code cycles is $2.35\times 10^{14}$. The classical security parameter is approximately 112 bits.}
\label{fgr:rsa2048b}

\subsection{RSA-3072}

\includegraphics[width=0.475\textwidth]{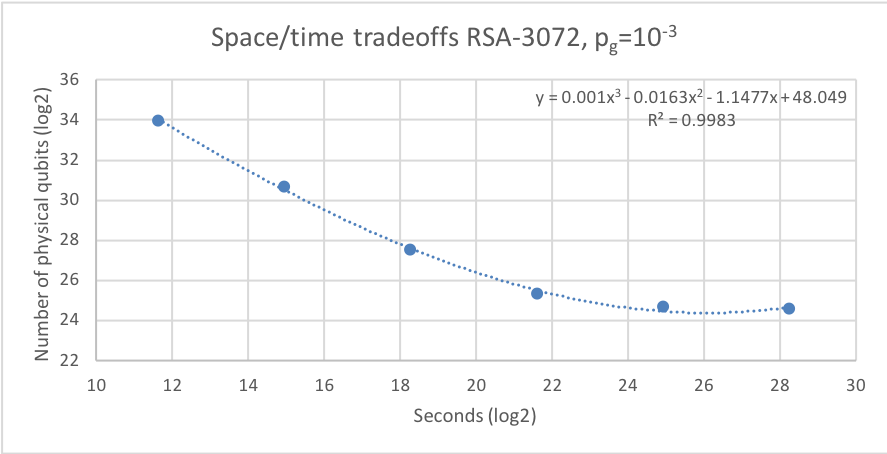}
\captionof{figure}{RSA-3072 space/time tradeoffs with physical error rate per gate $p_g=10^{-3}$. The scale is logarithmic (base 2). Approximately $y(16.3987) \approx 6.41\times 10^8$ physical qubits are required to break the scheme in one day (24 hours). The number of T gates in the circuit is $8.12\times 10^{12}$, the corresponding number of logical qubits is 6146, and the total number of surface code cycles is $1.58\times 10^{15}$. The classical security parameter is approximately 128 bits.}
\label{fgr:rsa3072a}

\includegraphics[width=0.475\textwidth]{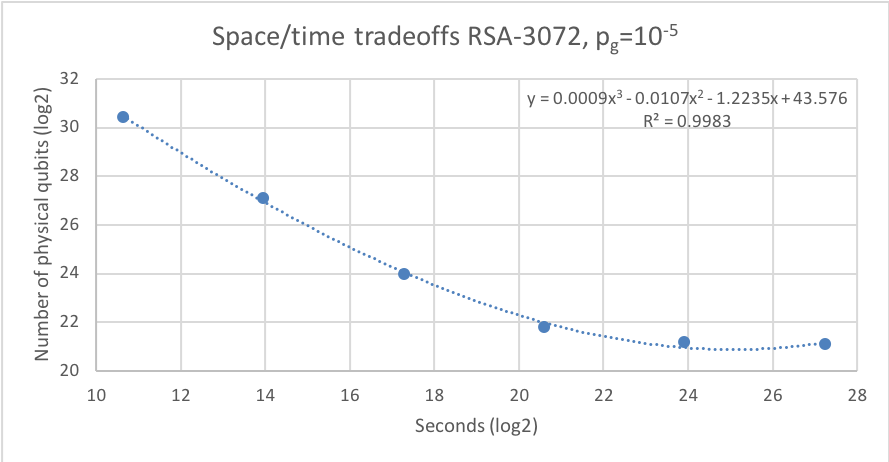}
\captionof{figure}{RSA-3072 space/time tradeoffs with physical error rate per gate $p_g=10^{-5}$. The scale is logarithmic (base 2). Approximately $y(16.3987) \approx 2.55\times 10^7$ physical qubits are required to break the scheme in one day (24 hours). The number of T gates in the circuit is $8.12\times 10^{12}$, the corresponding number of logical qubits is 6146, and the total number of surface code cycles is $7.91\times 10^{14}$. The classical security parameter is approximately 128 bits.}
\label{fgr:rsa3072b}

\subsection{RSA-4096}

\includegraphics[width=0.475\textwidth]{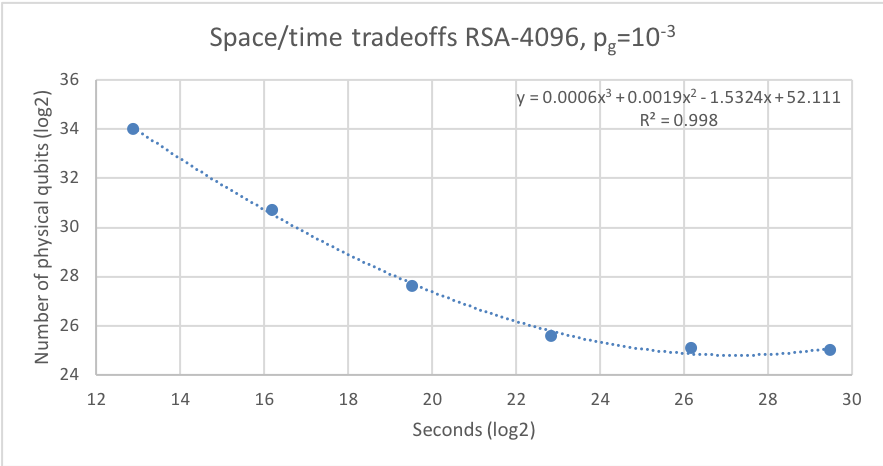}
\captionof{figure}{RSA-4096 space/time tradeoffs with physical error rate per gate $p_g=10^{-3}$. The scale is logarithmic (base 2). Approximately $y(16.3987) \approx 1.18\times 10^9$ physical qubits are required to break the scheme in one day (24 hours). The number of T gates in the circuit is $1.92\times 10^{13}$, the corresponding number of logical qubits is 8194, and the total number of surface code cycles is $3.75\times 10^{15}$. The classical security parameter is approximatively approximately 156 bits.}
\label{fgr:rsa4096a}

\includegraphics[width=0.475\textwidth]{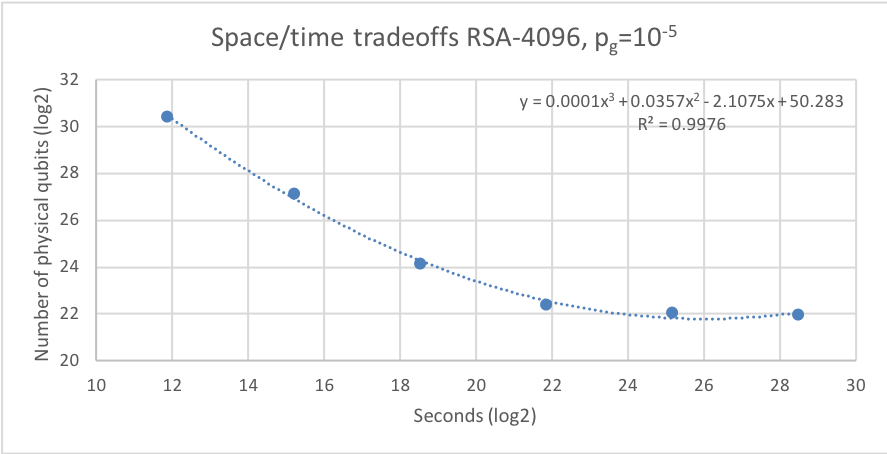}
\captionof{figure}{RSA-4096 space/time tradeoffs with physical error rate per gate $p_g=10^{-5}$. The scale is logarithmic (base 2). Approximately $y(16.3987) \approx 5.70\times 10^7$ physical qubits are required to break the scheme in one day (24 hours). The number of T gates in the circuit is $1.92\times 10^{13}$, the corresponding number of logical qubits is 8194, and the total number of surface code cycles is $1.88\times 10^{15}$. The classical security parameter is approximatively approximately 156 bits.}
\label{fgr:rsa4096b}

\subsection{RSA-7680}

\includegraphics[width=0.475\textwidth]{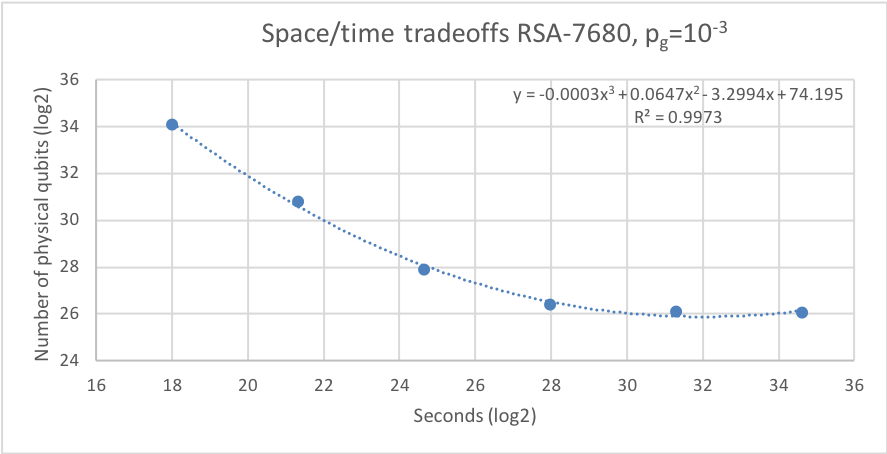}
\captionof{figure}{RSA-7680 space/time tradeoffs with physical error rate per gate $p_g=10^{-3}$. The scale is logarithmic (base 2). Approximately $y(16.3987) \approx 7.70\times 10^{10}$ physical qubits are required to break the scheme in one day (24 hours). The number of T gates in the circuit is $1.27\times 10^{14}$, the corresponding number of logical qubits is 15362, and the total number of surface code cycles is $2.64\times 10^{16}$. The classical security parameter is approximately 192 bits.}
\label{fgr:rsa7680a}

\includegraphics[width=0.475\textwidth]{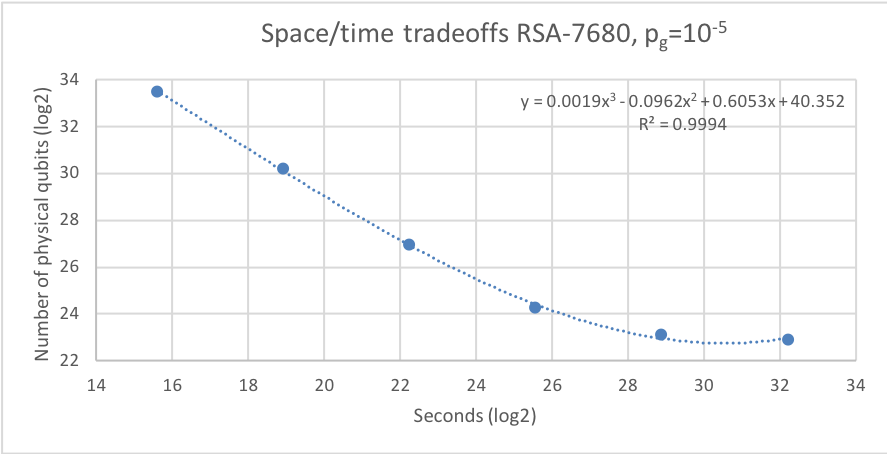}
\captionof{figure}{RSA-7680 space/time tradeoffs with physical error rate per gate $p_g=10^{-5}$. The scale is logarithmic (base 2). Approximately $y(16.3987) \approx 7.41\times 10^{9}$ physical qubits are required to break the scheme in one day (24 hours). The number of T gates in the circuit is $1.27\times 10^{14}$, the corresponding number of logical qubits is 15362, and the total number of surface code cycles is $2.47\times 10^{16}$. The classical security parameter is approximately 192 bits.}
\label{fgr:rsa7680b}

\subsection{RSA-15360}

\includegraphics[width=0.475\textwidth]{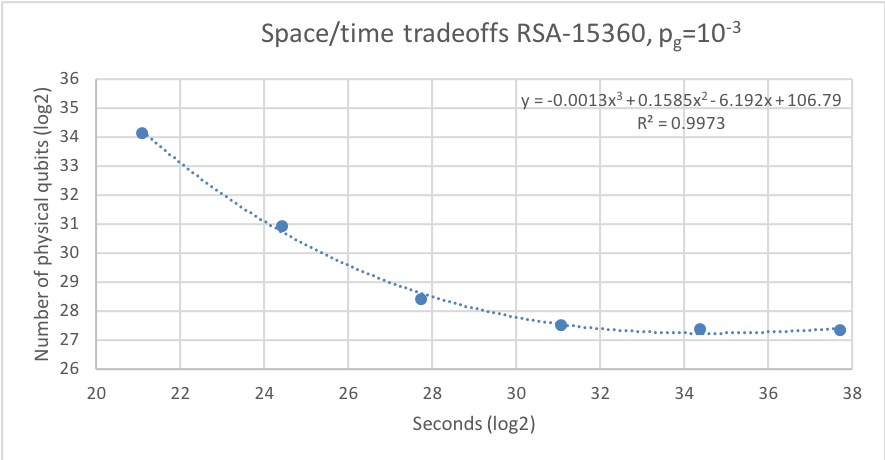}
\captionof{figure}{RSA-15360 space/time tradeoffs with physical error rate per gate $p_g=10^{-3}$. The scale is logarithmic (base 2). Approximately $y(16.3987) \approx 4.85\times 10^{12}$ physical qubits are required to break the scheme in one day (24 hours). The number of T gates in the circuit is $1.01\times 10^{15}$, the corresponding number of logical qubits is 30722, and the total number of surface code cycles is $2.24\times 10^{17}$. The classical security parameter is approximately 256 bits.}
\label{fgr:rsa15360a}

\includegraphics[width=0.475\textwidth]{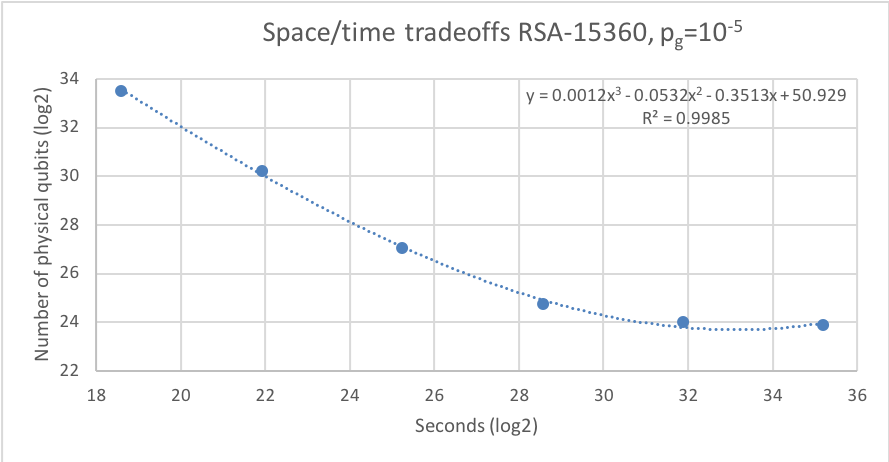}
\captionof{figure}{RSA-15360 space/time tradeoffs with physical error rate per gate $p_g=10^{-5}$. The scale is logarithmic (base 2). Approximately $y(16.3987) \approx 7.64\times 10^{10}$ physical qubits are required to break the scheme in one day (24 hours). The number of T gates in the circuit is $1.01\times 10^{15}$, the corresponding number of logical qubits is 30722, and the total number of surface code cycles is $1.98\times 10^{17}$. The classical security parameter is approximately 256 bits.}
\label{fgr:rsa15360b}

\section{Elliptic curve schemes\label{sct::ecc}}
In the following section we compute the space/time tradeoffs for attacking public-key cryptographic schemes based on solving the discrete logarithm 
problem in finite groups generated over elliptic curves, namely NIST P-160, NIST P-192, NIST P-224, NIST P-256, NIST P-384 and NIST P-521. For 
each scheme, we plot the space/time tradeoff points then fit it with a third degree polynomial, for $p_g=10^{-3}$ and $p_g=10^{-5}$, respectively. We 
used the logical circuits from~\cite{1706.06752}.

\subsection{NIST P-160}

\includegraphics[width=0.475\textwidth]{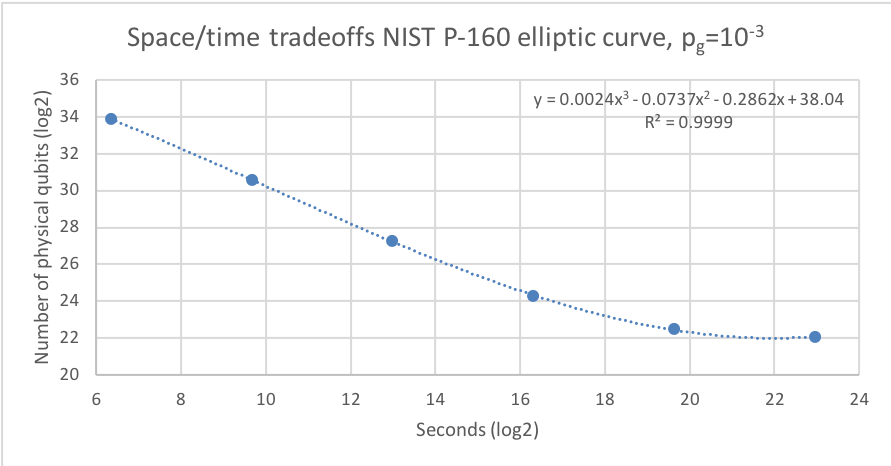}
\captionof{figure}{NIST P-160 elliptic curve space/time tradeoffs with physical error rate per gate $p_g=10^{-3}$. The scale is logarithmic (base 2). Approximately $y(16.3987) \approx 1.81\times 10^7$ physical qubits are required to break the scheme in one day (24 hours). The number of T gates in the circuit is $2.08\times 10^{11}$, the corresponding number of logical qubits is 1466, and the total number of surface code cycles is $4.05\times 10^{13}$. The classical security parameter is 80 bits.}
\label{fgr:p160a}

\includegraphics[width=0.475\textwidth]{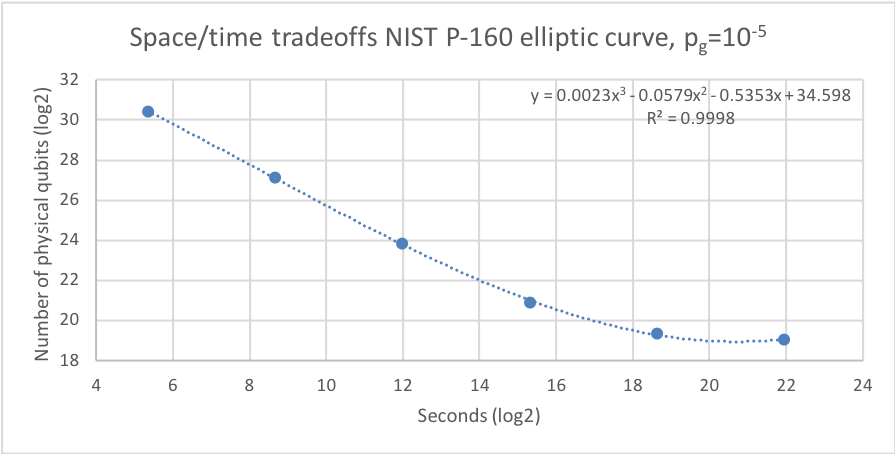}
\captionof{figure}{NIST P-160 elliptic curve space/time tradeoffs with physical error rate per gate $p_g=10^{-5}$. The scale is logarithmic (base 2). Approximately $y(16.3987) \approx 1.38\times 10^6$ physical qubits are required to break the scheme in one day (24 hours). The number of T gates in the circuit is $2.08\times 10^{11}$, the corresponding number of logical qubits is 1466, and the total number of surface code cycles is $2.03\times 10^{13}$. The classical security parameter is 80 bits.}
\label{fgr:p160b}

\subsection{NIST P-192}

\includegraphics[width=0.475\textwidth]{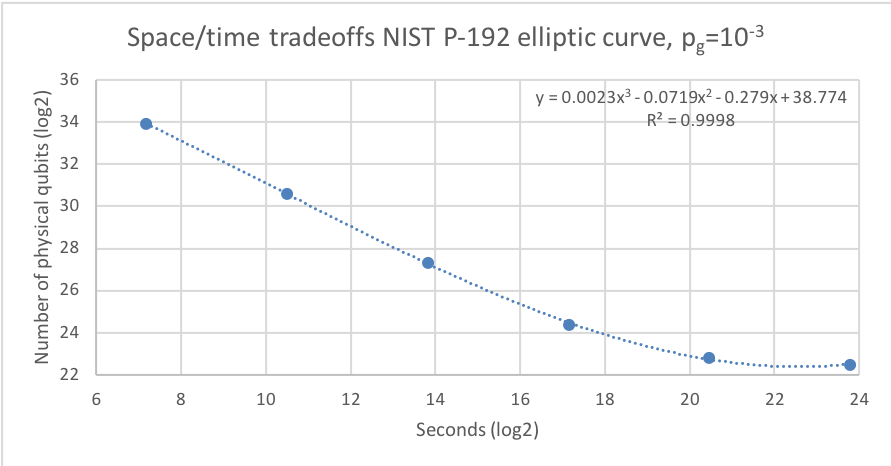}
\captionof{figure}{NIST P-192 space/time tradeoffs with physical error rate per gate $p_g=10^{-3}$. The scale is logarithmic (base 2). Approximately $y(16.3987) \approx 3.37\times 10^7$ physical qubits are required to break the scheme in one day (24 hours). The number of T gates in the circuit is $3.71\times 10^{11}$, the corresponding number of logical qubits is 1754, and the total number of surface code cycles is $7.23\times 10^{13}$. The classical security parameter is 96 bits.}
\label{fgr:p192a}

\includegraphics[width=0.475\textwidth]{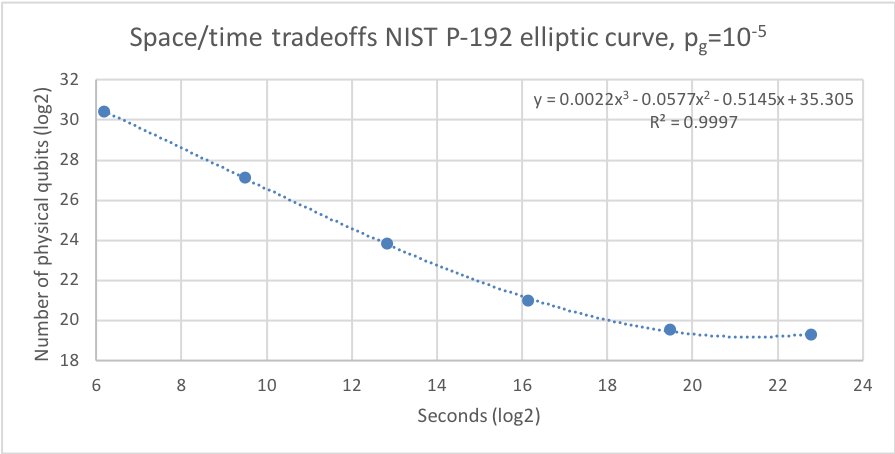}
\captionof{figure}{NIST P-192 space/time tradeoffs with physical error rate per gate $p_g=10^{-5}$. The scale is logarithmic (base 2). Approximately $y(16.3987) \approx 2.18\times 10^6$ physical qubits are required to break the scheme in one day (24 hours). The number of T gates in the circuit is $3.71\times 10^{11}$, the corresponding number of logical qubits is 1754, and the total number of surface code cycles is $3.62\times 10^{13}$. The classical security parameter is 96 bits.}
\label{fgr:p192b}

\subsection{NIST P-224}

\includegraphics[width=0.475\textwidth]{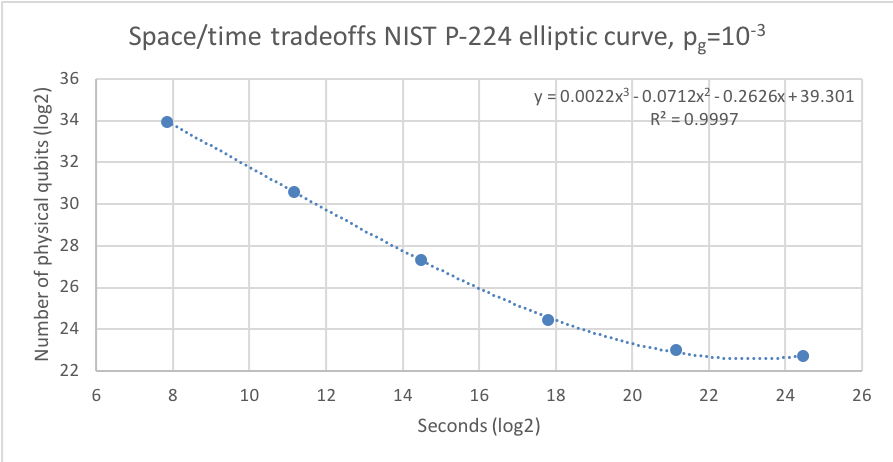}
\captionof{figure}{NIST P-224 elliptic curve space/time tradeoffs with physical error rate per gate $p_g=10^{-3}$. The scale is logarithmic (base 2). Approximately $y(16.3987) \approx 4.91\times 10^7$ physical qubits are required to break the scheme in one day (24 hours). The number of T gates in the circuit is $5.90\times 10^{11}$, the corresponding number of logical qubits is 2042, and the total number of surface code cycles is $1.15\times 10^{14}$. The classical security parameter is 112 bits.}
\label{fgr:p224a}

\includegraphics[width=0.475\textwidth]{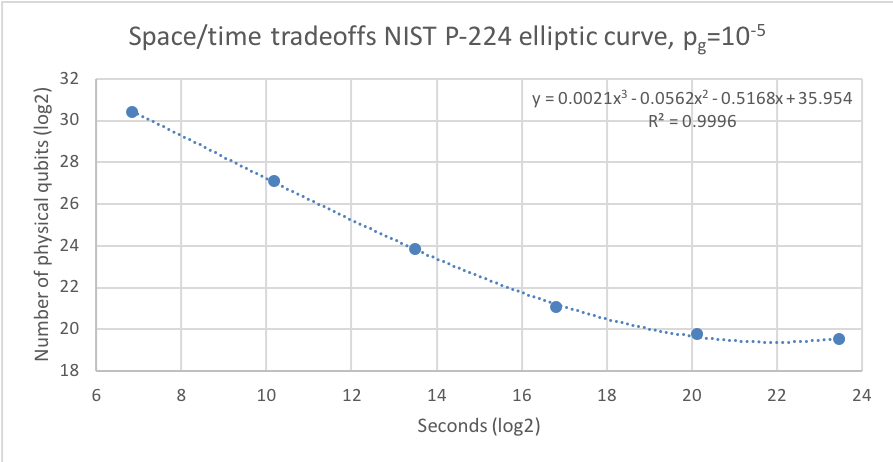}
\captionof{figure}{NIST P-224 elliptic curve space/time tradeoffs with physical error rate per gate $p_g=10^{-5}$. The scale is logarithmic (base 2). Approximately $y(16.3987) \approx 3.24\times 10^6$ physical qubits are required to break the scheme in one day (24 hours). The number of T gates in the circuit is $5.90\times 10^{11}$, the corresponding number of logical qubits is 2042, and the total number of surface code cycles is $5.75\times 10^{13}$. The classical security parameter is 112 bits.}
\label{fgr:p224b}

\subsection{NIST P-256}

\includegraphics[width=0.475\textwidth]{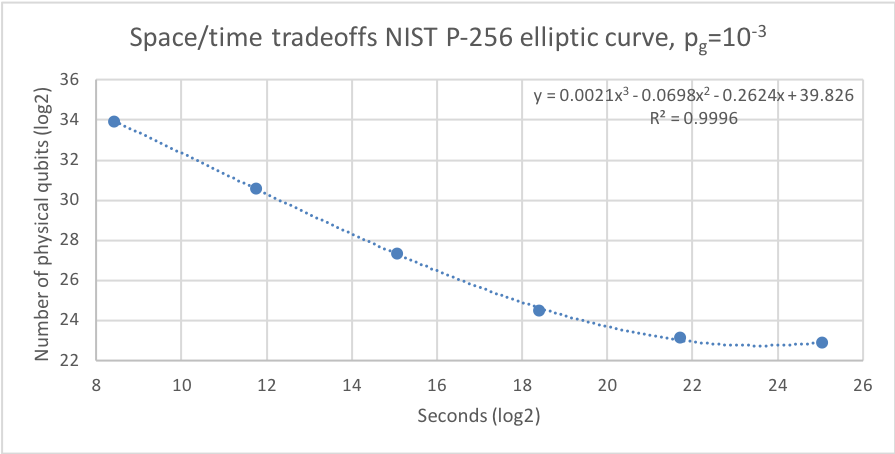}
\captionof{figure}{NIST P-256 elliptic curve space/time tradeoffs with physical error rate per gate $p_g=10^{-3}$. The scale is logarithmic (base 2). Approximately $y(16.3987) \approx 6.77\times 10^7$ physical qubits are required to break the scheme in one day (24 hours). The number of T gates in the circuit is $8.82\times 10^{11}$, the corresponding number of logical qubits is 2330, and the total number of surface code cycles is $1.72\times 10^{14}$. The classical security parameter is 128 bits.}
\label{fgr:p256a}

\includegraphics[width=0.475\textwidth]{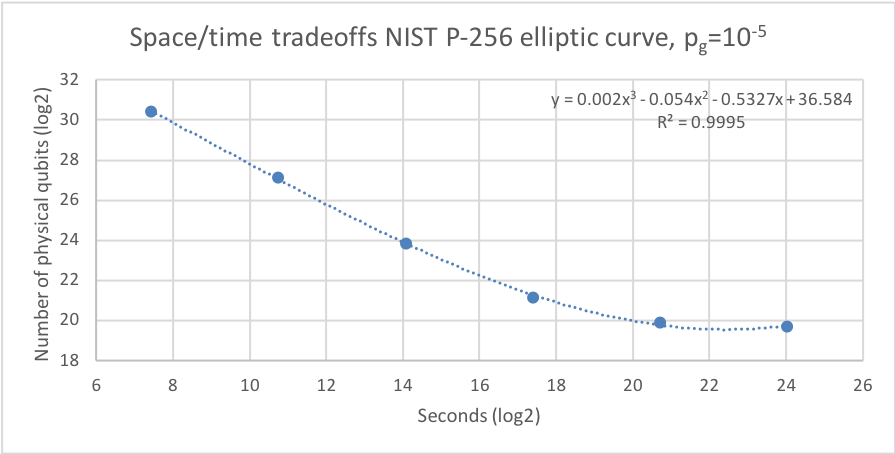}
\captionof{figure}{NIST P-256 elliptic curve space/time tradeoffs with physical error rate per gate $p_g=10^{-5}$. The scale is logarithmic (base 2). Approximately $y(16.3987) \approx 4.64\times 10^6$ physical qubits are required to break the scheme in one day (24 hours). The number of T gates in the circuit is $8.82\times 10^{11}$, the corresponding number of logical qubits is 2330, and the total number of surface code cycles is $8.60\times 10^{13}$. The classical security parameter is 128 bits.}
\label{fgr:p256b}

\subsection{NIST P-384}

\includegraphics[width=0.475\textwidth]{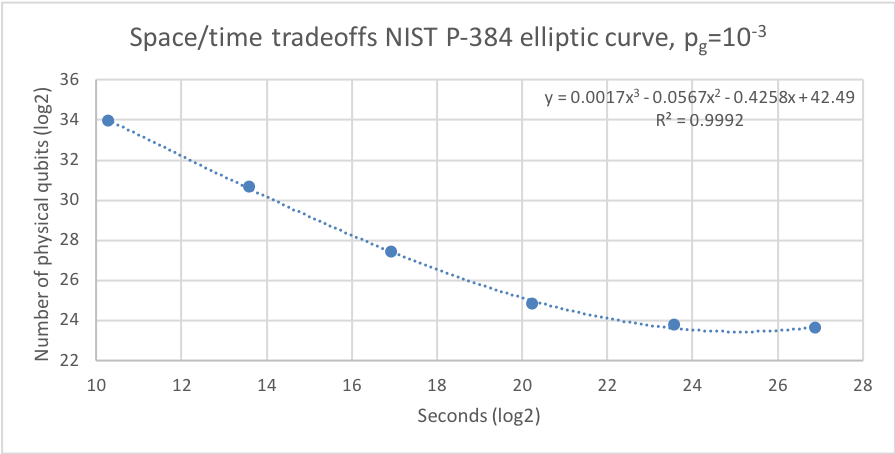}
\captionof{figure}{NIST P-384 elliptic curve space/time tradeoffs with physical error rate per gate $p_g=10^{-3}$. The scale is logarithmic (base 2). Approximately $y(16.3987) \approx 2.27\times 10^8$ physical qubits are required to break the scheme in one day (24 hours). The number of T gates in the circuit is $3.16\times 10^{12}$, the corresponding number of logical qubits is 3484, and the total number of surface code cycles is $6.17\times 10^{14}$. The classical security parameter is 192 bits.}
\label{fgr:p384a}

\includegraphics[width=0.475\textwidth]{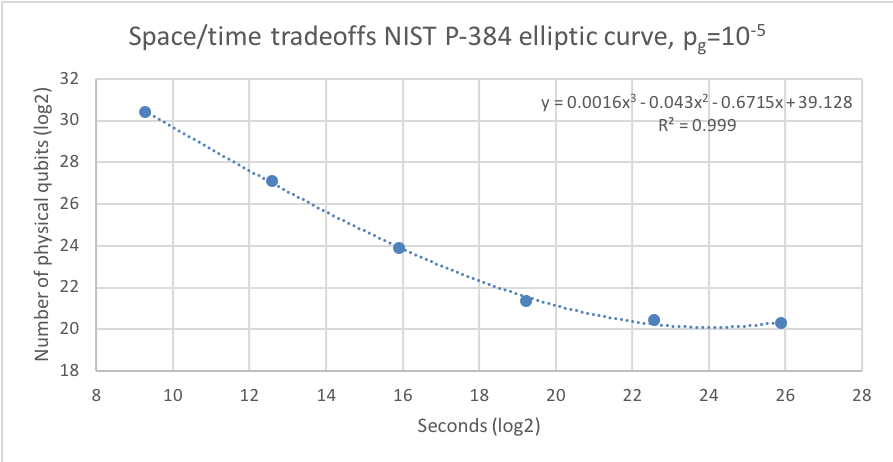}
\captionof{figure}{NIST P-384 elliptic curve space/time tradeoffs with physical error rate per gate $p_g=10^{-5}$. The scale is logarithmic (base 2). Approximately $y(16.3987) \approx 1.28\times 10^7$ physical qubits are required to break the scheme in one day (24 hours). The number of T gates in the circuit is $3.16\times 10^{12}$, the corresponding number of logical qubits is 3484, and the total number of surface code cycles is $3.08\times 10^{14}$. The classical security parameter is 192 bits.}
\label{fgr:p384b}

\subsection{NIST P-521}

\includegraphics[width=0.475\textwidth]{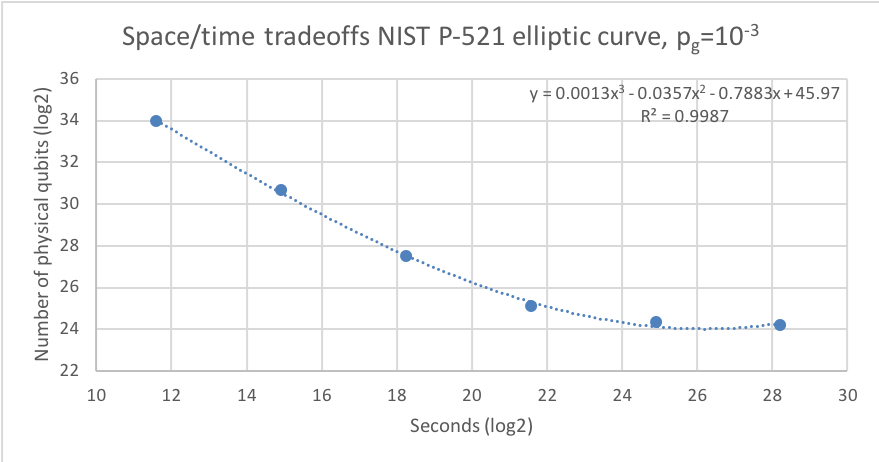}
\captionof{figure}{NIST P-521 elliptic curve space/time tradeoffs with physical error rate per gate $p_g=10^{-3}$. The scale is logarithmic (base 2). Approximately $y(16.3987) \approx 6.06\times 10^8$ physical qubits are required to break the scheme in one day (24 hours). The number of T gates in the circuit is $7.98\times 10^{12}$, the corresponding number of logical qubits is 4719, and the total number of surface code cycles is $1.56\times 10^{15}$. The classical security parameter is 256 bits.}
\label{fgr:p521a}

\includegraphics[width=0.475\textwidth]{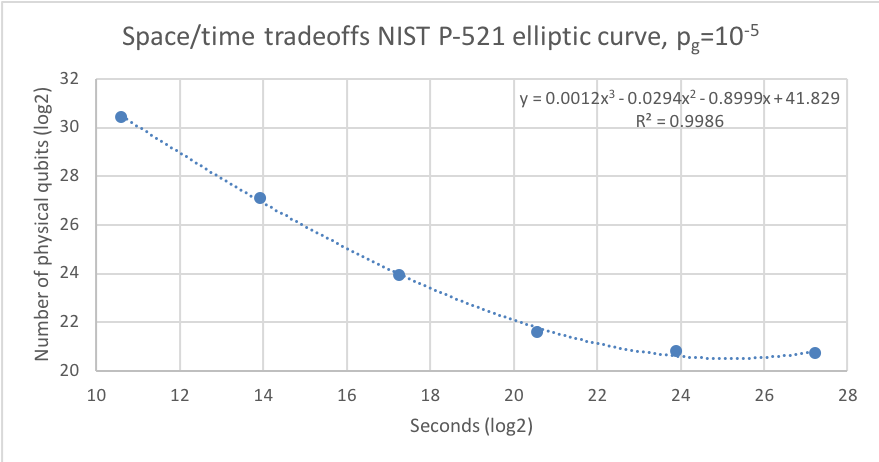}
\captionof{figure}{NIST P-521 elliptic curve space/time tradeoffs with physical error rate per gate $p_g=10^{-5}$. The scale is logarithmic (base 2). Approximately $y(16.3987) \approx 2.30\times 10^7$ physical qubits are required to break the scheme in one day (24 hours). The number of T gates in the circuit is $7.98\times 10^{12}$, the corresponding number of logical qubits is 4719, and the total number of surface code cycles is $7.78\times 10^{14}$. The classical security parameter is 256 bits.}
\label{fgr:p521b}

\section{Summary and conclusions}\label{sct::conclusion}
We analyzed the security of several widely used symmetric ciphers and hash functions against parallelized quantum adversaries. We computed the security parameter, wall-time and physical footprint for each cryptographic primitive. Our attack model was based on a brute force searching via a parallelized version of Grover's algorithm, assuming a surface-code fault-tolerant architecture based on defects and braiding techniques.

It is worth noting that throughout we are assuming that brute-force search where we treat the cryptographic function as a black-box is essentially the optimal attack against SHA and AES, which is currently believed to be the case.

Some symmetric key algorithms are susceptible in a model that permits ``superposition attacks''~\cite{quantph.1602.05973}. In most realistic instances, these attacks are not practical, however they do shed light on the limitations of certain security proof methods in a quantum context, and remind us that we shouldn't take for granted that non-trivial attacks on symmetric key cryptography may be possible.
For example, very recently, there have been several cryptanalysis results~\cite{1712.06239} and~\cite{1802.03856} that attempt to reduce breaking some symmetric algorithms to solving a system of non-linear equations.  Solving these non-linear equations is then attacked using a modified version of the quantum linear equation solver algorithm~\cite{PhysRevLett.103.150502}. The results are heavily dependent on the condition number of the non-linear system, which turns to be hard to compute (it is not known for most ciphers and hash functions such as AES or SHA). Provided the condition number is relatively small, then one may get an  advantage compared to brute-force Grover search. However at this time it is not clear whether this is indeed the case, and we do not have large-scale quantum computers to experiment with.

The quantum security parameter (based on our assumptions of using state-of-the-art algorithms and fault-tolerance methods) for symmetric and hash-based cryptographic schemes is summarized in Table~\ref{tbl1}. For more details about space/time tradeoffs achievable via parallelization of Grover's algorithm please see the corresponding Sec.~\ref{sct::ciphers}, Sec.~\ref{sct::hash} and Sec.~\ref{sct::bitcoin}, respectively.
\begin{table}[h!]
\begin{tabular}{ll}
\hline
Name    & qs  \\
\hline
AES-128 & 106 \\
AES-192 & 139 \\
AES-256 & 172 \\
\hline
SHA-256 & 166 \\
SHA3-256	 &167 \\
Bitcoin's PoW & 75\\
\hline
\end{tabular}
\caption{Quantum security parameter ($qs$) for the AES family of ciphers, SHA family of hash functions, and Bitcoin, assuming a conservative physical error rate per gate $p_g=10^{-4}$.}
\label{tbl1}
\end{table}

We also analyzed the security of asymmetric (public-key) cryptography, in particular RSA and ECC, in the light of new improvements in fault-tolerant 
quantum error correction based on surface code lattice surgery techniques. We computed the space/time tradeoff required to attack 
every scheme, using physical error rates of $10^{-3}$ and $10^{-5}$, respectively. We fitted the data with a third degree polynomial, which resulted in an analytical formula of the number of qubits required to break the 
scheme as a function of time.

The total number of physical qubits required to break the RSA schemes in 24 hours, together with the required number of $T$ gates, corresponding number of surface code cycles and corresponding classical security parameter is summarized in Table~\ref{tbl2}. For more details about possible space/time tradeoffs please see the corresponding Section~\ref{sct::rsa} of the manuscript.
\begin{table}[]
\begin{tabular}{lllll}
\hline
Name      & nq                    & Tc                    & scc           & s        \\
\hline
RSA-1024  & $3.01 \times 10^7$    & $3.01 \times 10^{11}$ & $5.86 \times 10^{13}$ & 80\\
RSA-2048  & $1.72 \times 10^8$    & $2.41 \times 10^{12}$ & $4.69 \times 10^{14}$ & 112\\
RSA-3072  & $6.41 \times 10^8$    & $8.12 \times 10^{12}$ & $1.58 \times 10^{15}$ & 128\\
RSA-4096  & $1.18 \times 10^9$    & $1.92 \times 10^{13}$ & $3.75 \times 10^{15}$ & 156\\
RSA-7680  & $7.70 \times 10^{10}$ & $1.27 \times 10^{14}$ & $2.64 \times 10^{16}$ & 192\\
RSA-15360 & $4.85 \times 10^{12}$ & $1.01 \times 10^{15}$ & $2.24 \times 10^{17}$ & 256\\
\hline
\end{tabular}
\caption{The total physical footprint ($nq$) required to break the RSA schemes in 24 hours, together with the required number of $T$ gates ($Tc$), the corresponding number of surface code cycles ($scc$), and the corresponding classical security parameter ($s$).
We assume a very conservative physical error rate per gate $p_g=10^{-3}$, more likely to be achievable by the first generations of fault-tolerant quantum computers.}
\label{tbl2}
\end{table}

The total number of physical qubits required to break the ECC schemes in 24 hours, together with the required number of $T$ gates, corresponding number of surface code cycles and corresponding classical security parameter is summarized in in Table~\ref{tbl3}. For more details about possible space/time tradeoffs please see the corresponding Section~\ref{sct::ecc} of the manuscript. As observed before in~\cite{1706.06752}, breaking RSA schemes demands more quantum resources in comparison with elliptic curve-based schemes, for the same level of classical security.
\begin{table}[]
\begin{tabular}{lllll}
\hline
Name  & nq                 & Tc                    & scc          & s          \\
\hline
P-160 & $1.81 \times 10^7$ & $2.08 \times 10^{11}$ & $4.05 \times 10^{13}$ & 80\\
P-192 & $3.37 \times 10^7$ & $3.71 \times 10^{11}$ & $7.23 \times 10^{13}$ & 96\\
P-224 & $4.91 \times 10^7$ & $5.90 \times 10^{11}$ & $1.15 \times 10^{14}$ & 112\\
P-256 & $6.77 \times 10^7$ & $8.82 \times 10^{11}$ & $1.72 \times 10^{14}$ & 128\\
P-384 & $2.27 \times 10^8$ & $3.16 \times 10^{12}$ & $6.17 \times 10^{14}$ & 192\\
P-521 & $6.06 \times 10^8$ & $7.92 \times 10^{12}$ & $1.56 \times 10^{15}$ & 260\\
\hline
\end{tabular}
\caption{The total physical footprint ($nq$) required to break the ECC schemes in 24 hours, together with the required number of $T$ gates ($Tc$), the corresponding number of surface code cycles ($scc$), and the corresponding classical security parameter ($s$). We assume a very conservative physical error rate per gate $p_g=10^{-3}$, more likely to be achievable by the first generations of fault-tolerant quantum computers.}
\label{tbl3}
\end{table}

Recent developments in the theory of fault-tolerant quantum error correction have great impact on evaluating the effective strength of cryptographic
schemes against quantum attacks, as the fault-tolerant layer of a quantum computation is the most resource-intensive part of running a quantum  
algorithm. Therefore, monitoring the advances in the theory of quantum error correction is of crucial importance when estimating the strength (or 
weakness) of a cryptographic scheme against a quantum adversary. This work serves as a benchmark against which the impact of future advances can be compared.

\begin{acknowledgments} 
Most of this work is based on research supported by the Global Risk Institute for its members.
We also acknowledge support from NSERC and CIFAR. IQC and the Perimeter Institute are supported in part by the 
Government of Canada and the Province of Ontario.  Vlad Gheorghiu thanks Austin Fowler for helpful discussions 
and clarifications regarding lattice surgery methods.
\end{acknowledgments}

\bibliographystyle{aipnum4-1} 

%

\end{document}